# A self-heating electrochemical cell with nine decades of programmable linear resistance

*Adam L. Gross[1†], Sangheon Oh[1†], Minseong Park[1,2], T. Patrick Xiao[3], François Léonard[1], Wyatt Hodges[3], Joshua D. Sugar[1], Jacklyn Zhu[1], Sritharini Radhakrishnan[1], Sangyong Lee[4], Jolie Wang[1], Adam S. Christensen[3], Sam Lilak[3], Patrick S. Finnegan[3], Patrick Crandall[1], Christopher H. Bennett[3], William Wahby[3], Robin Jacobs-Gedrim[3], Matthew J. Marinella[5], Yiyang Li[4], Su-in Yi[2], Nad Gilbert[3], Sapan Agarwal[1], A. Alec Talin[1]* and Elliot J. Fuller[1]**

[1]Sandia National Laboratories, 7011 East Ave. Livermore, CA 94550, US

[2]Department of Electrical and Computer Engineering, Texas A&M University, College Station, TX 3127, US

[3]Sandia National Laboratories, 1515 Eubank SE. Albuquerque, NM 87185, US

[4]Materials Science and Engineering, University of Michigan, Ann Arbor, MI, US

[5]Electrical, Computer and Energy Engineering, Arizona State University, Tempe, AZ, US

†These authors contributed equally

*Corresponding authors: ejfull@sandia.gov, aatalin@sandia.gov

## ABSTRACT

A programmable linear resistor with a compact footprint would have profound implications for microelectronics, enabling efficient in-sensor analog signal processing and in-memory computing. Non-volatile memory offers a potential solution but suffers from limitations due to the programming mechanisms that confine switching to nanoscale constrictions or field-sensitive semiconductor junctions, leading to non-linear current-voltage relationships and errors. Here, we introduce a tunable resistor that is programmed into non-volatile, high-precision resistance states spanning nine orders of magnitude, with linear current-voltage characteristics across the entire range – significantly improving the performance and widening the application space of resistive memory. A key advance is an electrothermal gate that simultaneously spreads heat and electrochemical reactions during programming to enable large, bulk composition modulation. The volumetric modulation can host thousands of linear resistance states with 100× lower conductance errors than other memory. This enables direct processing of analog signals with high fidelity, and we demonstrate variable-gain amplification, division, and multiplication. Integration with CMOS is used to show resilience to electrical and thermal disturb in arrays and to demonstrate retention of analog levels at <1% average loss for more than 2 months across 100 devices. Simulations indicate matrix multiplication efficiency could approach >1,000 TOPS/W.

## INTRODUCTION

Analog signal processing has seen a resurgence due to increased demand for energy-efficient edge processing, in-sensor processing, and artificial intelligence (AI)[1-8]. A linear resistor is one of the fundamental building blocks of analog electronics and is responsible for linear signal transformations through Ohm's law ($V = I \times R$). For example, linear resistors are used to trim current sources in integrated circuits, serve as feedback in amplifiers and sensors, and are used in analog filters. Linear resistors used for analog processing are mostly considered static

components – in integrated circuits, doped silicon channels or thin films are fabricated near the required resistance values and then laser trimmed to meet tolerances. However, the ability to change the resistance after fabrication is highly desirable. For example, compact tunable resistors could be closely integrated with any sensor that routinely encounters signals that span a large dynamic range – e.g., LiDAR[9-12], phased-array systems[13-16], and ultrasound imaging[17-19] – to efficiently implement otherwise cumbersome variable-gain amplifiers or to reduce the large areal and energy overhead of high-resolution analog-to-digital conversions. Further, analog AI inference engines also require tunable resistors which serve as weights in neural networks and hold significant promise to improve energy efficiency relative to their digital counterparts[20]. However, using multiple resistors to implement a variable resistor (e.g., a digital potentiometer) incurs severe limits for these applications: large footprints, small bandwidths, low precision and significantly reduced efficiency.

Non-volatile resistive memory presents a potential solution as it is both compact and reprogrammable with low circuit complexity. Despite this, most resistive memory can only execute linear transformations accurately over small ranges in voltage (< 200 mV) and at large conductance (typically >100 μS) which constrains efficiency and limits applications. This is because physical programming mechanisms confine switching to field-sensitive semiconductor junctions (i.e., flash), nanoscale constrictions (i.e., memristors, conductive bridge memory) or exhibit highly non-equilibrium dynamics (i.e., phase change memory)[21-23] which introduces non-linearity and errors. For example, with flash memory, the low conductance regime is in the sub-threshold region of a field-effect transistor where current has an exponential dependence on the source-drain voltage[24]. One consequence of memory non-linearities is that AI inference engines can only accept binarized voltage inputs that must be processed serially to achieve high accuracy which introduces latency[23,25]. Altogether, these effects lead distortions and errors that reduce accuracy and reliability which severely constrain analog processing applications.

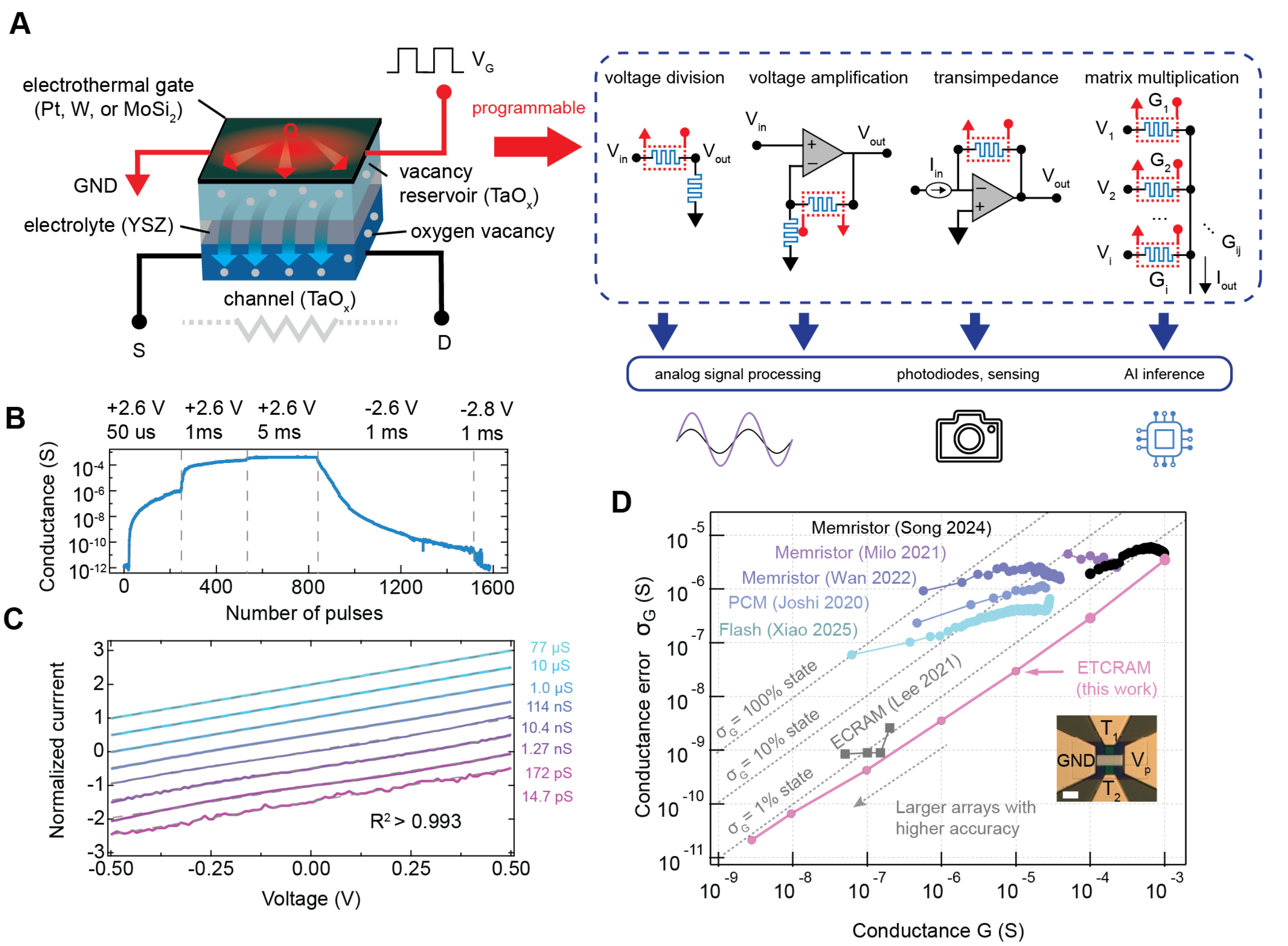

**Fig. 1. Electro-thermo-chemical random-access memory (ETCRAM) device operation, functional circuits, and comparisons with existing non-volatile memory. (A)** A schematic of the ETCRAM device with a $TaO_x$ channel layer that acts as a tunable resistor (blue), YSZ electrolyte (gray), a $TaO_x$ vacancy reservoir (light blue), and an electrothermal gate (dark gray). Depending on the device size, different materials such as Pt, W, or $MoSi_2$ are used to achieve the required programming temperature (300-400 °C). Q indicates heat spreading. Highlighted Box: Circuit diagrams of analog signal processing with ETCRAM. The ETCRAM programming electrodes are indicated in red, and the resistance terminals are indicated in black. Amplifiers are indicated in gray. (**B**) Conductance modulation of the device pictured in the optical image (inset of Fig. 1D). Programming pulses with different durations and voltages were applied based upon the overall conductance range and are indicated with black text and separated by gray lines. (**C**) Current-voltage (*I*-*V*) characteristic curves of the channel layer after programming a device to different conductance values ranging from about 10 pS to 100 µS. Current is normalized to values at 0.5 V for all curves to compare linearity and are offset vertically from zero current for clarity. A linear dashed line is shown to guide the eye. Over the range shown, the squared Pearson coefficients are > 0.993 from linear. (**D**) Comparison of combined programming and read error between ETCRAM and other memory as reported in the literature.[26-31] The scale bar in the inset is 10 µm.

To overcome these limitations, we present a device we call electro-thermo-chemical random-access memory (ETCRAM) which is depicted schematically in Fig. 1A. ETCRAM is a hybrid between electrochemical random-access memory (ECRAM)[26,32-37] and PCM, combining the electric potential driven programming of ECRAM with the thermal programming of PCM to yield analog properties that are inaccessible by either approach. Filamentary memristors also combine electric and thermal activation for programming, but these devices often suffer from excessive noise and stochasticity due to the unpredictive nature of the filament formation mechanisms. ETCRAM solves this issue by sufficiently decoupling the thermal and electronic degrees of freedom, enabling bulk (non-filamentary) modulation of the conductance. First, the self-heating overcomes kinetic barriers to enable $\times 10^9$ reversible tunability, about 5-6 orders of magnitude greater than ECRAM. When this is combined with bulk programming, it enables 100$\times$ lower conductance error than non-volatile memories such as memristors, flash, or PCM. A key component of ETCRAM is an electrothermal that gate acts simultaneously as a heater, heat spreader, and source of electrochemical currents and voltages. In the schematic in Fig. 1A, the functional layers are shown: a $TaO_x$ channel that functions as a tunable resistor (dark blue), a yttria-stabilized zirconia (YSZ) electrolyte layer for oxygen vacancy transport (gray), and a $TaO_x$ reservoir layer to store unused vacancies (light blue). Here, the electrothermal gate (dark gray) has three critical features. First, the application of voltage to the $V_G$ terminal passes both Joule heating current across the gate to a ground terminal and electrochemical current vertically through the electrolyte to the channel to change its resistance. This enables simultaneous heating and electrochemistry from a single pulse. Second, the electrothermal gate impedance and thermal conductivity are engineered through material selection (Pt, W, or $MoSi_2$) and layer thickness (10-100 nm) to target ideal programming temperatures for reaction kinetics via resistive heating (300-400 °C).

In contrast, oxygen-based ECRAM requires furnaces[26,32,37,38] or externally coupled microcircuits for heating[34,35,39], while memristors or PCM generally do not have any controllable temperature due to thermal runaway. For example, graphene heaters have been coupled to ECRAM to accelerate AI training and demonstrated retention in ambient for one hour with a tunability of about 2×. In another example, ECRAM devices were investigated for AI training based on protons in $WO_x$ with a low activation energy and therefore programmed quickly at room temperature and demonstrated ambient retention for a few minutes[40]. Here, rather than address AI training, ETCRAM is designed for programmable processing tasks where non-volatility, precision, and dynamic range are desired and programming efficiency is not critical. Therefore, rather than protons, we use oxygen vacancies in $TaO_x$ due to their high activation energy for vacancy movement of roughly 1.5 eV[41]. Coupling $TaO_x$ to self-heating enables orders of magnitude improvement in retention (e.g., >2 months in ambient) compared with ECRAM AI training cells. In addition, we resolve the integration challenges associated with external heating circuits by using an existing layer as a heat source rather than using additional layers and circuitry. In ETCRAM, the self-heating source is in intimate contact with the electrochemical layers which improves efficiency and simulations indicate this approach has potential to scale to 100 nanometer dimensions. Further, we integrate ETCRAM with CMOS using an interposer and demonstrate a shared backplane between devices which reduces signal wires from five per device[34] down to three – a critical step for realizing memory arrays with high areal density. Due to its long retention, low circuit complexity, and high linearity, we demonstrate that ETCRAM can execute high-fidelity programmable signal processing and has strong potential to achieve record-efficiency in AI inference applications.

## RESULTS AND DISCUSSION

### Dynamic range, read linearity, and bulk programming

ETCRAM increases the analog dynamic range of non-volatile memory devices by orders of magnitude. Fig. 1B is a plot of programming where voltage pulses $V_G$ are applied to the electrothermal gate, and the channel conductance is read out across source-drain terminals labeled S and D. Here, pulses were chosen to traverse a $0.7\times10^9$ dynamic range of conductance from $1.0\times10^{-12}$ S to $0.7\times10^{-3}$ S using approximately 800 pulses in each direction. With the application of positive write voltages, oxygen vacancies are driven into the channel and out of the reservoir and the electronic conductance of the channel $G$ is increased. Oppositely, for negative write voltages, the vacancies are driven from the channel layer into the reservoir and $G$ is decreased. Intermediate regimes can be traversed with smaller voltages and much shorter durations (e.g., down to 15 ns), enabling thousands of programmable levels. For example, to enable pulsed programming of an ETCRAM device over a more confined 10× range between $G$ = 1 nS - 10 nS, smaller voltages $V_G$ = ±1.75 V and shorter duration pulses (2 μs down to 50 ns) can be used (Supplementary Note 3 and Fig. S5). Further, we found devices to endure > $10^5$ read-write operations without any read or write failure observed (Supplementary Note 5 and Fig. S7).

In addition to a large dynamic range, ETCRAM devices were found to have linear current-voltage relationships between S and D when programmed across the full conductance dynamic range, as defined by a linear fit with an $R^2$ coefficient >0.99. In Fig. 1C top panel, channel layer $I$-$V$ curves are plotted for the lower eight decades of conductance, with line fits indicated by a dashed line. Pure Ohmic current-voltage relationships were found up to ±2 V between 1 mS and 1 μS, and a linear relationship with minor deviations was observed up to ±500 mV between 100 nS to 100 pS, with pure Ohmic behavior returning at 10 pS. For more details of fitting and linearity see Supplementary Note 2. Other memory such as flash and memristors are typically linear only above 100 μS and up to ±200 mV[21,22], after which the nonlinearity is sufficient to introduce significant distortions. It should be emphasized that both PCM and memristors typically cannot support input voltages up to 2 V due to shared paths for write and read, causing higher voltage inputs to unavoidably alter the device conductance. ETCRAM does not have this limitation and preserves the device state up to 2V due to separated paths for write and read. Additionally, electrostatic memory (e.g., flash), generally cannot support linear *I-V* relationships at low conductance values due sub-threshold operating characteristics.

ETCRAM not only extends the dynamic range, but it also increases precision by at least 100× relative to other memories. To characterize the precision limits and number of states available in ETCRAM, we measured the analog conductance error $\sigma_G$ for a device across six decades of conductance. We define the conductance error $\sigma_G$ as the total standard deviation associated with resetting and writing to a target state 10 times and then subsequently reading the target state 100 times for each conductance. Therefore, $\sigma_G$ contains errors from both read and write operations (i.e., due to noise, short term drift, etc.). In Fig. 1D, $\sigma_G$ is plotted as a function of the mean conductance value $G$ and contrasted across multiple technologies. To guide the eye, dashed lines indicating 100%, 10%, and 1% error are indicated. Over the range in Fig. 1D, the ETCRAM device (black) has significantly lower $\sigma_G$ (<1%) and higher precision. For example, the ETCRAM device has a lower $\sigma_G$ by 440× compared to memristors[29], 200× compared to flash[27], and 130× compared to PCM[28] for $G$ < 1 μS (Supplementary Note 7 and Supplementary Table 2). At higher conductance ($G$ > 10 μS), the ETCRAM $\sigma_G$ is consistently lower, but converges to memristor values at $G$ = 1.0 mS. Our reported ETCRAM $\sigma_G$ baseline is limited by write noise as indicated by power spectral density measurements of read noise which sit well below this baseline (see Supplementary Note 6 and Fig. S8). A simple state density analysis by division of the range by the conductance error $\sigma_G$ yields 3,180 distinguishable analog levels from 1 nS – 1 mS. Notably, the lower left portion of Fig. 1D represents regions that can support both higher precision and larger arrays (due to lower current draw). As we discuss in detail later, larger arrays of devices with high-bit precision have fewer digital overheads per operation than multiple small arrays and are therefore critical to improving the efficiency of analog accelerators.

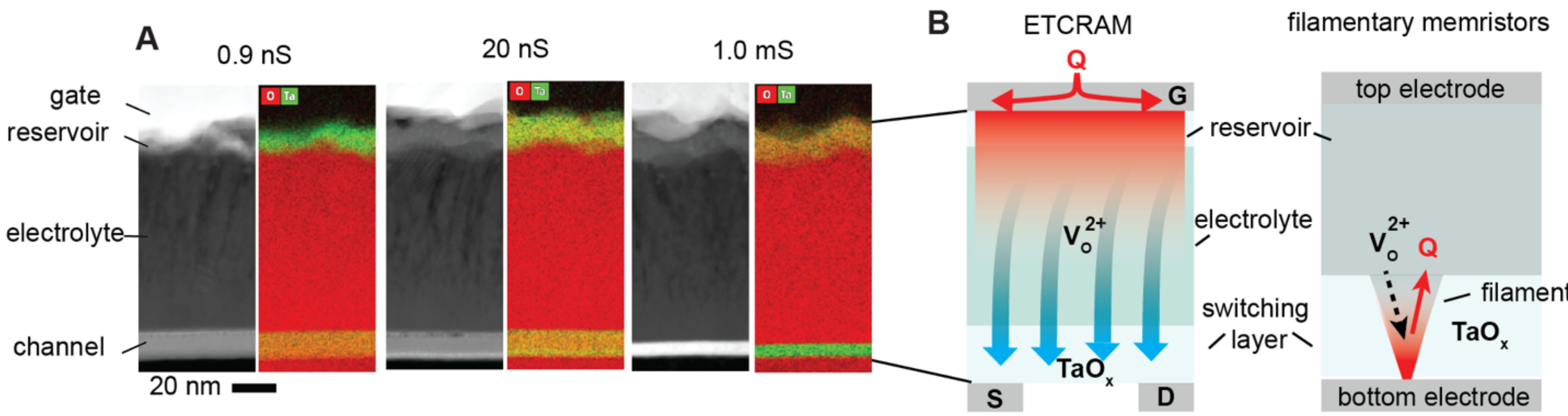

**Fig. 2. Bulk programming with ETCRAM.** **(A)** TEM cross-sections of three ETCRAM devices in three different programmed states, with corresponding EDS maps showing the distribution of Ta and O in the separate layers. Oxygen counts have been colored with red pixels and Ta counts with green. **(B)** Schematics depicting the working mechanisms of ETCRAM devices versus filamentary memristors. In the left ETCRAM schematic, bulk oxygen vacancy modulation is assisted by uniform reactions that occur across the electrolyte interface and uniform heat distribution *Q* due to heat spreading during programming. Comparatively, filamentary memristors (depicted in the right schematic) have a concentration of heat and vacancies around a filament.

We used transmission electron microscopy (TEM) to reveal the physical processes responsible for the dynamic range and precision of ETCRAM devices. TEM images combined with elemental composition (energy-dispersive X-ray spectroscopy) were taken of devices programmed into several different states (Fig. 2A). The differences in the channel composition clearly articulate the bulk nature of the device: the least conductive device channel ($G$ = 0.9 nS) is nearly completely oxidized, with a suboxide mix of Ta and O in the channel, and in contrast the most conductive device channel ($G$ = 1.0 mS) is nearly metallic, with little O content seen in the channel. In addition to changes observed in the channel, we find the trends are reversed in the reservoir: the reservoir becomes more O-deficient when the channel is oxidized. Intermediate regimes, for example at 20 nS, also follow this trend (Fig. 2A, center panel). A schematic comparing programming mechanisms between ETCRAM and memristors is depicted in the schematic in Fig. 2B. With memristors (Fig. 2B right panel), the electrically conductive path from the reservoir layer to the channel layer leads to thermal runaway and the concentration of heat and oxygen vacancies; it has been proposed that this limits the ultimate dynamic range and precision. For example, at low conductance values, a nearly one-dimensional filament forms which is highly sensitive to atomic motions of a few atoms or charge traps (e.g., noise sources)[42]. In contrast, ETCRAM promotes uniform, bulk reactions from the electrolyte surface due to efficient heat spreading and the prevention of dielectric breakdown by the electrically insulating electrolyte.

**ETCRAM in analog signal processing**

Due to its programming precision, linearity, and nonvolatility, ETCRAM can be widely deployed as a programmable analog signal processing element. To demonstrate this, we integrated ETCRAM into fundamental building blocks of analog electronics: voltage dividers, voltage amplifiers, transimpedance amplifiers, and matrix multipliers. An overview of the circuits we tested are shown in the highlighted schematic in Fig. 1A (blue box). First, we discuss the most basic example of voltage division. An ETCRAM device was integrated into a divider, a 1 V sine wave applied to the input, and the output voltage was measured. The programming terminals were grounded between programming steps and during signal analysis. Sine waves were chosen to simplify harmonic analysis and $f$ = 100 Hz was chosen to most clearly quantify higher frequency harmonics introduced by nonlinearities. The divided voltage output was measured after programming the device to various values spanning division factors from 1 - 100×, and the results are plotted in Fig. 3A. The plots demonstrate that finely spaced analog values of division can be programmed across a wide range (100×) and process voltages up to 1V without distortion. Next, variable-gain amplification was demonstrated with the amplified voltage output plotted in Fig. 3B. Here, a 1 mV amplitude sine wave ($f$ = 100 Hz) is amplified by 1 - 2,000× by repeatedly programming ETCRAM and measuring the output of the amplifier. Signals with much higher frequency components were also processed with high fidelity (See Supplementary Note 10). Analysis of the total harmonic distortion (THD) is presented in Fig. 3C for the pure tone sine wave.

After amplification, the THD relative to the fundamental frequency ranged between 1.7% – 3.3% for gains between 10 dB – 66 dB, which was primarily due to circuit parasitic elements other than ETCRAM (in this case, breadboard capacitance). However, this THD is within the tolerances of most visual and sensor signal processing[43] and provides an upper bound on the total THD introduced by ETCRAM. Similarly, transimpedance amplification and voltage encoded dot products were demonstrated with high accuracy and without significant harmonic distortions (see Supplementary Notes 9-12).

Altogether, these circuit demonstrations constitute a substantial improvement over programmable signal processing based upon memristors or CMOS. For example, memristors introduce frequency-dependent gain and signal distortions due to non-linearity[44]. When comparing to CMOS approaches such as digital potentiometers, we consider the ETCRAM improvements to area, precision and parasitic capacitance. The devices in Figs. 3A-C had an active area of 2 μm × 2 μm, with a capacitance to the gate estimated at 10 fF based upon the thickness of the YSZ dielectric. Comparatively, digital potentiometers have millimeter dimensions with parasitic capacitance >1 pF[45]. Similarly, on-chip approaches, such as ladders of resistors coupled directly to amplifiers, suffer from volatility and similarly large areas that grow as the number of discretized values or resistors are increased[46]. In contrast, ETCRAM is non-volatile and can be programmed with high precision over finely spaced levels. Due to its small footprint, ETCRAM can be directly integrated at the pixel level for imaging technologies such as LIDAR. Further downscaling in dimensions should be possible for denser sensor arrays, and we discuss this in detail later.

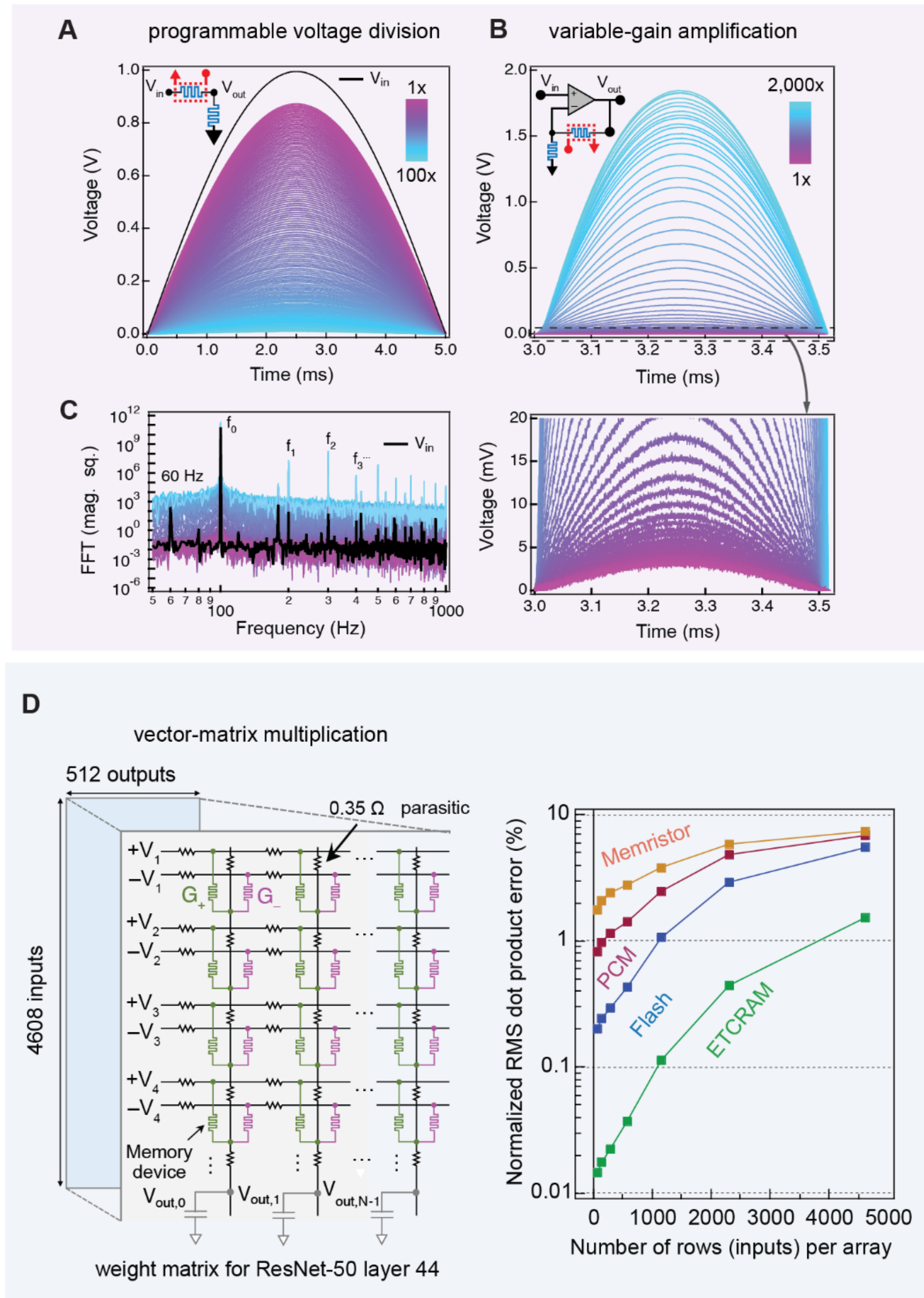

**Fig. 3. Analog signal processing with ETCRAM.** (**A)** An example of programmable voltage division of a 100 Hz sine wave is plotted for divisions of 1 - 100x using the circuit in panel A. (**B)** Variable-gain voltage amplification of a 1 mV amplitude sine wave is shown for gains of 1 - 2000× using ETCRAM integrated with an operational amplifier. The lower panel shows <20 mV amplified signals. **(C)** The power spectral density of the voltage amplification from panel b with fundamental frequency $f_0$ and harmonics $f_1$, $f_2$, $f_3$+... labeled in black text. **(D)** Top: A schematic of matrix vector multiplication using ETCRAM in a crossbar array. Bottom: Simulated error versus array size for different memory devices when processing analog MVMs,

using weights and inputs from one layer of ResNet-50. The simulation includes a parasitic interconnect resistance of 0.35 Ω[47] and the error is separately normalized to the signal range for each array size (Supplementary Note 14 and Fig. S17). For all memories other than ETCRAM, we assumed bit-serial application of digital inputs to avoid signal distortions due to *I-V* nonlinearity.

Next, we used architectural simulations to study how ETCRAM would scale to large arrays for matrix-vector multiplication (MVM). Using the array schematic depicted in Fig. 3D, we compared ETCRAM to different technologies with increasing array size. The simulations use a weight matrix from one layer of the ResNet-50 convolutional neural network trained on the ImageNet dataset[48] and model the published conductance ranges and errors for devices shown in Fig. 1D, including the effect of parasitic *IR* drops along the array interconnects. ETCRAM achieves the best performance in both small and large arrays due to its low errors and low conductance, which minimizes *IR* drop errors. The ability to scale to large arrays is critical for energy efficiency, since the output peripheral circuits (e.g., operational amplifiers and analog-to-digital converters) dominate the energy consumption of analog systems[49-51], and large arrays can amortize these costs across more operations. With the same MVM error, ETCRAM can scale to a larger array than flash[52], and a significantly larger array than PCM[28] or memristors[47]. The utility of ETCRAM in MVM is further enabled by linear current-voltage characteristics. Voltage-mode MVMs have been previously implemented with memristors[47], but had limited accuracy due to nonlinear *I-V* characteristics over multiple decades of conductance. Due to ETCRAM's unique linearity, we found that arrays can process large MVMs with an efficiency of 1,360 TOPS/W with 8-bit peripheral circuits when the technology is scaled to a moderately advanced node, e.g. 22 nm (Supplementary Note 17). This efficiency is 10-100× higher than in other non-volatile memories at a similar computational precision.[20] Beyond energy efficiency, the precision of ETCRAM enables it to achieve software-equivalent accuracy without any hardware-aware training on complex, commercially relevant AI workloads such as COCO object detection (ETCRAM accuracy was 0.08% less than ideal), where all other technologies in Fig. 3D lose significant accuracy compared to digital (i.e., flash accuracy was 5.39% less than ideal). More details can be found in Supplementary Note 20.

**Operating limits and integration with CMOS**

ETCRAM can be programmed with fewer operations compared to state-of-the-art memory, with nanosecond speeds, long retention, and with power efficiency that scales favorably for tasks such as programmable inference. First, we demonstrate that only a few programming steps are required to reach critical precision thresholds. As shown in Fig. 4A, devices can be accurately tuned to within 0.6% error within six write cycles using closed loop programming. This can work both within a decade of conductance (left panels) and between decades of conductance (right panels). Notably, memristors have recently demonstrated high precision (limited to >50 μS), but required sophisticated modeling, five cycles of coarse tuning, additional fine tuning steps, and modeling with external processing[22]. When utilizing fast pulsing electronics, ETCRAM programming speeds can be sub-100 ns (see Supplementary Note 15). In Fig. 4B, programming with 45 ns pulse widths is shown, with decreasing modulation strength as 15 ns is approached.

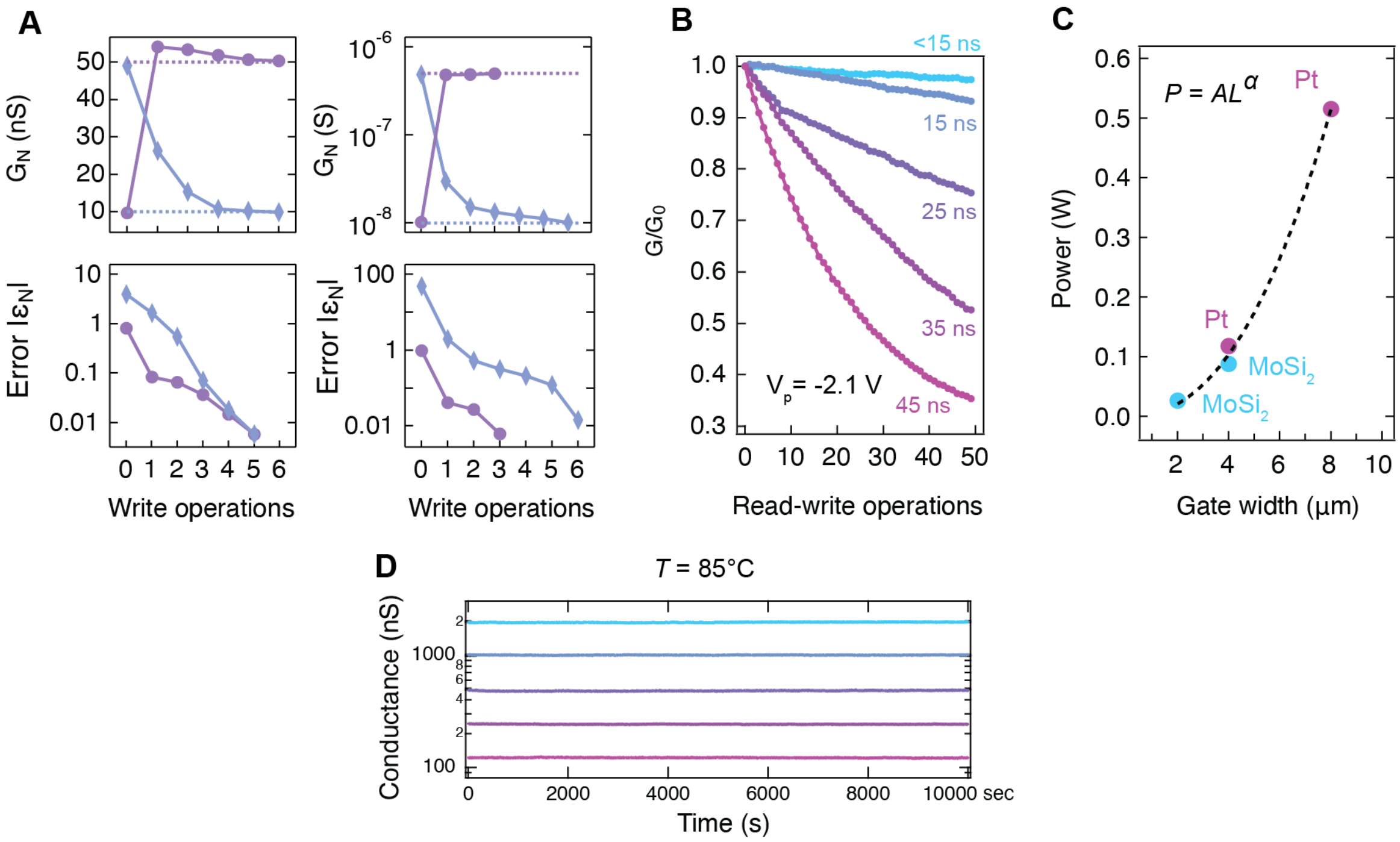

**Fig. 4. ETCRAM state selection, speed, and power consumption during operation. (A)** Closed loop programming between target values $G$ = 50 nS and 10 nS (left panels) and $G$ = 10 nS and 0.5 μS (right panels). The conductance value $G_N$ and error $\varepsilon_N$ are plotted against the number of write operations. Programming voltages varied between 2.0 – 2.2 V. **(B)** Nanosecond programming for a 4 μm device, applying constant $V_p$ = -2.1 V pulses with varying write pulse durations. **(C)** Power consumption of ETCRAM devices with different dimensions and electrothermal gate widths: $L$ = 8 μm (Pt), 4 μm (Pt and $MoSi_2$), and 2 μm ($MoSi_2$). Dashed line indicates power law fit ($P = AL^{\alpha}$) to data points. **(D)** Retention of a 2 μm ETCRAM device over the course of $10^4$ seconds, at 85°C.

Next, we discuss the power scaling in ETCRAM. The power required to program devices was studied as a function of device area. We found a temperature rise of $\Delta T_C$ = 300-400 °C was ideal for programming devices, as determined by using the Pt electrothermal gate resistance as a temperature indicator (Supplementary Note 18). A simple physical model to capture the power efficiency at micron-scale dimensions is $\Delta T=P/(GA)$, where $\Delta T$ is the temperature rise of the device, $P$ is the electrical power dissipated by the electrothermal gate, $G$ is the areal thermal conductance loss, and $A$ is the electrothermal gate area. The critical power $P_C$ then depends quadratically on gate width $L$, as in $P_C \propto L^2 G \Delta T_C$ to meet $\Delta T_C$. In Fig. 4C, the power required for programming ($\Delta T_C \sim 350$ °C) is plotted against feature size, indicating that reducing the lateral dimension by ½ leads to roughly ¼ the required power. Because both heat and voltage are required to drive electrochemistry, lowering the voltage to reduce power at smaller dimensions would inhibit electrochemical reactions. Therefore, to keep voltage fixed but lower the power, the electrothermal gate resistance must be engineered according to $P = V^2/R$. For example, in Fig. 4C, larger devices with $A = 8 \times 24$ μm$^2$ required Pt or W heaters ($R \sim$ 10-20 Ω) whereas smaller devices with $A = 2 \times 6$ μm$^2$ required a $MoSi_2$ heater ($R \sim 180$ Ω) to program at sufficient voltage (> 1.5 V) and temperature. A power law fit to the data yields an exponent of $\alpha = 2.31 \pm 0.154$, yielding a roughly quadratic dependence.

To assess power scaling of ETCRAM at nanometer dimensions, we performed finite element simulations of the steady-state temperature distributions in the devices (Supplementary Note 19). The simulations indicate that programming currents of around 160 μA and voltages of about 2 V (approximately 320 μW) can be achieved with appropriate selection of the heater thermal conductivity ($\kappa = 10$ W/mK), geometry ($A = 100 \times 100$ nm$^2$ device area), and electrical resistance ($R = 12.5$ kΩ). Such metrics are achievable with common materials: for example, polysilicon heaters have a thermal conductivity of approximately 14 W/mK and can be doped to achieve the required electrical resistance. Therefore, we anticipate that programming ETCRAM with sub-micron lateral dimensions will have similar power consumption to PCM cells, which also rely upon localized heating. Such power consumption can meet the requirements for tasks such as programmable gain or inference.

We next evaluate ETCRAM retention characteristics which shows significant improvements over memory used for AI training. As an example, Figure 4D is a plot of retention at 85 °C over the course of $10^4$ seconds (~2 hours) for an ETCRAM device programmed to levels between 100 nS and 2 μS. Over these timescales we find no observable memory loss within the error of the test setup. In fact, below 200°C, memory loss in ETCRAM was not observed at these durations. However, by increasing the temperature to 215 - 245°C small losses could be observed over approximately a 1-hour timeframe. Assuming an Arrhenius activated loss process, we obtained an activation energy of 1.59$\pm$0.35 eV (Supplementary Note 16 and Fig. S20). Such a large energy is consistent with previous measurements of oxygen vacancy migration in $TaO_x$[41] and suggests TaOx-based ETCRAM has potential for long analog retention times for 10 years at 85°C.

To characterize retention across more devices and for longer durations we integrated ETCRAM into CMOS using an interposer. One hundred ETCRAM devices were programmed across a 60 - 340 μS range. The range was selected based upon the resolution limits of the peripheral CMOS control. After two months in ambient conditions, the devices had an average decrease of 0.87% ± 1.6% of the programmed conductance (Supplementary Note 22 and Fig. S29), which is similar-to or better-than the retention characteristics of state-of-the-art memristors or PCM arrays[51,53,54]. We note that the temperature-accelerated retention measurements of single devices suggest we should not observe any memory loss during this period in ambient. We therefore we attribute the average decrease to either long term drift or environmental oxidation due to oxygen ingress. The average drop in conductance is consistent with ingress. Therefore, we propose that improvements to passivation could be used to mitigate such effects but would require efforts outside the scope of this work. We note the large standard deviation is due to conductance loss values that approach the noise limits of the CMOS control circuitry.

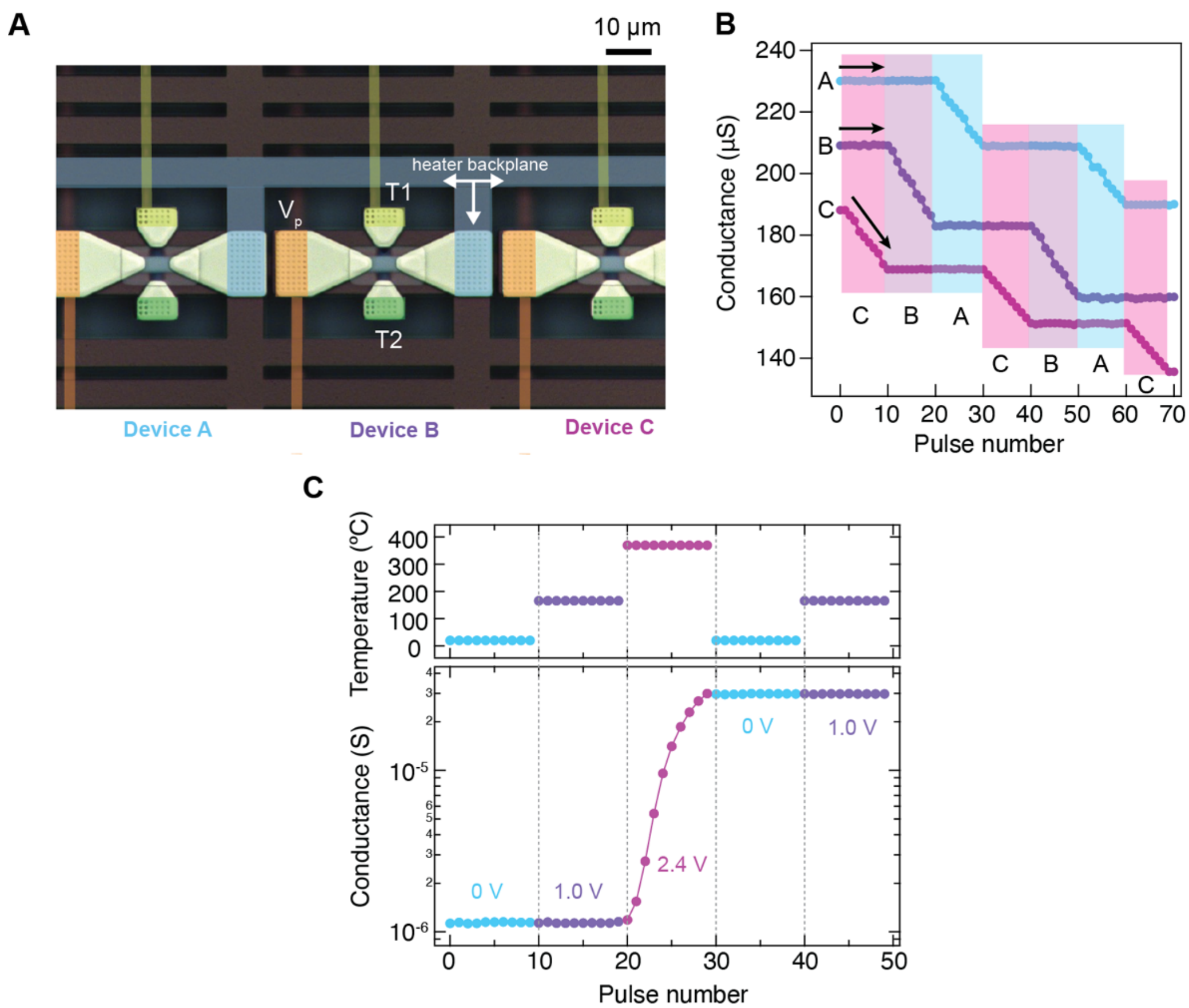

**Figure 5. Circuit considerations for CMOS integration of ETCRAM**. **(A)** An optical image of ETCRAM devices electrically connected through a global backplane (blue) with shading of readout terminals T1 (yellow), T2 (green), and programming terminal $V_p$ (orange). **(B)** Programming using the backplane without disturb of adjacent cells is shown, with device conductance A, B, and C plotted as dots and programming regions of each cell highlighted by the shaded regions. For programming devices, $V_p$ = -2.5 V, t = 200 ms pulses were used. **(C)** Measured conductance during thermal disturbance of a single device. Thermal disturbance is simulated by applying $V_p$ = 1.0 V to the heater prior to readout. The temperature scale is estimated based upon the measured temperature thresholds for programming ETCRAM in other devices.

Next, we consider ETCRAM integration into CMOS, including topics of circuit complexity and resilience to electrical and thermal disturb. To test this, our CMOS interposer was made using a backplane architecture to reduce circuit complexity, as depicted in the optical image in Fig. 5A. Here, the GND terminal from Fig. 1A is shared across every cell as a global backplane (blue) and does not require a dedicated signal line for each device. We note this does not constitute a thermal backplane[35] as joule heating in ETCRAM is still highly localized. Our backplane reduces the number of signal lines per device from four to three in an array. Together with self-heating, this constitutes a total reduction from five lines to three when compared to externally heated circuits. Fig. 5B plots independent programming through the backplane which shows no electrical disturb between devices. For example, in the first 10 pulses device C is programed, but devices A and B are unaffected.

Lastly, we consider the situation of thermal disturb from neighboring cells which could potentially arise for highly scaled devices (<100 nm) and aggressively pitched memory arrays. To experimentally simulate this effect in a real device, we program a single ETCRAM below the voltage threshold for programming, but with enough temperature (~165°C) to simulate a 'disturb' event. In Fig. 5C, we observe no measurable effects from the sub-voltage-threshold programming at 1V with an estimated 165°C temperature rise (purple). To summarize, ETCRAM requires both heat and voltage to drive electrochemistry which makes it inherently robust to electrical and thermal disturbances.

## CONCLUSION

A programmable linear resistor opens new possibilities across many areas. For example, ETCRAM could replace precision on-chip resistors for low-voltage signal processing and eliminate laser trimming of patterned resistors; a cumbersome and costly manufacturing process[55]. ETCRAM may additionally find utility in dense sensor arrays, such as LIDAR, where programmable gain is desirable across a small, low capacitance footprint. Similarly, it could enable tunable filters for wireless communications that provide programmable frequency selectivity over a wide range[56]. Due to volumetric programming and separate programming terminals, ETCRAM can host applications with larger voltages (up to 2V) than memory elements such as PCM or memristors, which would self-program under these conditions. For analog AI inference, ETCRAM provides a striking improvement with 100× lower conductance noise in the critical low-conductance range, and our simulations indicate the potential for >1,000 TOPS/W in MVM efficiency. Such improvements could tip the value-proposition for analog systems to replace more traditional approaches such as graphics processing units. Beyond analog electronics, we have shown that an electrothermal gate enables engineerable, uniform, and reversible reactions over a wide composition range. The core scientific principles behind ETCRAM operation can be applied to other electrochemical systems and to modulate other physical properties: for example, dielectric response and thermal conductivity.

## METHODS

### Device fabrication

ECRAM devices were fabricated on 4" Si wafers with a 200 nm dry thermal $SiO_2$ oxide. A separate set of devices to demonstrate CMOS-compatibility and retention was fabricated on a custom-built interposer with a 1 um standoff $SiO_2$ thickness. Bilayer photoresist was employed for liftoff, using positive-tone liftoff resists (LOL2000, LOR5B) and top imaging layers (AZ5214, S1813). Patterning of all layers was performed with a Heidelberg MLA 150 maskless aligner. All layers were deposited using an AJA ATC 2200 sputter system with 150 W RF or DC power in 3 mTorr Ar. Typical layer thickness are 75 nm (gate metal), 10 nm ($TaO_x$ reservoir), 80 nm (YSZ), and 10 nm ($TaO_x$ channel). Contacts were fabricated using sputtered Ta-Au or Ta-Pt and are around 150 nm thick. Prior to testing, the final wafers were sonicated in NMP, diced, and stored in an Ar flow glovebox. To enable programming in air, a epoxy passivation layer (EPO-TEK 301) was applied to devices. Devices were not otherwise annealed prior to testing. More details can be found in Supplementary Note 1.

## Electrical measurements

For the signal processing demos and devices on CMOS, devices were passivated with epoxy and programmed in air. For all other data, electrical measurements were taken in vacuum with a base pressure of $5\times10^{-6}$ mTorr. A National Instruments DAQ and Keysight B1500 were used to source the read and write pulses, with standard read voltages of V = 0.2 V (NI DAQ) and V = 0.1 V (B1500). Custom integrated circuits were used to buffer signals between the DAQ and device (OPA-551 with 3 MHz bandwidth). For the nanosecond pulsing measurements a Keithley 4200A-SCS with a 4225 Pulse measure unit (PMU) and 4225 remote amplifiers (RPM) were used. Load line effect compensation (LLEC) is employed during programming in the case the resistance of the heater was smaller than the output impedance of the PMU.

## TEM measurements

Devices programmed to different conductance states were thinned to electron transparency using standard FIB/SEM techniques. A Thermo Fisher Helios 660 NanoLab DualBeam instrument was operated at 30 kV for trenching and thinning, with successively lower Ga ion beam currents used until the final thinning step, during which 5 kV was applied. STEM measurements were performed using a Thermo Fisher Titan Themis Z instrument operated at 300 kV. EDS and EELS measurements were conducted simultaneously, utilizing a large solid-angle SDD EDS detector (Super-X) and a Gatan 969 Quantum GIF equipped with a K2 CMOS detector. Typical beam currents ranged from 150 to 300 pA, and pixel times of 10 to 50 ms were used for data acquisition.

## ACKNOWLEDGMENTS

This work was supported by the Sandia Laboratory-Directed Research and Development (LDRD) Program. Sandia National Laboratories is a multimission laboratory managed and operated by National Technology and Engineering Solutions of Sandia, LLC., a wholly owned subsidiary of Honeywell International, Inc., for the US Department of Energy's National Nuclear Security Administration under contract DE-NA-0003525. The views expressed in the article do not necessarily represent the views of the US Department of Energy or the United States Government. Additionally, S.L. and Y.L. were supported by the Defense Advanced Project Research Agency under a Young Faculty Award grant no. D24AP00328.

## CONFLICT OF INTEREST

We, the authors, declare that US Patent 12,446,479 is related to this manuscript. We declare no other competing interests.

# Supplementary Information

## A self-heating electrochemical cell with nine decades of programmable linear resistance

*Adam L. Gross[1†], Sangheon Oh[1†], Minseong Park[1,2], T. Patrick Xiao[3], François Léonard[1], Wyatt Hodges [3], Joshua D. Sugar[1], Jacklyn Zhu[1], Sritharini Radhakrishnan[1], Sangyong Lee[4], Jolie Wang[1], Adam S. Christensen[3], Sam Lilak[3], Patrick S. Finnegan[3], Patrick Crandall[1], Christopher H. Bennett[3], William Wahby[3], Robin Jacobs-Gedrim[3], Matthew J. Marinella[5], Yiyang Li[4], Su-in Yi[2], Nad Gilbert[3], Sapan Agarwal[1], A. Alec Talin[1*] and Elliot J. Fuller[1*]*

[1]Sandia National Laboratories, 7011 East Ave. Livermore, CA 94550, US

[2]Department of Electrical and Computer Engineering, Texas A&M University, College Station, TX 3127, US

[3]Sandia National Laboratories, 1515 Eubank SE. Albuquerque, NM 87185, US

[4]Materials Science and Engineering, University of Michigan, Ann Arbor, MI, US

[5]Electrical, Computer and Energy Engineering, Arizona State University, Tempe, AZ, US

† Authors contributed equally

* Corresponding authors: ejfull@sandia.gov, aatalin@sandia.gov

## Supplementary Note 1

ETCRAM device design and fabrication

Several different ETCRAM designs have been employed in the present manuscript, with the main differences between them being the dimensions and the gate material. Below in Supplementary Fig. S1 are CAD schematics of the Pt gate based-ETCRAM devices at the *L* = 8 µm and *L* = 4 µm node sizes. Our convention to indicate device dimension or node size in the main text is to use the minimum channel/heater dimension (*L* = 2 µm, *L* = 4 µm, *L* = 8 µm, etc.). The devices consist of a thin Ta channel layer on the substrate, YSZ, Ta reservoir, and Pt on top of the reservoir. The gate has a 3:1 aspect ratio, and the active device area corresponds to the square region of the intersection of the channel and heater footprints. The reservoir is a thin Ta 10 nm layer underneath and parallel to the gate. Thin Au 10 nm source/drain contacts have been deposited on the bottom near edge of the active device area, and larger 610 nm-thick Au pads were deposited for all the contacts in the final step. The substrate is a 4" wafer of dry 200 nm $SiO_2$ on Si.

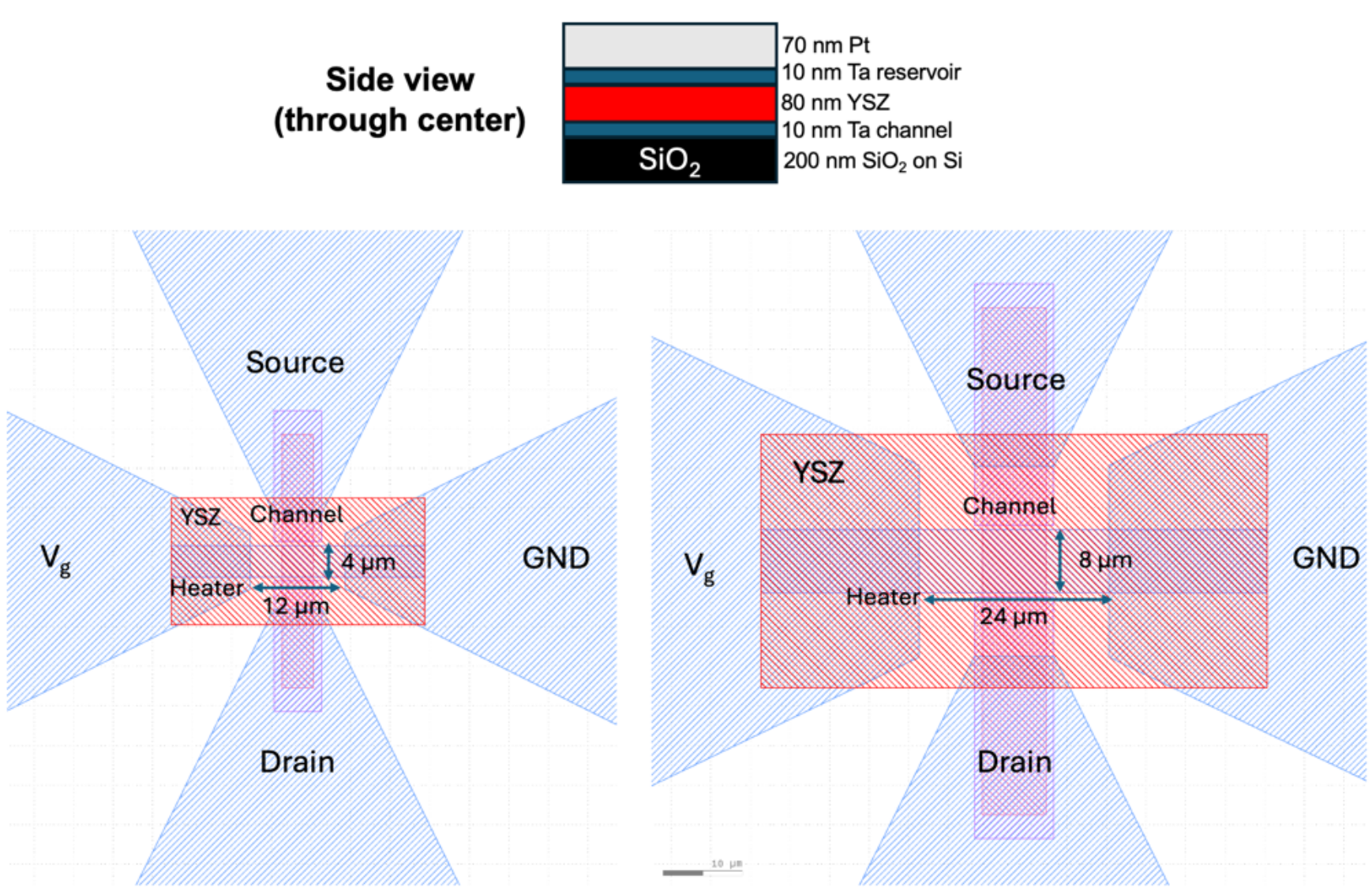

**4 µm Pt device**
Areas:
Heater/reservoir: 4 µm x 12 µm
YSZ: 16 µm x 32 µm
Channel: 4 µm x 5 µm

**8 µm Pt device**
Areas:
Heater/reservoir: 8 µm x 24 µm
YSZ: 32 µm x 64 µm
Channel: 8 µm x 9 µm

**Supplementary Figure 1.** CAD schematics of 4 μm and 8 μm ETCRAM devices with a Pt gate and larger YSZ. The area of the different layers area indicated below, excluding the portions beneath the contacts. A side view through the device center is provided, showing the thickness of each layer.

Another set of ETCRAM devices were fabricated with W (4 μm) and $MoSi_2$ (2 μm) gates, and the CAD schematics for these are shown in Supplementary Fig. S2. The W gate was used for some measurements due to the higher stability of W and similar impedance as to Pt. $MoSi_2$ was used for the 2 μm device due to its higher R ~ 180 Ω impedance to enable electrothermal programming at small dimensions. In addition, these devices contain a smaller area YSZ layer, which was observed to improve stability during programming. Supplementary Table 1 lists the different devices variants.

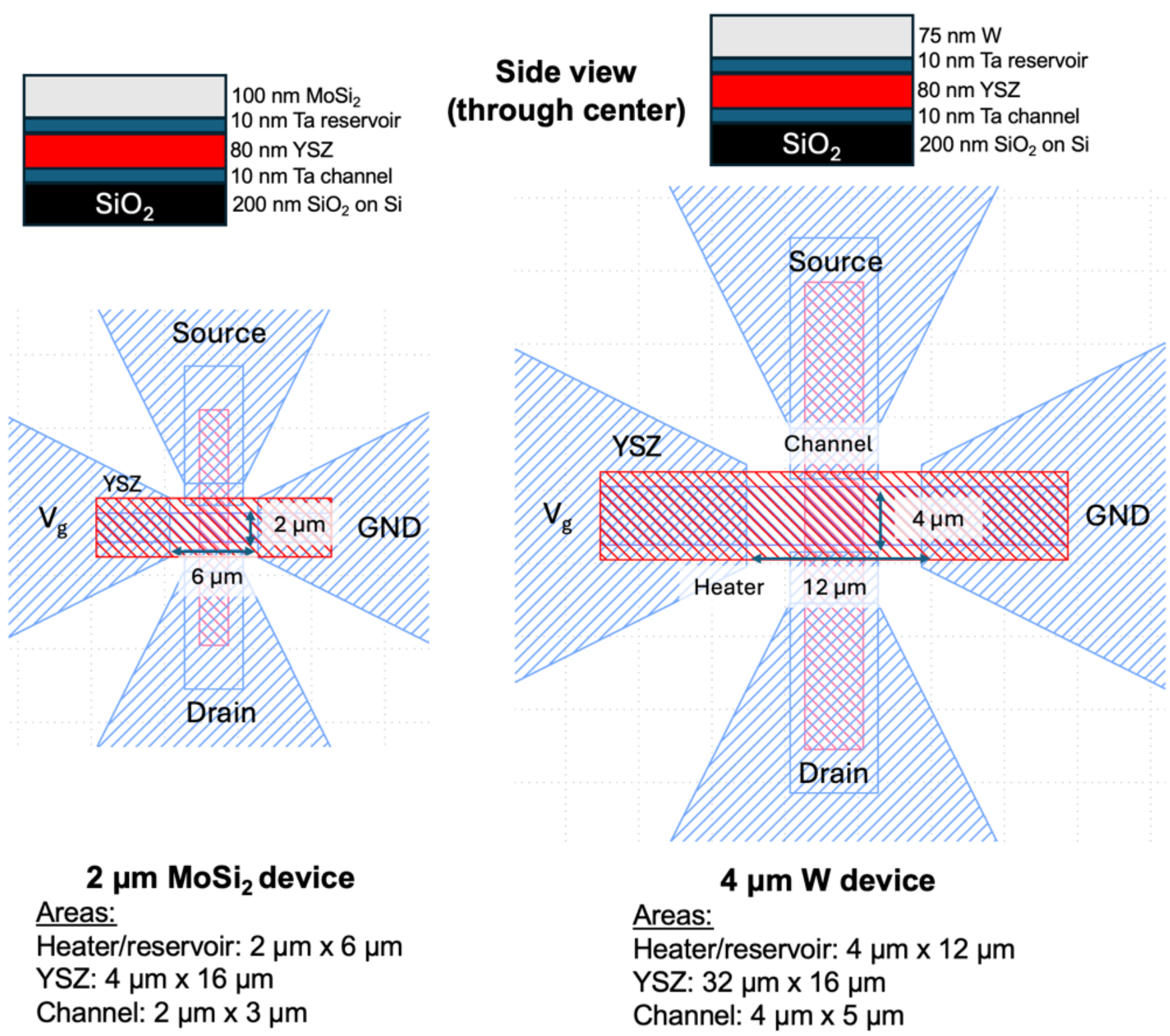

**Supplementary Figure 2.** CAD schematics of 2 μm and 4 μm ETCRAM devices with $MoSi_2$ and W gates, respectively. The YSZ area is reduced compared to the Pt ETCRAM devices, with the areas of the different layers indicated below, excluding the portions beneath the contacts. A side view through the device center is provided, showing the thickness of each layer.

| Wafer ID | Gate material | YSZ area (80 nm) | Heater Pads | S/D Pads |
|---|---|---|---|---|
| TaOx#10, #11 | Pt | Large | Ta-Au (sputter) | Ta-Au (sputter) |
| TaOx#13, #19 | $MoSi_2$ | Small | Ta-Pt (sputter) | Ta-Pt (sputter) |
| TaOx#15 | W | Small | W (underlayer) Au (e-beam) | Ta-Pt (sputter) |

**Supplementary Table 1.** List of the different ETCRAM wafers fabricated for this study

**Supplementary Note 2**

Full ETCRAM current-voltage characteristic curves up to 2 V

In the main text, Fig. 1c presents normalized ETCRAM channel I-V curves for states spanning G = 77 μS – 14.7 pS. Below in Figs. S3-S4 are the unnormalized ETCRAM channel I-V curves for different programmed states, showing the decrease in the magnitude of the current as the conductance decreases. The Keysight B1500A SMUs were used to program devices to target conductance spaced approximately 10x apart and to take the I-V curves. Overlaying each I-V curve are best fit lines and their corresponding $R^2$ coefficients in the interval ± 500 mV. The slopes of the best fit lines define the conductance reported in Fig. 1c. Ohmic current-voltage relationships were found to up to 2 V between 1 mS and 1 μS (Supplementary Note 3, Fig. S4). Small, correlated residuals appear near the end points in the IV curves between 1 μS and 100 pS but at less than 10% deviation from linear. The full range may have acceptable linearity up to 2 V for applications where as much as 150% deviation is acceptable.

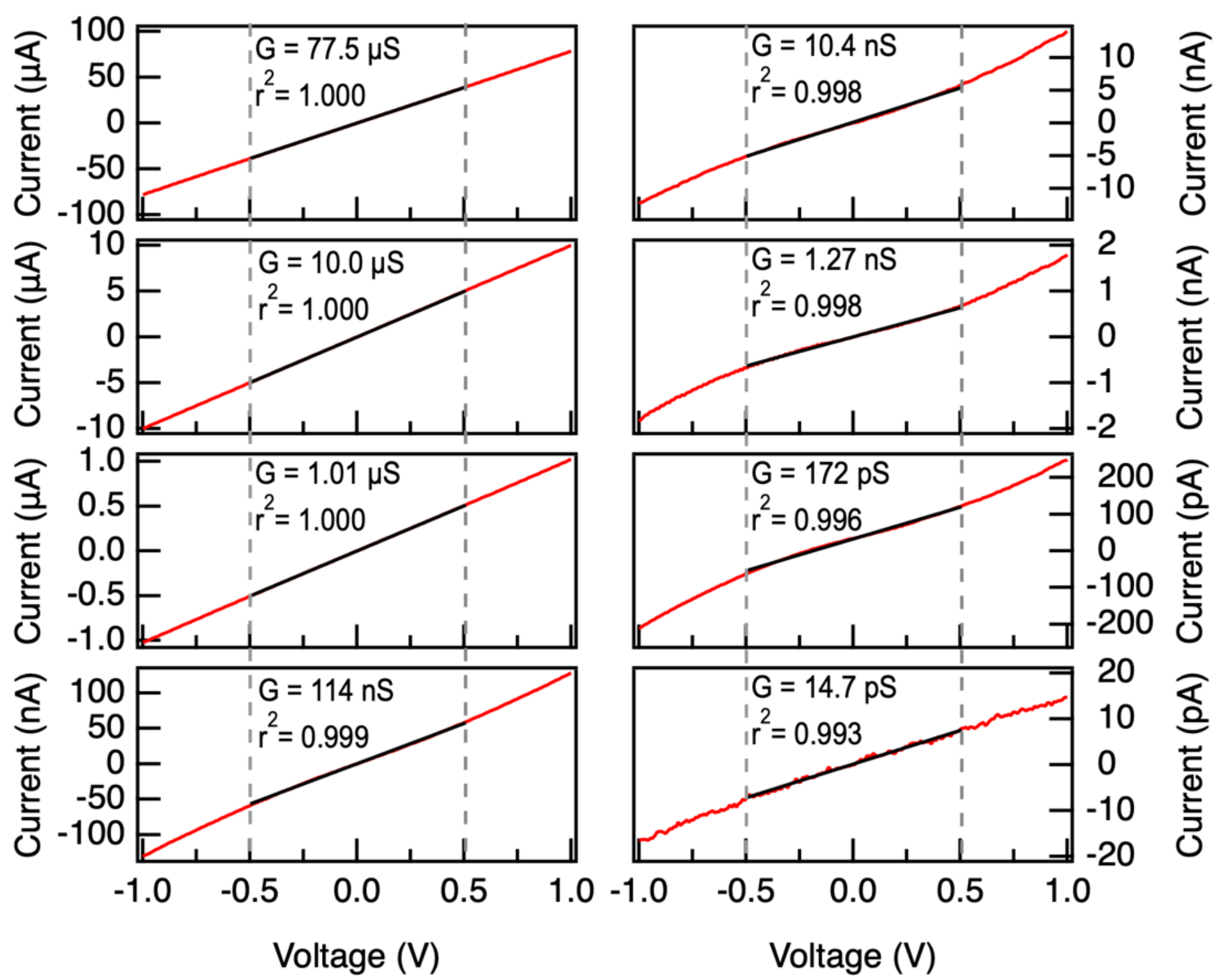

**Supplementary Figure 3.** ETCRAM channel I-V curves up to ± 1.0 V for different conductance states.

Additional I-V curves were taken up to ± 2.0 V for the states shown in Fig. S4 (same states as in Fig. S3). The best fit lines from Fig. S3 are overlayed for comparison. For G = 77 μS, 10.0 μS, and 1.01 μS, Ohmic behavior is maintained up to ± 2.0 V.

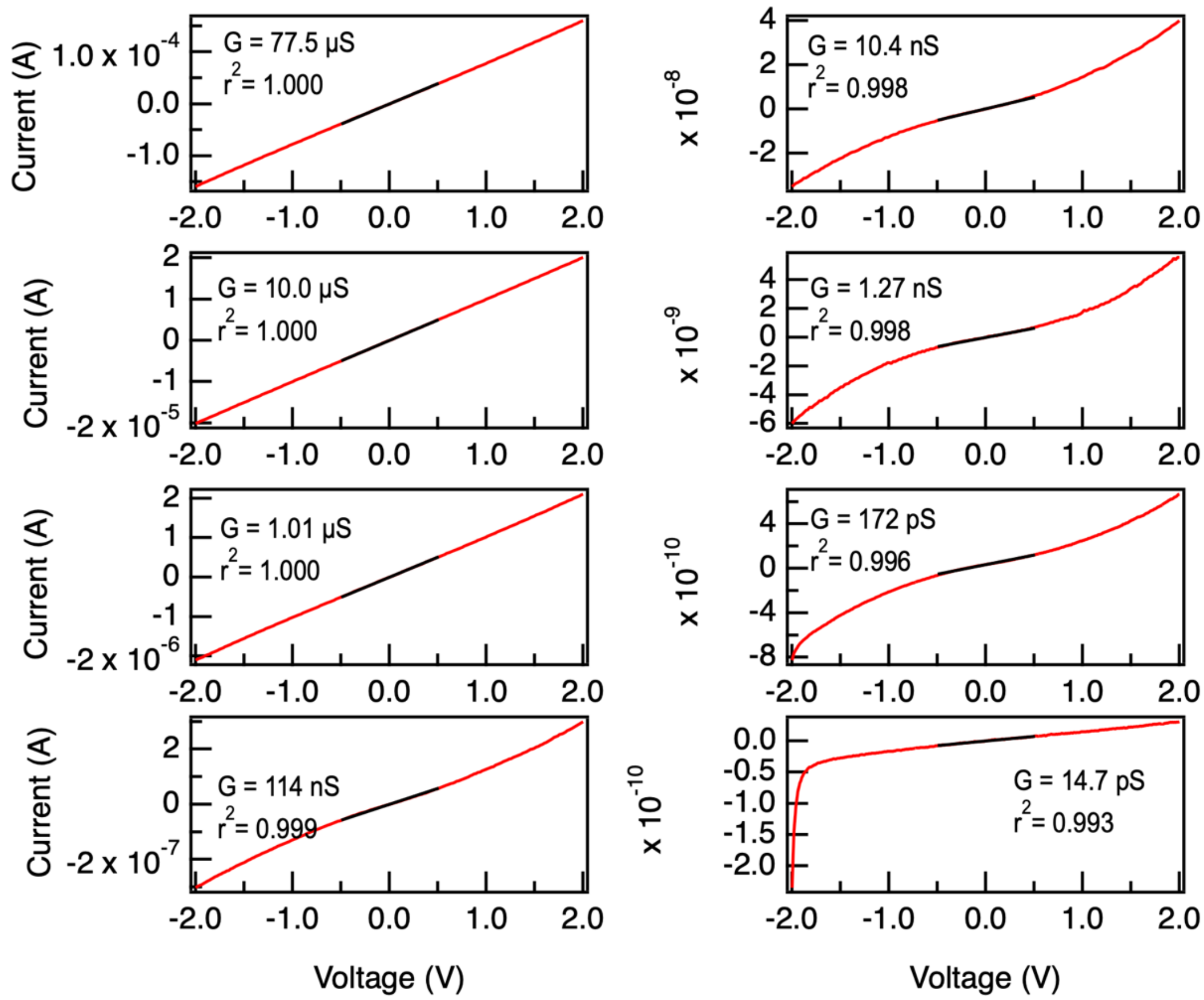

**Supplementary Figure 4.** ETCRAM channel I-V curves up to ± 2.0 V for different conductance states.

**Supplementary Note 3**

Range and tunability of ETCRAM devices

Fig. S5a shows an ETCRAM device being switched and cycled over a ~10x range in conductance, showing the effect of reversing the programming pulse sign. By continued application of successive programming pulses, a wider range in conductance can be accessed. This is shown in Fig. S5b, where the ETCRAM channel resistance is tuned across several orders of magnitude, subdivided over several ramps each with different write voltage and pulse duration conditions. The pulse magnitudes and durations were chosen to maintain an update strength sufficient to cover each decade in about 100 programming pulses.

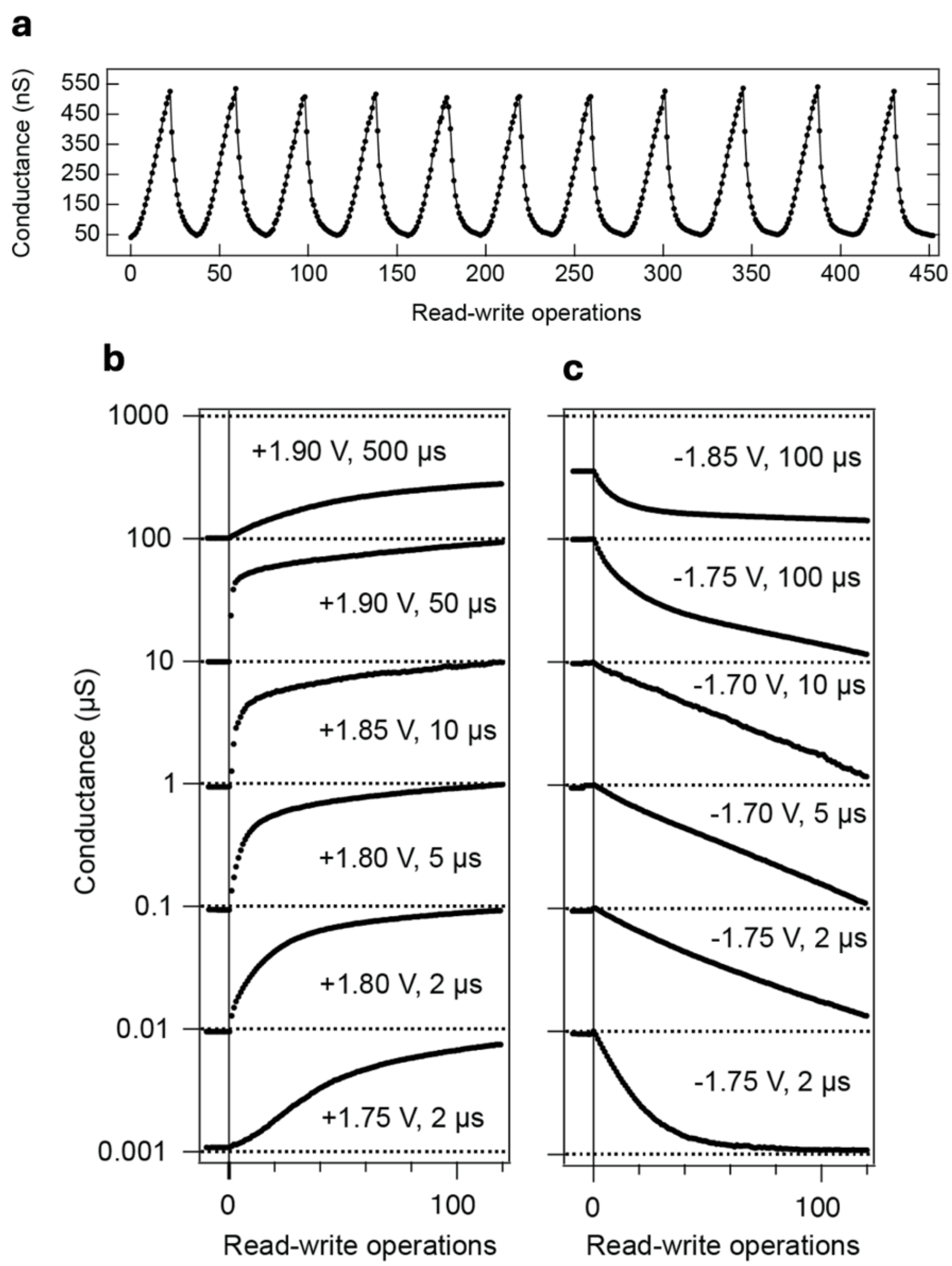

**Supplementary Figure 5.** (a) Cycled ETCRAM device programming over a 10x range, using ±1.8 V, 20 µs pulses. (b) ETCRAM programming across six decades in conductance, with different positive programming voltages and durations (potentiation). (c) ETCRAM programming across six decades with different negative programming voltages and durations (depression).

**Supplementary Note 4**

State selection in ETCRAM devices

The conductance of ETCRAM devices can be tuned deterministically and precisely by application of programming pulses. Fig. S6a shows a demonstration of a ladder of states being accessed. Each level is separated from the previous one by three programming pulses for potentiation and depression ($V_g$ is plotted below), showing non-volatility in between programming. By employing a simple write-verify control loop (Fig. S6b, inset), devices can be programmed to specific states with high levels of precision (<0.01% from target). A similar degree of precision is seen both within a decade of conductance (Fig. S6b) and across different decades (Fig. S6c).

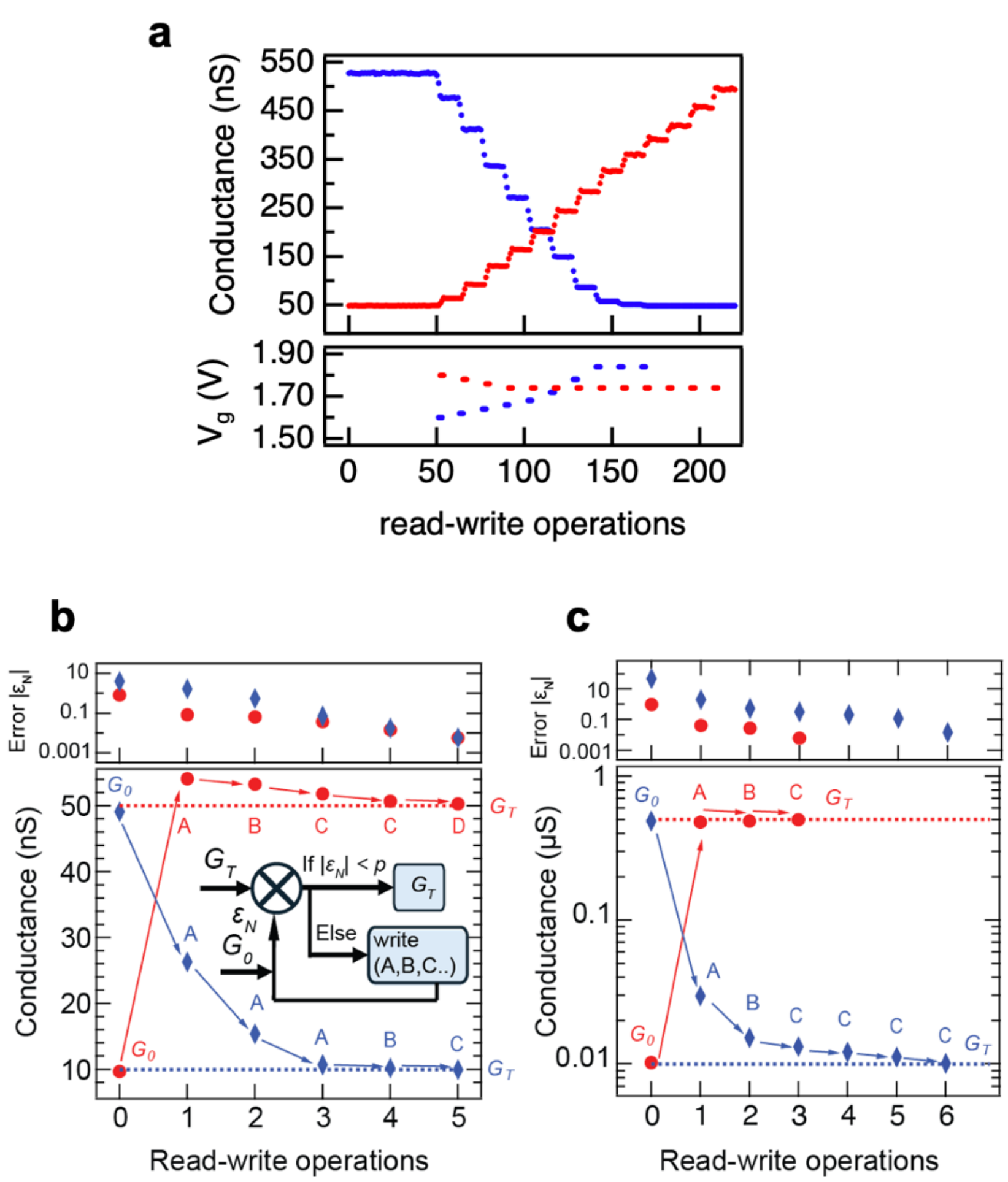

**Supplementary Figure 6.** State ladder and state selection in ETCRAM. (a) Ladder of states surveyed using $V_g$ pulses of varying magnitude (below). (b) State selection within a decade showing the employed control loop (inset) and (c) state selection over a wider range, between two decades. Error is the percent away from target and is plotted on top.

## Supplementary Note 5

### Endurance of ETCRAM devices

Endurance tests were performed on a 4 μm W heater ETCRAM device cycling between $G$ = 5 nS – 50 nS (10x range). Supplementary Fig. S7 shows device cycling initially and after >$10^5$ programming pulses. The programming conditions were kept constant over the course of the measurement: $V_g$ = +1.8 V, 100 $\mu$s (potentiation) and $V_g$ = -1.72 V, 100 $\mu$s (depression) using the NI-DAQ and a Lakeshore probe station in vacuum. The channel state was read out with $V_{read}$ = 0.2 V with $V_g$ = 0 V using the NI DAQ. The device was able to be reliably tuned through the different states 122,654 times (Supplementary Fig. S7, bottom). No failure in the heater was observed over the course of the test, and the device continued to operate.

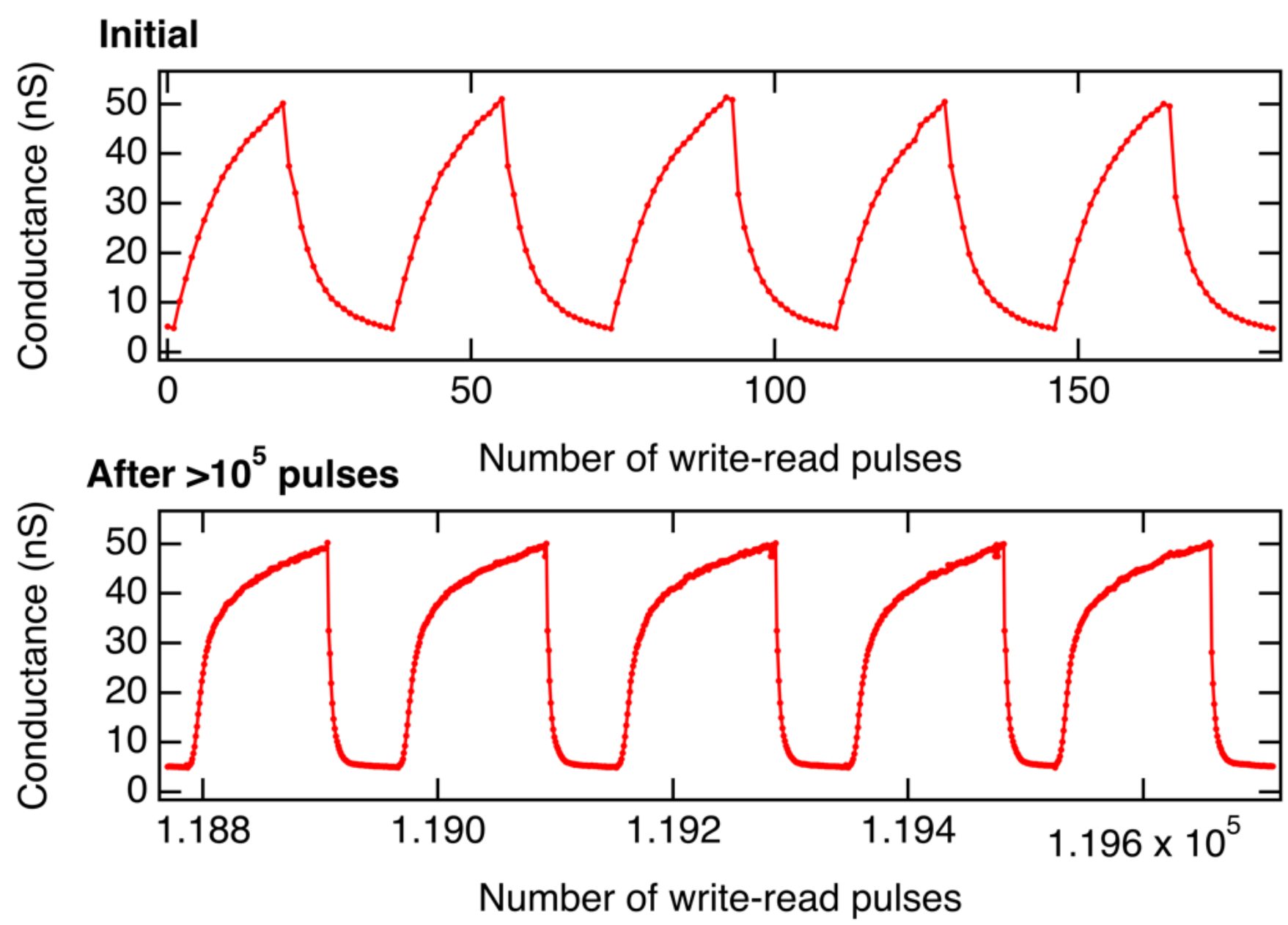

**Supplementary Figure 7.** Endurance data for a 4 μm W gate ETCRAM device. Data is shown for the initial five cycles (top) and cycling after >110,000 write pulses (bottom).

**Supplementary Note 6**

Quantification of noise floor of ETCRAM devices

In the main text, Fig. 1d shows the conductance error $\sigma_G$ for several different channel conductance values for ETCRAM and various memory technologies. Here, $\sigma_G$ is defined as the standard deviation associated with writing to a target state 10 times and reading the target state 100 times for each conductance *G* and includes read and write noise. To perform the read, 100 spot measurements were taken with the Keysight B1500, with a voltage of $V_{read}$ = 0.5 V and pulse duration of about <1 ms. Reads were done in this manner due to unavailable auto-ranging fast pulsing units for the Keysight B1500, and the need to closely integrate with the NI DAQ for programming the devices. In the other works we are comparing the error/noise of ETCRAM to, the read pulse duration is substantially shorter, down to 10 ns. Compared to a 1 ms read, a 10 ns read is expected to be noisier due to additional frequencies in the range 1 kHz – 100 MHz (1 ms – 10 ns) affecting the reading of *G*. Thus, we must quantify the amount of that additional noise and demonstrate that it is not larger than our reported $\sigma_G$.

To verify this, we have collected noise spectra in an ETCRAM device programmed to six different conductance values ranging from 4.76 nS – 20.5 μS with a semiconductor parameter analyzer (Keithley 4200A-SCS). Supplementary Fig. S8 shows unnormalized noise power spectral density (NSD) curves for the six different conductance values, with the thermal noise floor shown as dashed lines from 1 kHz to 100 MHz. The spectra were obtained from fast Fourier transforms of time series data of the channel current (1 s, 10 samples each) to calculate a noise amplitude spectral density (ASD), which was squared to yield the PSD/NSD. The initial part of the NSD from DC to about 200 Hz is assumed to be dominated by 1/*f* noise, after which the NSD levels off to the noise floor, which is dominated by Johnson-Nyquist noise. Additional contributions to the noise floor are also expected from the measurement apparatus, which were not possible to separate in this analysis. The noise floor plotted in Supplementary Fig. S8 was projected from the mean NSD near the end of the spectra (1 kHz – 1.598 kHz) out to 100 MHz, corresponding to a 10 ns read.

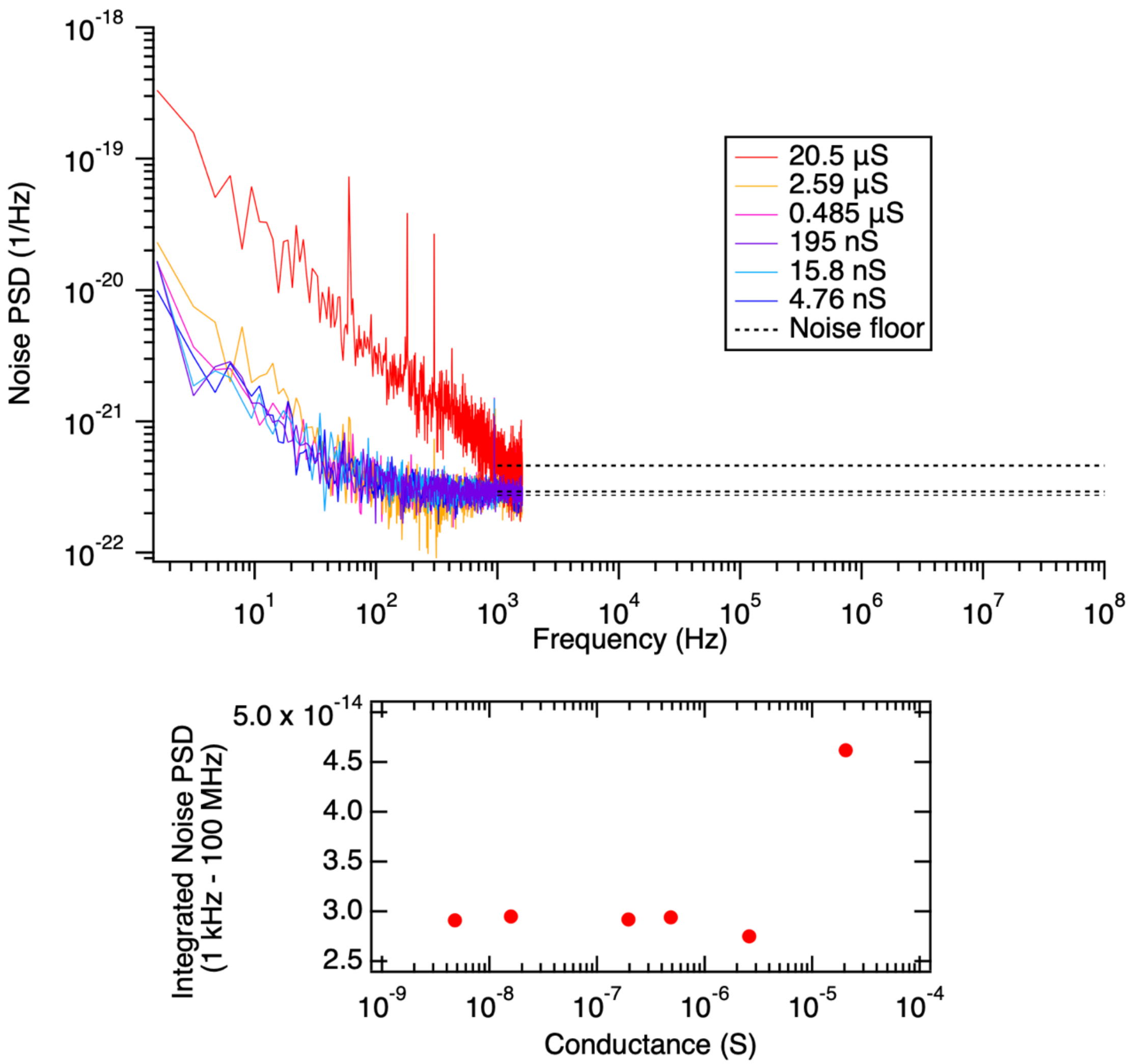

**Supplementary Figure 8.** Noise power spectral density (NSD) spectra of an ETCRAM device at six different channel conductance values. The noise floor (dashed lines) was assumed from the average noise PSD between 1 kHz – 1.598 kHz and extended to 100 MHz. Bottom: plot of the integrated noise PSD over 1 kHz – 100 MHz.

Integration of the NSD from 1 kHz – 100 MHz yields values for $\sigma_{NSD}$ plotted in Supplementary Fig. S3 (bottom). The integrated noise is on average $\sigma_{NSD} = 3\times10^{-14}$ for the six conductance values, which is 700x lower than the lowest $\sigma_G$ in reported in Fig. 1d ($\sigma_G = 2.1384\times10^{-11}$). The total integrated noise PSD, including the 1/*f* component up to 1 kHz, is ~$10^{-18}$ and is negligible compared to $\sigma_G$. Thus, it is unlikely for a faster *t* = 10 ns measurement to introduce more noise to shift the $\sigma_G$ curve in Fig. 1d, and our comparatively slower ETCRAM measurements can be accurately compared to other examples in the literature. We have also recollected the $\sigma_G$ data in Fig. 1d on a second ETCRAM device (Supplementary Fig. S9) with 5 writes/50 reads for three different *G*, which tracks closely with the $\sigma_G$ data for the first device.

**Supplementary Note 7**

Comparison of noise characteristics across different memory technologies

To quantify the difference in the ETCRAM noise floor to the different memory technologies shown in Fig. 1d and Supplementary Fig. S9, the reported $\sigma_G$ of different technologies can be compared to the equivalent $\sigma_G$ of ETCRAM (here denoted '$\sigma_{ETCRAM}$') at the same conductance. Because of the large conductance range being traversed in Supplementary Fig. S9, the ratio between $\sigma_G$ and $\sigma_{ETCRAM}$ varies significantly from one end of the range to the other, and even within the reported range for each technology. This is evident when looking at the dashed lines bounding the SONOS data denoted "lower noise" and "higher noise".

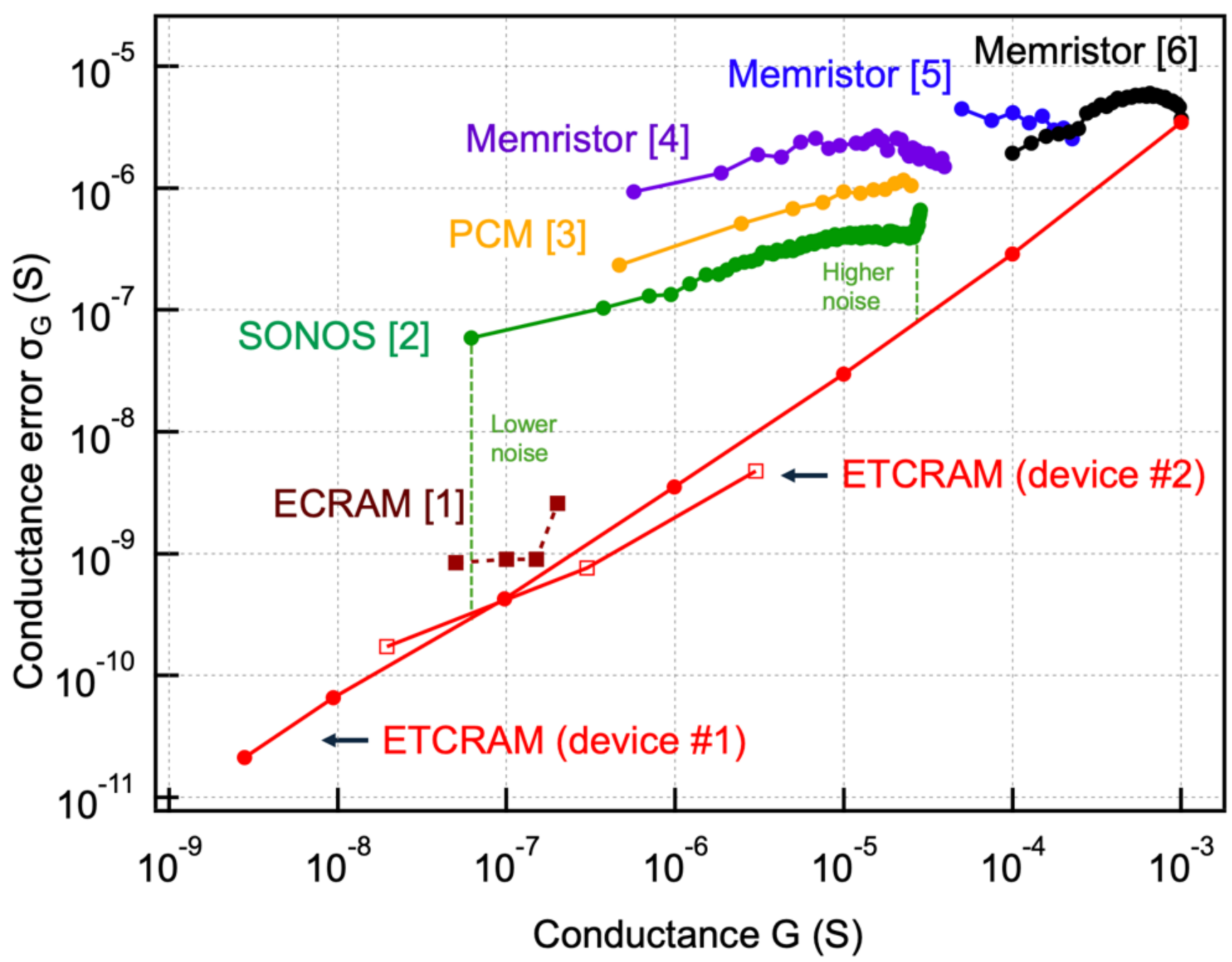

**Supplementary Figure 9.** Noise comparison across different technologies for different *G*. Data is the same as in Fig. 3c, with the inclusion of additional data for a second ETCRAM device.

To quantify these differences, Supplementary Table 2 shows the $\sigma_G/\sigma_{ETCRAM}$ ratios for each technology at the same mean conductance. Interpolation was used to estimate the $\sigma_{ETCRAM}$ between the empirical values. Below $G$ < 1 μS, ETCRAM remains substantially lower noise than other technologies: up to 130x for PCM, 200x for SONOS, and 440x for memristors. The advantage of ETCRAM diminishes at higher conductance values ($G$ > 1 μS) with for example, SONOS coming within 5.2x of ETCRAM at $G$ = 5.2 μS and memristors converging to ETCRAM near $G$ = 1.0 mS.

| Technology | Lower ETCRAM noise | | Higher ETCRAM noise | | Reference |
|---|---|---|---|---|---|
| | **G (μS)** | **$\sigma_G/\sigma_{ETCRAM}$** | **G (μS)** | **$\sigma_G/\sigma_{ETCRAM}$** | |
| ECRAM | 0.050 | 3.4 | 0.15 | 1.4 | [1] Lee et al. (2021) |
| SONOS | 0.062 | 200 | 25.74 | 5.2 | [2] Xiao et al. (2025) |
| PCM | 0.466 | 130 | 24.98 | 14 | [3] Joshi et al. (2020) |
| Memristor | 0.566 | 440 | 39.26 | 13 | [4] Wan et al. (2022) |
| Memristor | 50.000 | 31 | 225.00 | 3.6 | [5] Milo et al. (2021) |
| Memristor | 99.858 | 6.7 | 999.73 | 1.1 | [6] Song et al. (2024) |

**Supplementary Table 2.** Comparison of noise floor of ETCRAM to different memory technologies[1-6]. The maximum and minimum $\sigma_G$ relative to the corresponding $\sigma_G$ for ETCRAM devices (‘$\sigma_{ETCRAM}$’) are obtained by comparison to interpolated ETCRAM $\sigma_G$ values in Fig. 3c.

We note that the reported values of $\sigma_G$ for the different devices in Fig. 1d (and Supplementary Fig. S9) do not have precisely the same definition. This is due to the lack of a standardized method to benchmark the precision of analog conductance in the literature, as well as significant differences in the scale of device characterization (i.e. isolated devices vs large integrated arrays). We note the differences below:

- The errors for SONOS, PCM, memristor (Wan 2022), and memristor (Milo 2021) are based on the variability in conductance of many devices programmed to the same target conductance: 8192 devices/target for SONOS, 10000 devices/target for PCM, and 1,024 devices/target for memristors (Milo 2021). All of these utilized a write-verify routine. This type of characterization accounts for device-to-device variability and also (implicitly) accounts for write noise and read noise.

- The errors for memristor (Song 2024) represent the mean read noise standard deviation of 8192 devices programmed to the same state using a write-verify routine. The device-to-device variability in the read noise is not included in the plot but is reported[6].
- The errors for ECRAM (Lee 2021) measure the standard deviation of write noise of a single device, not programmed using a write-verify routine.

**Supplementary Note 8**

Noise power spectrum of scaled ETCRAM devices

Additional NSD data has been collected on different sized ETCRAM devices (*L* = 2 μm – 64 μm), which were initially pristine and not annealed (high conductance). The low frequency noise measurements ETCRAM requires a measurement system with a noise floor lower than the devices' intrinsic noise, which is not achievable with either Keithley 4200A-SCS or NI-DAQ. For the noise power spectrum measurement in Supplementary Fig. S10, we used our custom noise measurement setup with SR785 dynamic signal analyzer and a FEMTO DLPCA-200 transimpedance amplifier with a gain of $10^5$. The device under test and DLPCA-200 are shielded separately with metal boxes. A $10^5$ gain of DLPCA-200 is used for the noise measurements. 0.2 V of DC bias is applied by the DLPCA-200 and maintained during noise measurements. Each noise spectrum is averaged over 50 noise spectrums measured over a 2.0 to 1.6 kHz frequency range and is normalized by the average current flow for each device for comparison.

The normalized NSD curves for the channels of these devices are shown in Supplementary Fig. S10a, showing the $1/f$ and $1/f^2$ components. The NSD was measured with a DC bias of $V$ = 0.2 V from the amplifier with the gate grounded. The normalized NSD at $f$ = 10 Hz is plotted in Fig. S10b for the devices. All scaled devices show much less NSD than the memristor device, whose device area is 20 × 20 μm and is made of the same switching material (i.e., $TaO_x$). In Fig. S10c, the read noise of a 4 μm device at different programmed channel resistances is shown. To measure the read noise, 50 pulses (V = 0.5 V, t = 150 μs) sourced from PMUs are used to measure the resistance of the channel at each conductance, with the standard deviation of the resistance distribution plotted on the *y*-axis. The read noise of the device is at least one order of magnitude less than its nominal resistance regardless of its resistance states.

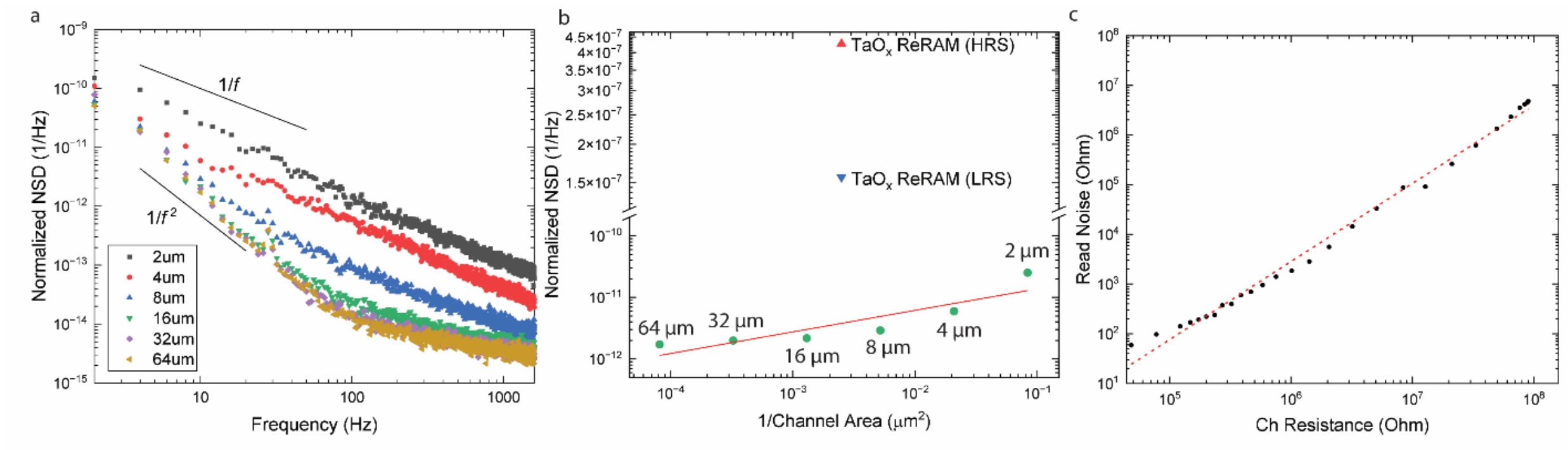

**Supplementary Figure 10.** Channel read noise of $TaO_x$ ETCRAM devices. (a) The normalized noise spectral density (NSD) of ETCRAM devices with different channel dimensions, with the 1/*f* and $1/f^2$ slopes indicated. (b) The normalized NSD at *f* = 10 Hz for the different devices, with an interrupted *y*-axis. The normalized NSD increases as the inverse of channel area increases, with a line shown as a guide to the eye. Comparison to memristor (ReRAM) $1.56 \times 10^{-7}$ 1/Hz (LRS) and $4.27 \times 10^{-7}$ 1/Hz (HRS) is included. (c) The read noise of a programmed 4 μm $TaO_x$ ETCRAM device measured at various resistance states.

## **Supplementary Note 9**

Dynamic gain voltage amplification

The demonstration of dynamic voltage amplification in Fig. 3b is described below in further detail. Fig. S11a shows the essential components of the circuit constructed, where the ETCRAM device (red box) takes the role of the feedback resistor ($R_f$) in a non-inverting voltage amplifier. For each gain setting, a one-second-long sinusoidal input signal with a fundamental frequency of $f_0$ = 100 Hz and V = 1.0 mV amplitude (after the voltage divider) is sourced from a NI DAQ to test the response of the amplifier. A commercial Analog Device AD820 operational amplifier is used to provide active negative feedback. The ETCRAM device is based on a 2 μm $MoSi_2$ gate and two independent DAQ outputs connected to source and drain (with drain held at GND) are used to tune $R_f$ of the device to achieve the desired amplification: $V_{out} = \left(1 + \frac{R_f}{R_g}\right) * V_{in}$ , or $G = 1 + \frac{R_f}{9000\ \Omega}$ . The voltage applied to the gate is $V_g = V_S - V_D$, with $V_D$ = 0 V. After programming the device to the desired state, the DAQ simultaneously inputs $V_{in}$ and measures $V_{out}$ over a one second window, with the ETCRAM gate grounded during measurement. Fig. S11b shows $V_{out}$ for increasing gain settings over three cycles, each cropped from the full one second measured signal. Programming pulses were constant ($V_g$ = -2.2 V, 20 μs) and a single pulse was used to tune the amplifier to each subsequent gain level. Each $V_{out}$ was DC voltage offset and phase corrected by performing fits to the function $y(t) = y_0 + A \sin(2\pi f t + \phi)$. The gains were calculated using the amplitude *A* from the fits and are plotted in Fig. S11c. The fast Fourier transforms shown in Fig. 3c are taken using the full one second signals, without any corrections or modifications of the raw signals. Additionally, we demonstrate that the amplifier gain can be tuned in the opposite direction (i.e., in direction of decreasing gain) using opposite sign programming pulses ($V_g$ = +2.2 V, 50 μs) which is shown in Fig. S11d for half-cycles.

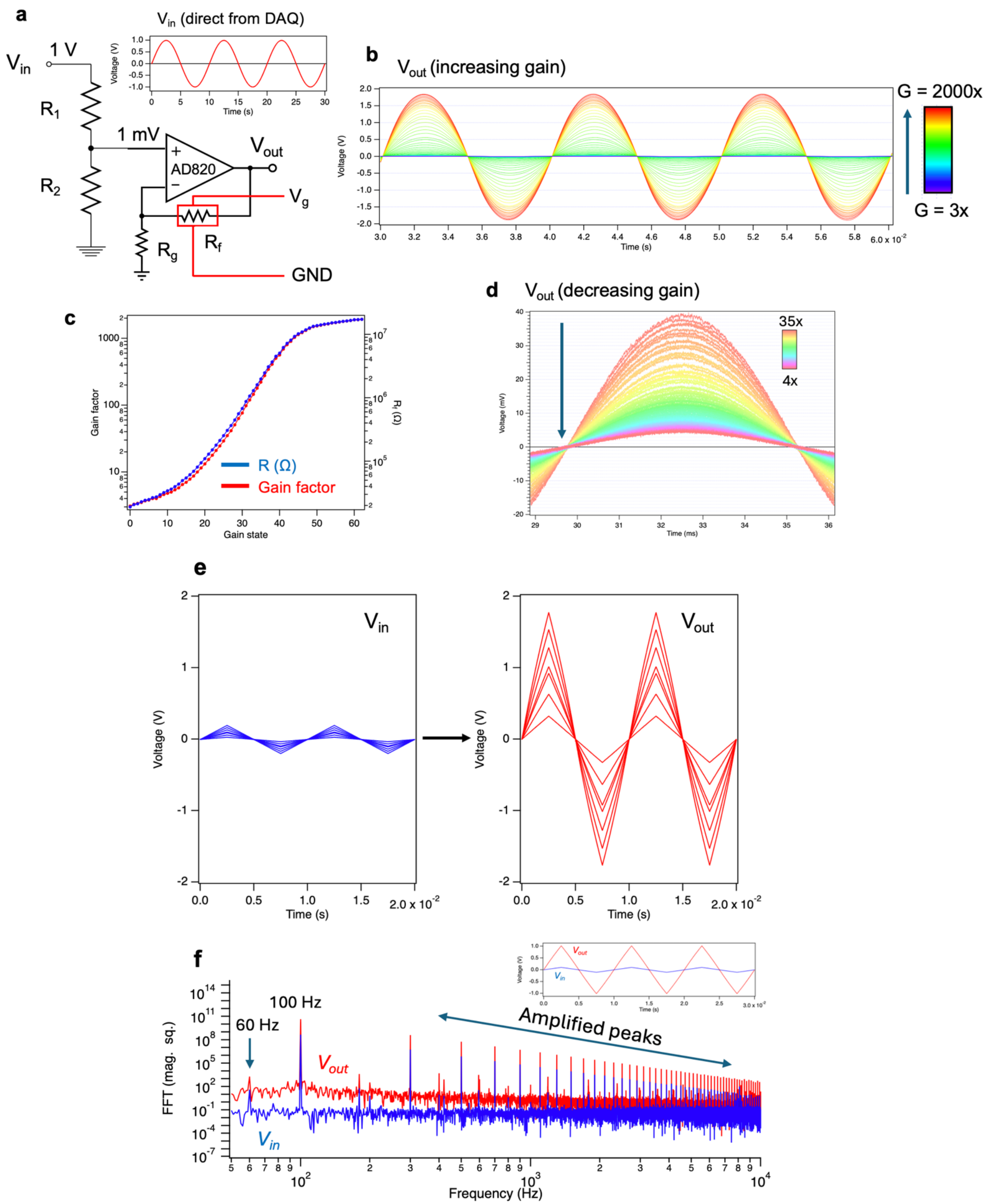

a
$V_{in}$ (direct from DAQ)
1 V
$V_{in}$
$R_1$
$R_2$
1 mV
AD820
$V_{out}$
$V_g$
$R_g$
$R_f$
GND
b
$V_{out}$ (increasing gain)
G = 2000x
G = 3x
Voltage (V)
Time (s)
c
Gain factor
Gain state
$R_f$ (Ω)
R (Ω)
Gain factor
d
$V_{out}$ (decreasing gain)
35x
4x
Voltage (mV)
Time (ms)
e
$V_{in}$
$V_{out}$
Voltage (V)
Time (s)
f
100 Hz
60 Hz
Amplified peaks
$V_{out}$
$V_{in}$
FFT (mag. sq.)
Frequency (Hz)

**Supplementary Figure 11.** Dynamic voltage amplification circuit. (a) Circuit diagram showing voltage input ($V_{in}$ from DAQ) at voltage divider input. $V_{in}$ = 1 V (amplitude), with f = 100 Hz. The voltage divider output is connected to an operational amplifier (Analog Devices AD820) in the non-inverting configuration. The tunable ETCRAM device ($R_f$) is wired up to the op-amp and is indicated by the red box. Static resistor values: $R_1$ = 10 kΩ, $R_2$ = 10 Ω, and $R_g$ = 9 kΩ. (b) Output of the voltage amplifier ($V_{out}$) as a function of increasing gain factors, shown for three cycles. (c) Gain factors and corresponding $R_f$ values for the signals plotted in (b). (d) Output of the voltage amplifier ($V_{out}$) as a function of decreasing gain factors, zoomed in showing a half-cycle. (e) $V_{in}$ and $V_{out}$ for 100 Hz triangle waves for seven different $V_{in}$ with fixed gain. (f) FFT of triangle wave $V_{in}$ (blue) and $V_{out}$ (red) for a gain factor of 10x. Corresponding waveforms shown in the inset.

Additionally, we also tested voltage amplification with triangle waves, which contain more frequency components in their Fourier series other than the fundamental $f_0$ = 100 Hz in the sine waves. Seven different input voltages up to a maximum of $V_{in}$ = 200 mV are shown in Fig. S12a (left, blue) along with their corresponding amplified wave forms (right, red). The corresponding FFTs of $V_{in}$ and $V_{out}$ are plotted in Fig. S12b for 10x gain, showing that the additional peaks present in the triangle waves are preserved in the amplified output.

**Supplementary Note 10**

Dynamic voltage division demo

Below are additional details regarding the dynamic voltage division demo in Fig 3a. Fig. S12a shows the main components of the voltage divider circuit, where an ETCRAM device (red box) replaces $R_1$ in the divider. The second resistor is held constant ($R_2$ = 50.5 kΩ). Like the dynamic gain voltage amplifier, one second sinusoidal signals are sourced from the DAQ (V = 1.0 V amplitude, f = 100 Hz) and simultaneously measured at the output of the divider ($V_{out} = \frac{50\text{ k}\Omega}{R_1 + 50\text{ k}\Omega} V_{in}$). A buffer follower (Analog Devices AD820) is included to minimize loading effects when measuring the output voltage with the DAQ. The ETCRAM device resistance $R_1$ was tuned using constant $V_g$ = -2.3 V, 1 ms pulses (Fig. S12b). The measured output signals are shown in Fig. S12c for three cycles. Fig. S12d shows the corresponding fast Fourier transforms of the $V_{out}$ signals, each showing a main peak at f = 100 Hz with smaller harmonics.

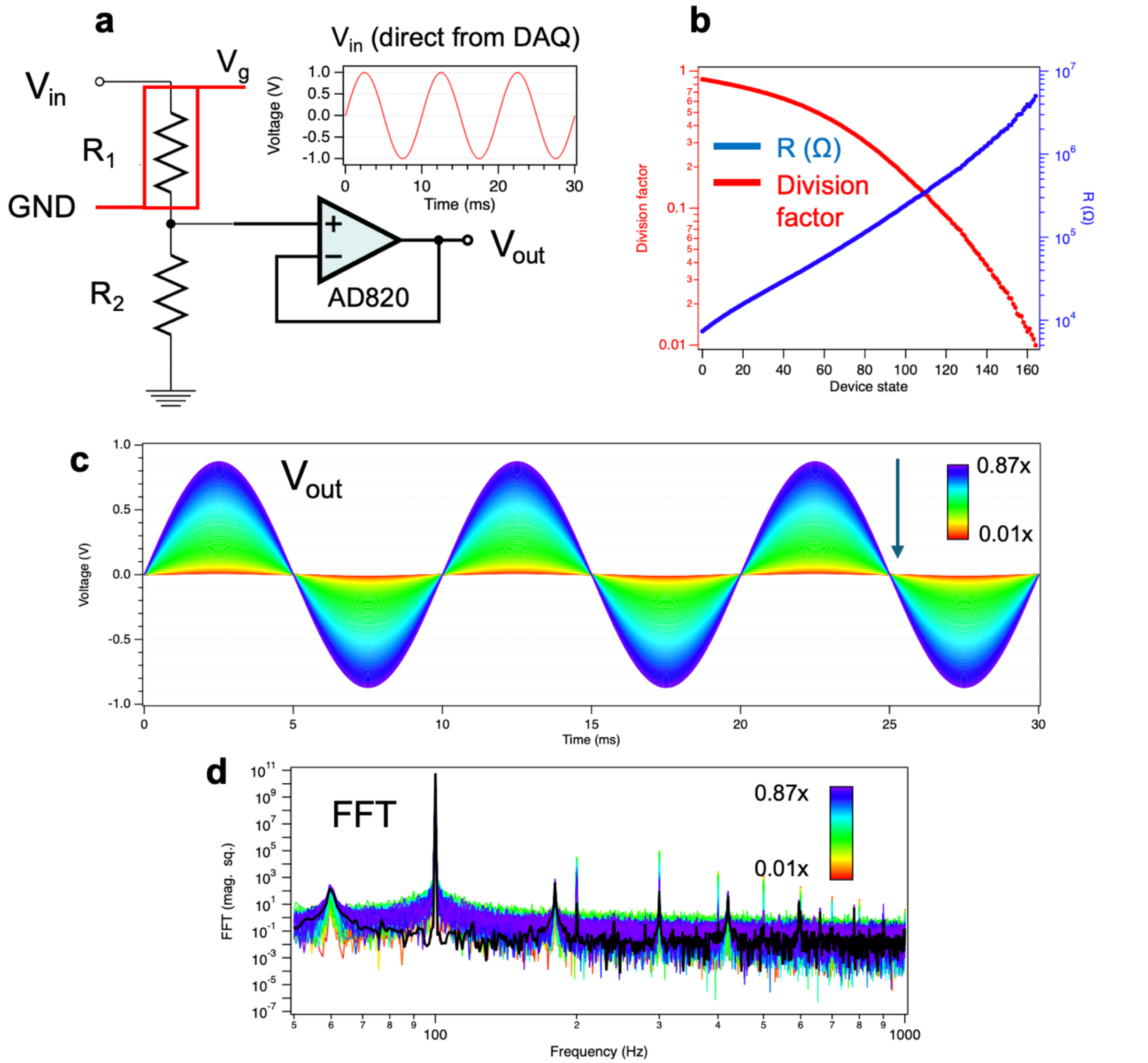

**Supplementary Figure 12.** Dynamic voltage division circuit. (a) Circuit diagram showing voltage input ($V_{in}$ from DAQ) at voltage divider input. $V_{in}$ = 1 V (amplitude), with f = 100 Hz. The tunable ETCRAM device ($R_1$) is indicated by the red box. The static resistor is $R_2$ = 50.5 kΩ. The voltage divider output is connected to an operational amplifier (Analog Devices AD820) to buffer $V_{out}$ to the DAQ. (b) Division factors and corresponding $R_1$ values when programming the ETCRAM device. (c) Output of the voltage divider ($V_{out}$) as a function of decreasing division factors, shown for three cycles. (d) Overlaid fast Fourier transforms of the voltage divider output signals. The FFT of $V_{in}$ is in black.

## **Supplementary Note 11**

Transimpedance amplification demo

In additional to voltage amplification and voltage division, ETCRAM devices can also be used as the feedback resistor $R_f$ in a transimpedance amplifier (i.e., current-to-voltage converter). The basic circuit diagram is shown below in Fig. S13a. Like the voltage amplifier, we used an Analog Devices AD820 op-amp to provide active negative feedback, with the ETCRAM device (red box) taking the role of the feedback resistor ($R_f$). A sinusoidal input current ($I_{in}$) was sourced from a Keithley 6221 AC/DC current source, with an amplitude of I = 1.0 nA and frequency f = 100 Hz. The ETCRAM device was programmed to resistance values between $R_f$ = 80 – 250 MΩ (Fig. S13b) and the DAQ was used to measure the output voltage $V_{out} = -I_{in}R_f$. The output waveforms are shown in Fig. S13c for selected $R_f$ values over three cycles, and the corresponding fast Fourier transforms are shown in Fig. S13d.

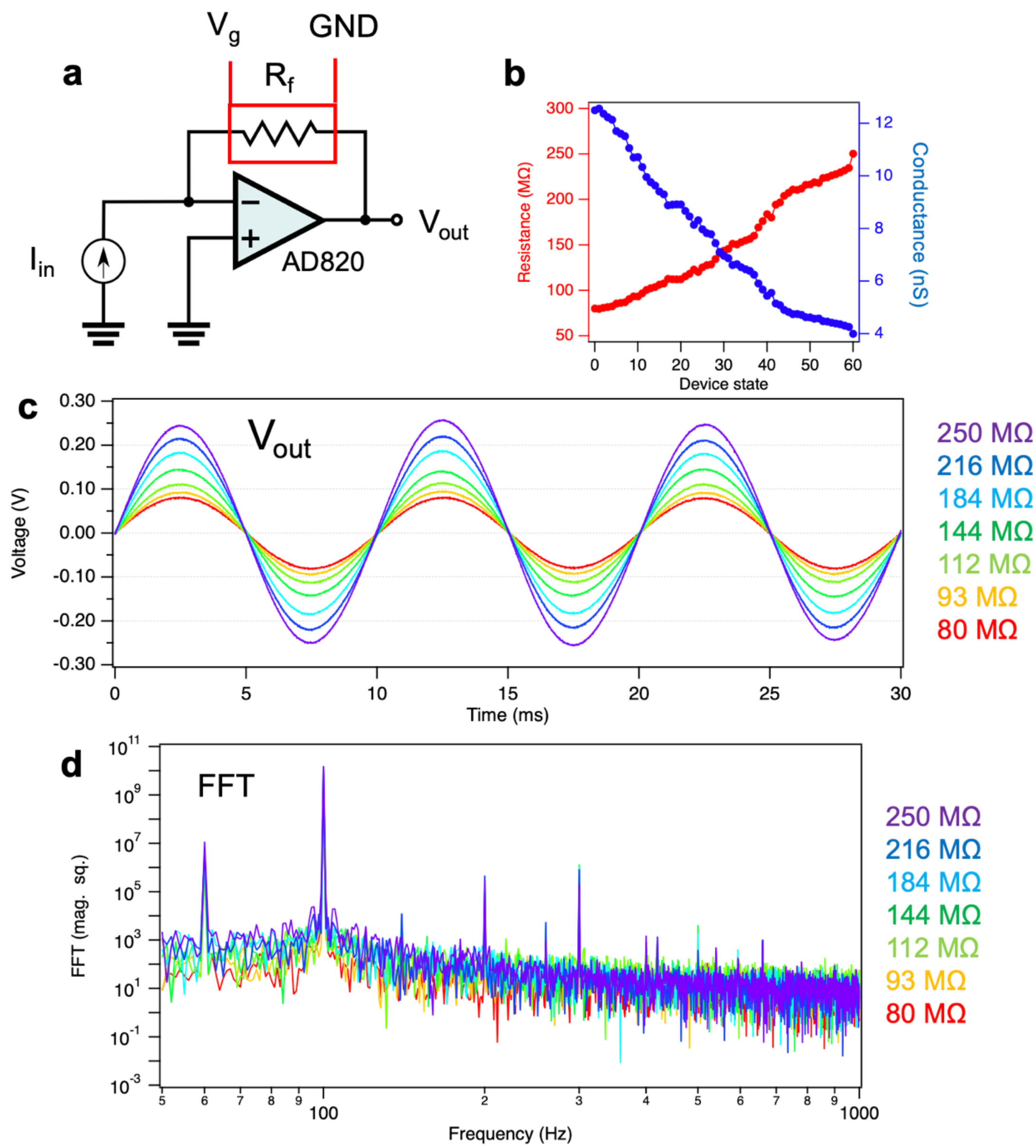

**Supplementary Figure 13.** Transimpedance amplification circuit. (a) Circuit diagram showing current input ($I_{in}$) to the operational amplifier (Analog Devices AD820) inverting input. The input current is $I_{in}$ = 1 nA (amplitude), with f = 100 Hz. The tunable ETCRAM device ($R_f$) is indicated by the red box. (b) $R_f$ values from programming the ETCRAM device. (c) Output of the transimpedance amplifier ($V_{out}$) as a function of different $R_f$ values, shown for three cycles. (d) Overlaid fast Fourier transforms of the output signals.

**Supplementary Note 12**

Column dot product demo

A primary exemplar that takes advantage of the inherent precision and linearity of ETCRAM is analog vector-matrix multiplication. To demonstrate the viability of using ETCRAM for analog VMM, we constructed a small three device circuit to compute the analog dot product of a column of ETCRAM devices. Fig. S14a shows a basic circuit diagram of the demo, where a vector of input voltages ($V_1$, $V_2$, $V_3$) is applied to three ETCRAM devices programmed to a set of conductances ($G_1$, $G_2$, $G_3$) (i.e., the matrix), and the resultant current $I_{out}$ is read out at the bottom of the column. A NI DAQ was used both to program devices to their target conductances and to apply voltages to the device channels to perform the analog dot product. A FEMTO DLPCA-200 transimpedance amplifier (biased at ground) was used in combination with the DAQ for measuring $I_{out}$. To demonstrate robustness, we program the devices to the three different randomized conductance values in Fig. S14b for each trial (A, B, C). Fig. S14c shows the I-V characteristic of each device after they are programmed to different conductances, showing a high degree of linearity up to ±1.0 V. For each trial, five different randomized voltage vector inputs are applied to the column of devices, and the dot product $I_{out}$ is measured for each vector. Fig. S14d tabulates these values, which are based on the level spacing of a 4-bit ADC. The left table shows the amplitude-encoded voltage values, and the right shows their corresponding binary representation.

**a**

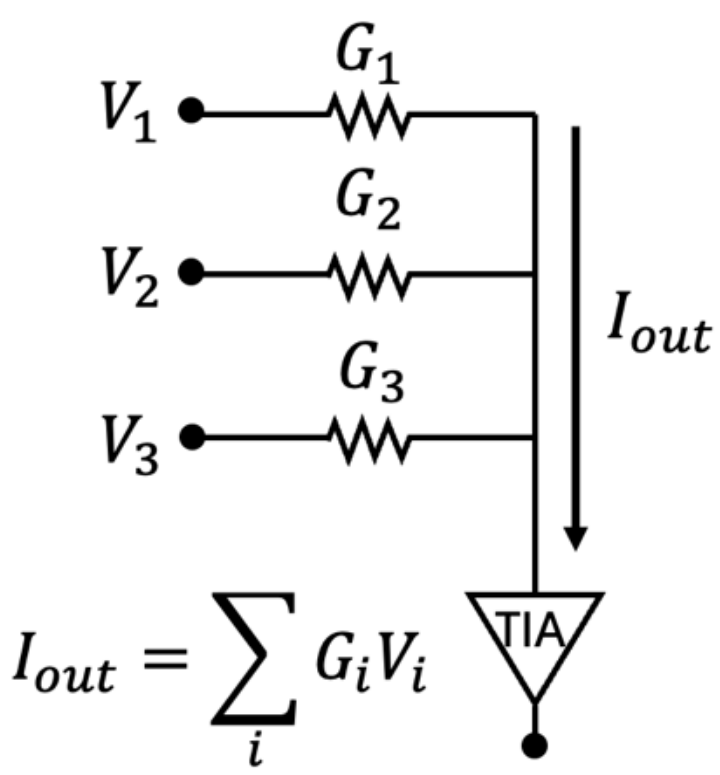

**b**

Randomized conductance vectors:

| | Set A (S) |
|---|---|
| $G_1$ | 3.37E-08 |
| $G_2$ | 4.00E-07 |
| $G_3$ | 9.10E-07 |

| | Set B (S) |
|---|---|
| $G_1$ | 6.51E-08 |
| $G_2$ | 2.65E-07 |
| $G_3$ | 1.22E-06 |

| | Set C (S) |
|---|---|
| $G_1$ | 1.19E-07 |
| $G_2$ | 7.88E-08 |
| $G_3$ | 3.17E-07 |

**c**

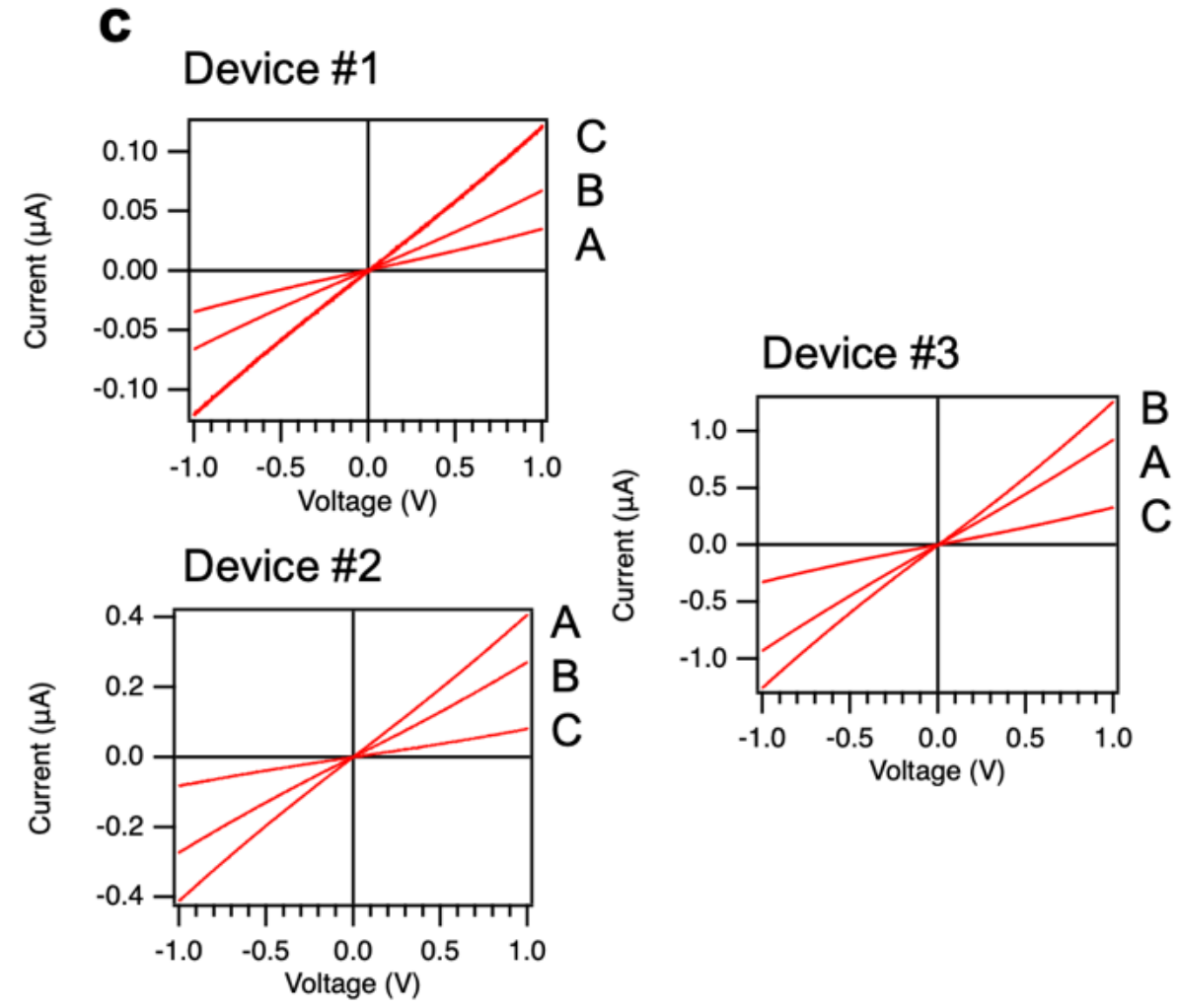

**d**

Randomized voltage vectors:

| | V1 (4-bit) | V2 (4-bit) | V3 (4-bit) |
|---|---|---|---|
| Vector A | 0.25 | 0.6875 | 0 |
| Vector B | 0.4375 | 0.125 | 0.5625 |
| Vector C | 0.625 | 0.1875 | 0.75 |
| Vector D | 0.375 | 0.5625 | 0.75 |
| Vector E | 0.3125 | 0.0625 | 0 |

Binary representation:

| V1 (4-bit) | V2 (4-bit) | V3 (4-bit) |
|---|---|---|
| 0 1 0 0 | 1 0 1 1 | 0 0 0 0 |
| 0 1 1 1 | 0 0 1 0 | 1 0 0 1 |
| 1 0 1 0 | 0 0 1 1 | 1 1 0 0 |
| 0 1 1 0 | 1 0 0 1 | 1 1 0 0 |
| 0 1 0 1 | 0 0 0 1 | 0 0 0 0 |

**Supplementary Figure 14**. Summary of dot product demonstration. (a) Circuit diagram showing a column of ETCRAM devices connect to compute the analog dot product. (b) Values of the randomized conductance vectors used for the dot product. (c) Characteristic I-V curves of each device in each of its different states (A, B, C). (d) Values of the randomized voltage vectors used for the dot product. Left: 4-bit amplitude-encoded vectors. Right: Corresponding binary vectors.

To compute the analog dot product in the amplitude-encoded representation, a single vector ($V_1$, $V_2$, $V_3$) is applied and $I_{out}$ is measured instantly in one step. Across all attempts, the average percent difference from the true dot product value was -1.94% (N = 15). While this level of accuracy of the dot product is acceptable for many applications, we note that a small systematic voltage error in the TIA is likely responsible. All 15/15 readouts are systematically offset from the true value indicating this does not arise from noise. For comparison, we have also computed the analog dot product in a binary-encoded manner. Here, a single bit at time is fed into the devices in serial fashion, with devices being either in the ON state (V = 0.1 V) or OFF state (V = 0 V) depending on the value of the significant digit. To compute $I_{out}$, the measured current was exponentially weighted using the formula $Y_j = \sum_{b=0}^{B_{in}-1} \frac{1}{2^{B_{in}-b}} \left( \sum_{i=0}^{N_r-1} G_{ij} V_i^{(b)} \right)$ in the post-analysis[7]. Across all attempts, the average percent difference from the true dot product value was -5.39% (N = 14) in the binary representation. The higher degree of error is due to nonidealities of the measurement setup and successive rounds of input and measurement required for binary-encoding.

**Supplementary Note 13**

ETCRAM 4x4 array programming and statistics

To demonstrate multiple device statistics, 16 ETCRAM devices (*L* = 8 μm) in a 4x4 array were programmed to target channel conductance states spanning 1 nS – 100 μS. The devices as fabricated are electrically isolated from each other (i.e., in a disconnected array) on 200 nm $SiO_2$, and placed into a Lakeshore probe station where programming was performed by positioning four probes on the pads of each device.

Initial measurement of the channels of the 16 devices show that all devices are initially conductive (Fig. S15, left), with $\bar{G}$ = 529 μS (*R* = 1.9 kΩ). The array was initialized to a target conductance value of *G* = 5.0 nS using a series of depression pulses ($V_g$ = -2.8 V, *t* = 1 ms). The distribution is shown in Supplementary Fig. S15, right.

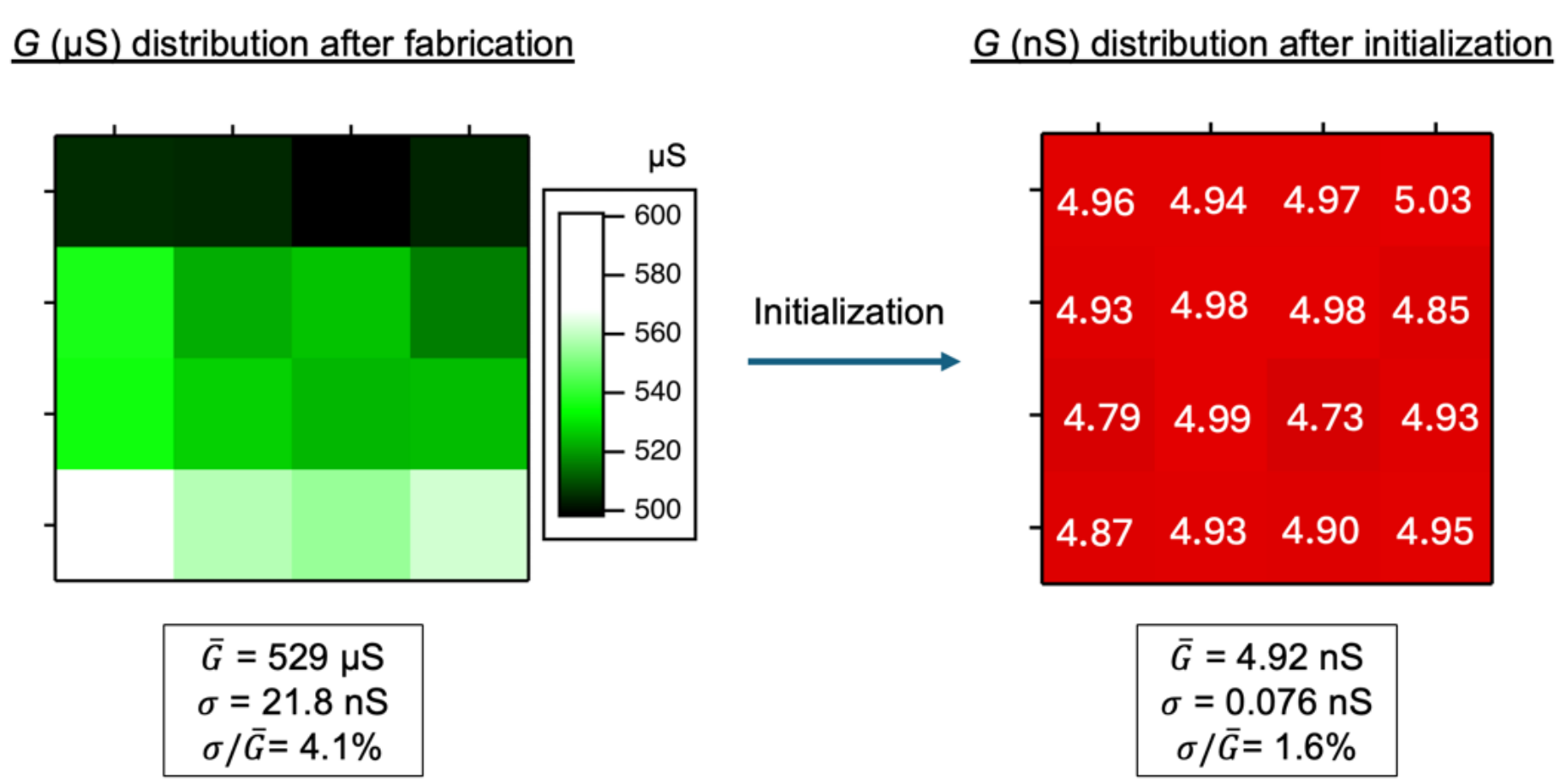

**Supplementary Figure 15.** Initialization of 4x4 array of 16 ETCRAM devices. The channel conductance distribution is measured before (left) and after (right) initialization. Mean $\bar{G}$, standard deviation σ, and σ/$\bar{G}$ are indicated in the boxes.

To explore the full extent of tunability and reliability of ETCRAM, each device was programmed to a different conductance value spanning a 10x range in conductance. To achieve the target state, write pulses ranging from $V_g$ = ±2.5 – 2.8 V, and $t$ = 10 μS – 10 ms were applied until the target state was reached with acceptable precision. Across the full $10^5$x range, the update strength was varied by altering the magnitude of $V_g$ and $t$. The gain of the channel readout was manually adjusted using a FEMTO DLPCA-200 transimpedance amplifier from $10^9$ to $10^4$, decreasing by a factor of 10 with each increasing decade of conductance. To verify the final programmed state of each device, a linear IV curve of the channel was obtained using the NI DAQ, with the slope of the line indicating the final target conductance (IV parameters: $V$ = 0.1 V, 1.0 s ramp up/down, three cycles). After the target conductance is reached in one device, the probes are lifted and repositioned on the adjacent device, and the process is repeated. Fig. S16 shows the distribution of programmed $G$ of the devices (left), the target $G$ (center), and $\Delta G$ error (right) defined as the absolute difference of $G$ from target. In addition, one device (R4C4) was programmed up to 999.65 μS, within 0.035% (350 nS) away from the target value. In this set of devices, the full dynamic range spanned over the course of the experiment was 200,000x.

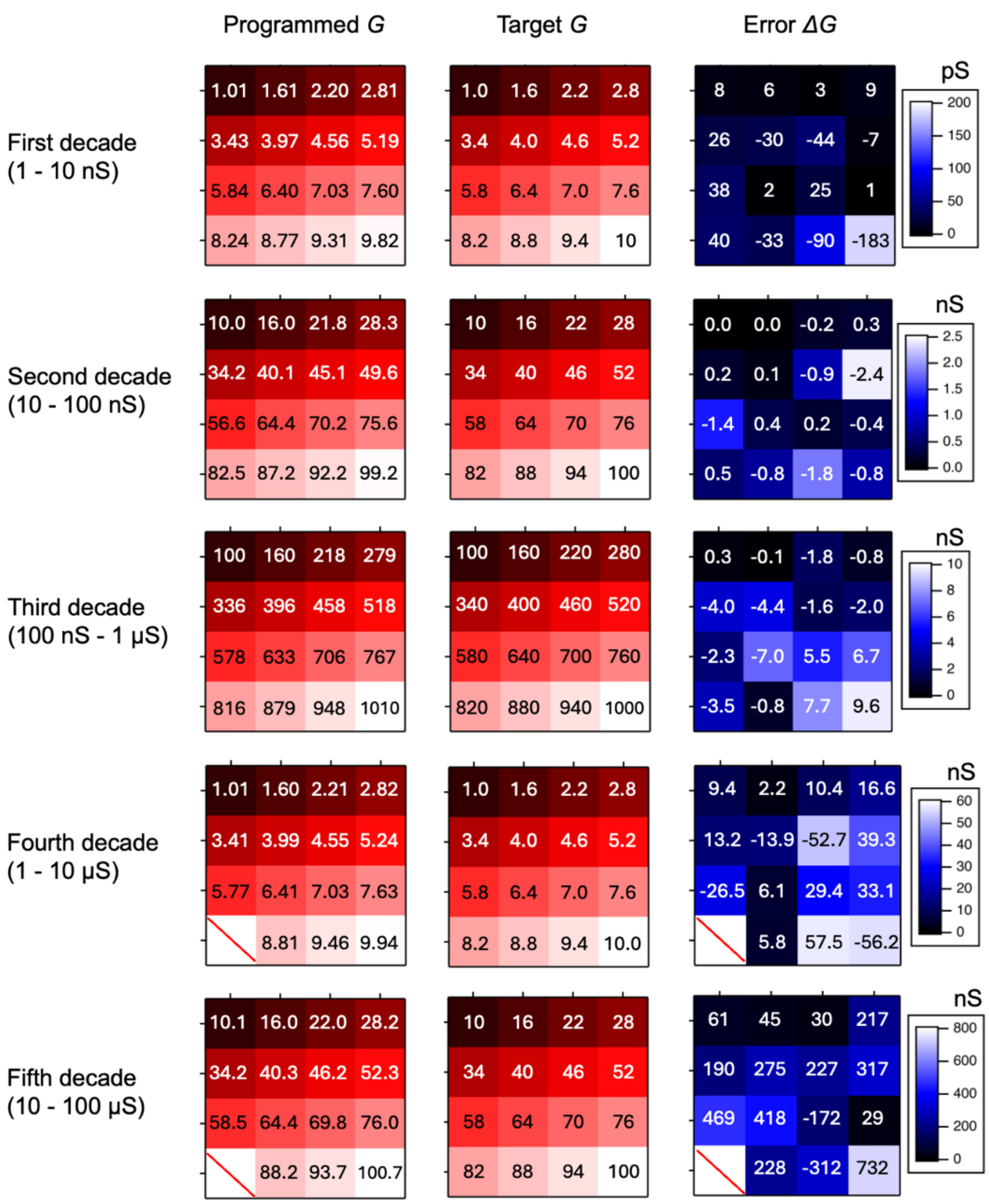

**Supplementary Figure 16.** Programming of 4x4 array over five decades in conductance. Left: programmed *G* for each device. Center: target *G* for each device. Right: *ΔG* for each device, defined as the absolute difference from the target *G*.

Supplementary Table 3 summarizes the programming statistics of all 16 devices across five decades in conductance. Out of all programming attempts, 97.5% were successful with the target conductance reached within 1.1% accuracy from target. Averaging across all trials, the average accuracy is 0.64% across all five decades. Over the course of programming, one device (R4C1) failed when attempting to increase *G* from 816 nS to 8.2 μS and unfortunately the conductance could not be further altered.

| Conductance range | *N* devices | Ave. error ($\overline{\Delta G}$) | Ave. % error |
|---|---|---|---|
| 1 – 10 nS | 16 | 0.034 nS | 0.56% |
| 10 – 100 nS | 16 | 0.65 nS | 1.1% |
| 100 nS – 1 μS | 16 | 3.64 nS | 0.62% |
| 1 μS – 10 μS | 15 | 23.3 nS | 0.46% |
| 10 μS – 100 μS | 15 | 233 nS | 0.45% |

**Supplementary Table 3.** Programming statistics of 4x4 array of 16 ETCRAM devices, listing the average absolute error $\Delta G$ from target for *N* devices and average absolute percent error from target.

## Supplementary Note 14

### Analog matrix-vector multiplication accuracy simulations

Simulations of analog matrix-vector-multiplications (MVMs) in Fig. 3g were conducted using the CrossSim tool for analog in-memory computing accuracy simulations[8]. For this evaluation, we extracted the weight matrix from layer 44 of the ResNet-50 convolutional neural network (CNN)[9], using weights from the MLPerf Inference Benchmark implementation (called ResNet50-v1.5)[10]. This convolutional layer has a 3×3 kernel size, 512 input channels, and 512 output channels, which is shaped into a 2D matrix of dimensions 4608 × 512 (inputs × outputs) to map onto a resistive memory array[7]. Input vectors are generated by executing all previous layers on 100 ImageNet test images. Since there are 196 sliding window MVMs per image, we used a total of 19,600 input vectors, each of which has 4608 elements that are quantized to 8 bits of resolution using the same fixed range. All of the inputs are non-negative due to a preceding ReLU (rectified linear) activation function, which is typical in CNNs. This choice of both the matrix and vectors ensures that the evaluation represents a realistic AI exemplar workload.

In our simulations, each signed weight is mapped to exactly two devices, whose difference in conductance encodes the weight value. One of the two devices is set to the minimum programmable conductance ($G_{min}$), while the other conductance is set between $G_{min}$ and $G_{max}$ and encodes the magnitude of the weight. Random conductance errors are independently sampled and added to the conductance of every device in the array, then these conductances are kept static across MVMs (i.e. errors are modeled as programming error, not cycle-to-cycle read noise). The random errors are assumed to be normally distributed with zero mean, and the standard deviation of the error distribution depends on the target conductance and follows the state dependence shown in Fig. 1d for each of the device technologies being compared.

To ensure a fair device-level MVM accuracy comparison, we assumed the same array electrical topology for all four devices, even if the exact topology used in their associated published accelerator is different. This topology is shown in Fig. 3g (top): input voltages are applied at the ends of the rows, and the same interconnect that carries the input signal supplies the current that flows to the output. The bottoms of the columns are assumed to be held at a virtual ground (0 V) by peripheral circuitry. Metal interconnect resistances induce parasitic IR drops across every row and every column. CrossSim accurately simulates all of these distributed IR drops in every MVM, including their dynamic data dependence and spatial non-uniformity[11]. We use a value of 0.35 Ω, as a typical value of the metal interconnect resistance between two memory cells in both the row and column dimensions[12]. We assume that the devices that carry positive and negative currents are interleaved within the array, allowing local cancellation of currents that reduce IR drops. For arrays with fewer than 4608 rows, the weight matrix is partitioned across multiple arrays which each compute partial dot products (e.g., two partitions for 2304 rows, 64 partitions for 72 rows).

Additionally, the following device-specific settings are used:

| Device | $G_{min}$ | $G_{max}$ | Input signal (8-bit resolution) | Reference |
|---|---|---|---|---|
| ETCRAM | 1 nS | 1.6 μS | 4-bit voltage × 2 cycles | This work |
| SONOS | 10 pS | 16.0 μS | 1-bit voltage × 8 cycles | [13] |
| PCM | 0.47 μS | 25.0 μS | 1-bit voltage × 8 cycles | [3] |
| Memristor | 0.57 μS | 39.3 μS | 1-bit voltage × 8 cycles | [4] |

**Supplementary Table 4.** MVM simulation device parameters.

Notably, due to their nonlinear I-V characteristics, the in-memory computing accelerators based on the SONOS, PCM, memristor devices above applied only binary input voltages, and multi-bit inputs had to be applied either in a bit-serial manner or via pulse width modulation, as otherwise the nonlinearity would induce significant accuracy loss[3,4,13]. Therefore, for accuracy comparisons we simulate SONOS, PCM, and memristor devices assuming bit-serial application of the 8 input bits, with a separate MVM simulated for every input bit. ETCRAM has highly linear I-V characteristics across its entire dynamic range, as shown in Fig. 1c. This means that inputs can be encoded by voltage in the analog MVMs without losing accuracy, but the digital-to-analog converters (DAC) that would be needed to supply 8-bit analog voltages have high energy cost. Therefore, for ETCRAM we simulate the application of 4-bit voltages across two sequential cycles (see note on energy at end of this section).

The dot product errors are shown in Fig. 3g (bottom). These are errors purely in the analog-domain dot products, i.e., we did not include ADCs in the simulation. The errors are computed between the simulated partial dot product values and the ideal partial dot product values. The root-mean-square (RMS) is computed across all partial dot products for the 19,600 input vectors. For the unpartitioned case (4608 rows), error is evaluated over $1.0\times10^7$ dot products, while for the partitioned cases, error is evaluated over up to $6.42\times10^8$ partial dot products (for 72 rows).

To make the results more interpretable, we normalized the dot product errors to the useful signal range, separately for each array size. The useful signal range is defined here to be the range of ideal partial dot product values that contain the inner 99.9% of all values, enforced to be symmetric about zero. This type of method is typically used to set the optimal range of ADCs in analog in-memory computing systems[3]. Therefore, one can approximately say that the 0.4% line in Fig. 3g (right) would correspond to the level spacing of an optimized 8-bit ADC for each given array size.

The un-normalized dot product errors are shown in Supplementary Fig. S17(a), together with the normalized dot product errors in Supplementary Fig. S17(b), which have been reproduced from Fig. 3g.

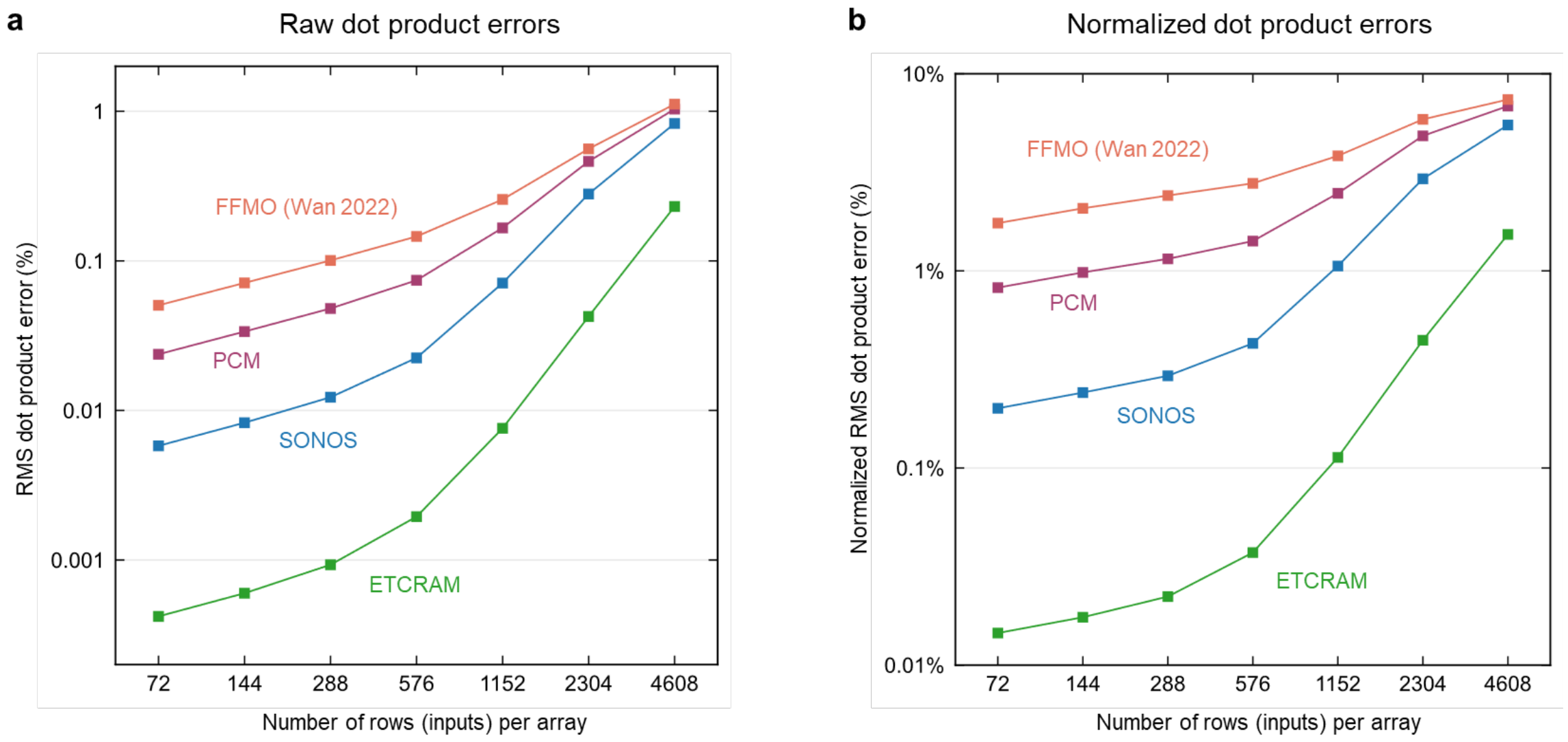

**Supplementary Figure 17.** (a) Comparison of raw (un-normalized) partial dot product errors, (b) Comparison of normalized partial dot product errors.

## Supplementary Note 15

Fast pulse programming

The fast pulse programming measurements shown in Fig. 4b were performed with a Keithley 4200A-SCS using 4225 pulse measure units (PMU) and 4225 remote amplifiers (RPM). Load line effect compensation (LLEC) was employed during fast pulse programming because the impedance of the heater (R ~ 15 Ω) is smaller than the output impedance of the PMU (50 Ω). While programming pulses are applied to the heater of a device, the channel is grounded. The conductance of the channel was measured with a source measure unit (SMU) with V = 0.5 V DC bias while the heater is grounded with another SMU.

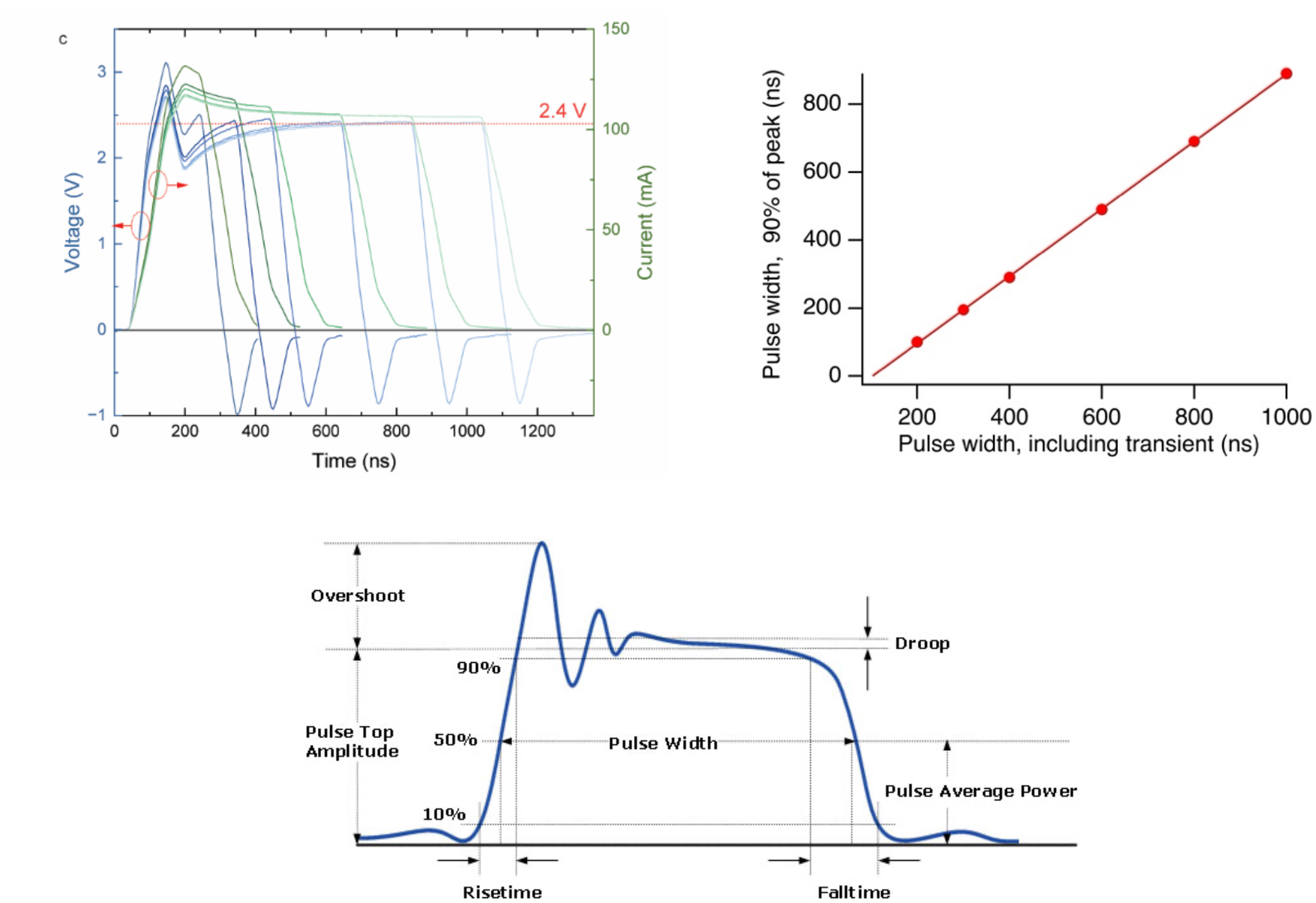

**Supplementary Figure 18.** Fast programming pulses sourced and measured using Keithley 4200A-SCS 4225 PMUs (left) showing voltages and currents of fast pulses of varying durations with 2.4 V of amplitude. Right: Linear fit relating the as-measured pulse width to the calculated pulse width using the 90% of current peak definition[14] (below).

An example of pulses applied to the gate for fast pulse programming is shown in Supplementary Fig. S18. The pulse amplitude is $V_g$ = +2.4 V and the pulse width is varied from 200 ns to 1 µs. Due to the impedance on the cables connecting the PMU and the devices, an initial transient voltage overshoot in the measured pulse is present. The true pulse width at the ETCRAM device can be obtained after correcting for the transient component in the first 100 ns of the pulse. Analysis of the full 200 ns - 1000 ns voltage and current waveforms in Supplementary Fig. S18 shows a linear $y = mx + b$ dependence ($a$ = -102, $b$ = 0.99) of the actual pulse width, defined as 90% of the peak current through heater (Supplementary Fig. S18, right), excluding the risetime and fall time. Extrapolation of this linear relationship to faster <200 ns pulses was used to correct for the initial transient and to better estimate the duration of the 15 ns – 45 ns pulses in Fig. 4b. The data in Fig. 4b is normalized to the initial conductance $G_0$ for each run and was on average about 9.4 µS across the five durations. It should be noted that the decreasing update strength in the Fig. 4b is consistent with nanosecond pulses of decreasing duration being applied to the gate, with the transient duration constant.

To explore this relationship in more detail, a matrix of programming voltages and pulse durations was tested in a device, and the channel conductance change was monitored. Heat maps showing the dependance of $\Delta G = (G-G_0)/G_0$ on voltage and duration are shown in Fig. S19. It generally observed that higher voltages and longer durations increase the magnitude of $\Delta G$ for potentiation and depression.

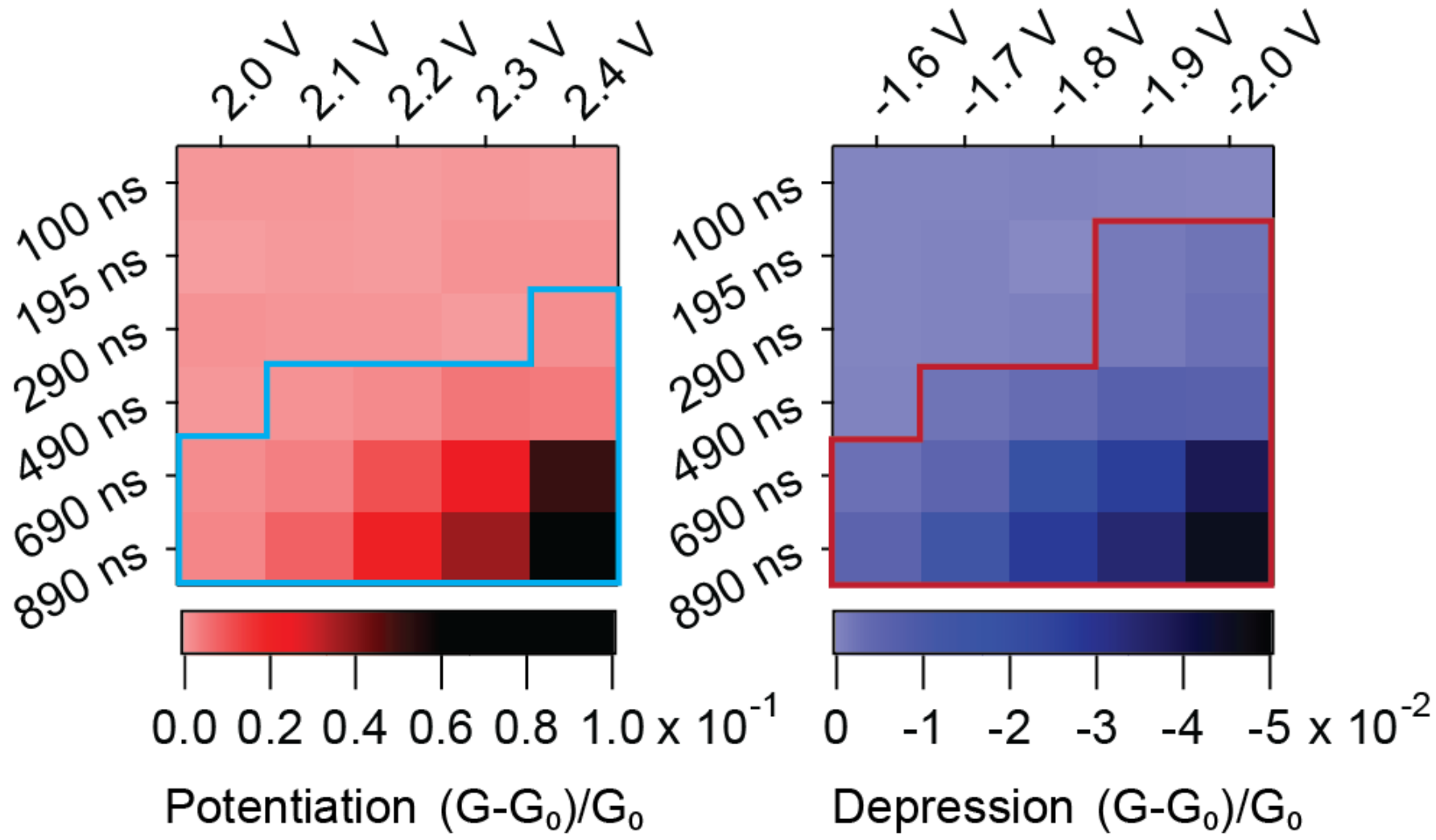

**Supplementary Figure 19.** Heat maps showing the update strength $\Delta G = (G-G_0)/G_0$ for different pulse conditions. Left is potentiation and right is depression. The colored outlines indicate values where $\Delta G$ is above the noise.

**Supplementary Note 16**

Retention of ETCRAM devices and determination of activation energy

Data showing the long-term stability of an ETCRAM device at *T* = 85 °C are shown in Supplementary Fig. S20(a) for two different programmed channel states: high (2 μS) and low (250 nS). This device is based on a 2 μm $MoSi_2$ heater, with the addition of a top epoxy coating for passivation. Devices were mounted to a metallic chip carrier in good thermal contact with the heating stage (Instec) and device pads were wire bonded to the chip carrier leads. All measurements were under high vacuum (P ~ 5 x $10^{-6}$ Torr). During state readout, V = 0.1 V read pulses were applied to the channel while the gate contacts were simultaneously held at V = 0 V (Keysight B1500). At *T* = 85 °C, no state loss is observed for any state in Fig. 4d.

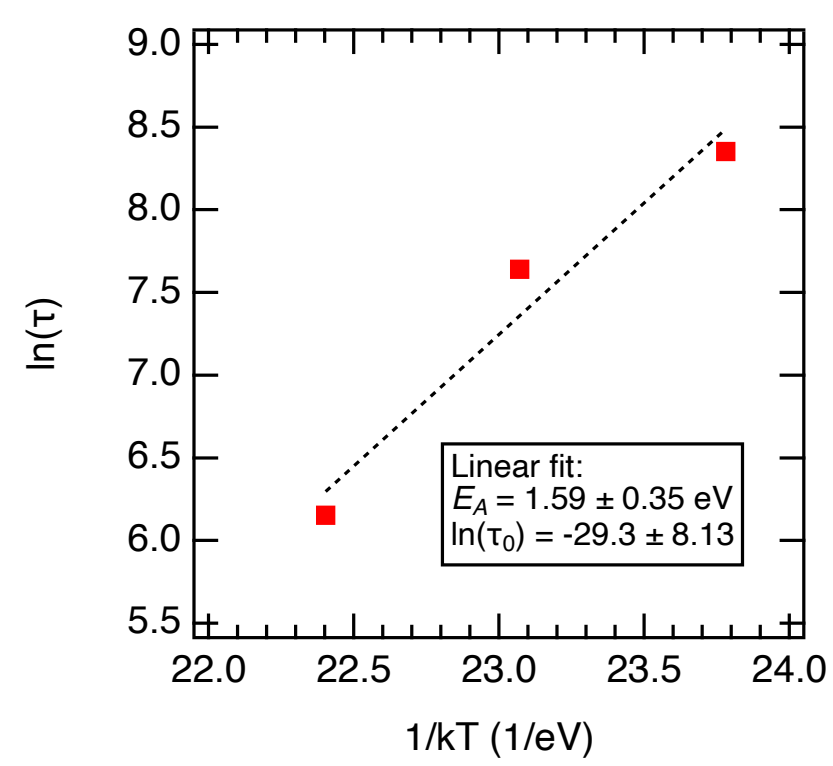

**Supplementary Figure 20.** Arrhenius plot showing estimate of the activation energy for these devices based of decay times at *T* = 215 °C, *T* = 230 °C, and *T* = 245 °C.

It is possible to observe detectable state loss only if devices are heated above *T* = 200 °C. In this regime, the temperature-dependent state loss from initial states at $G_0$ = 300 nS was recorded for three temperatures. We assume a thermally activated Arrhenius behavior between the decay time $\tau$ and the device temperature T: $\tau(T) = \tau_0 e^{E_A/kT}$. The decay time was defined as the time it takes for the state to decay to 5% of its initial value. Plotting the logarithm of $\tau(T)$ versus $1/k_B T$ allows determination of the activation energy of the loss process $E_A$ in this system: $\ln\tau(T) = \ln\tau_0 + \frac{E_A}{k_B T}$ (Fig. S20). A linear fit to the data yields $E_A = 1.59 \pm 0.35$ eV. This value is consistent with measurements of vacancy migration energies in the $TaO_x$ system with tracer diffusion experiments. In this case, $E_v$ = 1.5 ± 0.2 eV and $E_v$ = 1.6 ± 0.2 eV for the two suboxides $TaO_{0.7}$ and $TaO_{1.5}$, respectively[15]. Additionally, we observed that the devices were not damaged after extended heating up to $T$ = 300 °C during the retention measurements and continue to program normally after being cooled to room temperature.

## Supplementary Note 17

Energy advantage of ETCRAM

There are three aspects of ETCRAM that allow for extremely high-efficiency analog MVM operations while retaining high accuracy:

(1) Access to **very low conductance** (<100 nS), which enables ETCRAM to scale to larger arrays. This is critical, since the output peripheral circuits (e.g., operational amplifiers and ADCs) dominate the energy consumption of analog systems[12,16-18], and large arrays can amortize these costs across more operations. For the same MVM accuracy in Fig. 3g, ETCRAM can scale to an array size that is 3.2× larger than SONOS, 22× larger than PCM, and 64× larger than the memristor device[12].

(2) **Linear I-V characteristics**, which enables voltage-mode MVM operation rather than current-mode MVMs. This scheme is shown in Supplementary Fig. S21. In this configuration, the columns of the array are not held at virtual ground by operational amplifiers at the periphery. Rather, each column is effectively a voltage divider network with programmable resistors, where the voltage on the column will settle to an intermediate analog value that is between the maximum and minimum possible input voltages. For linear resistive memory devices, the output voltage on the j$^{th}$ column will be:

$$V_{\mathrm{out,j}} = \frac{\sum_{\mathrm{i=0}}^{N} G_{\mathrm{i,j}} V_{\mathrm{in,i}}}{\sum_{\mathrm{i=0}}^{N} G_{\mathrm{i,j}}}$$

where N is the number of rows in the MVM. The numerator represents the dot product, while the denominator that depends on the weights in each column but does not change dynamically with the inputs; therefore, the dot product can be recovered with a scaling step by a per-column constant factor after the ADC. A major advantage of voltage-mode MVM compared to current-mode MVM is that an operational amplifier is not needed on every column to hold its voltage at a virtual ground, which leads to significant MVM energy savings.

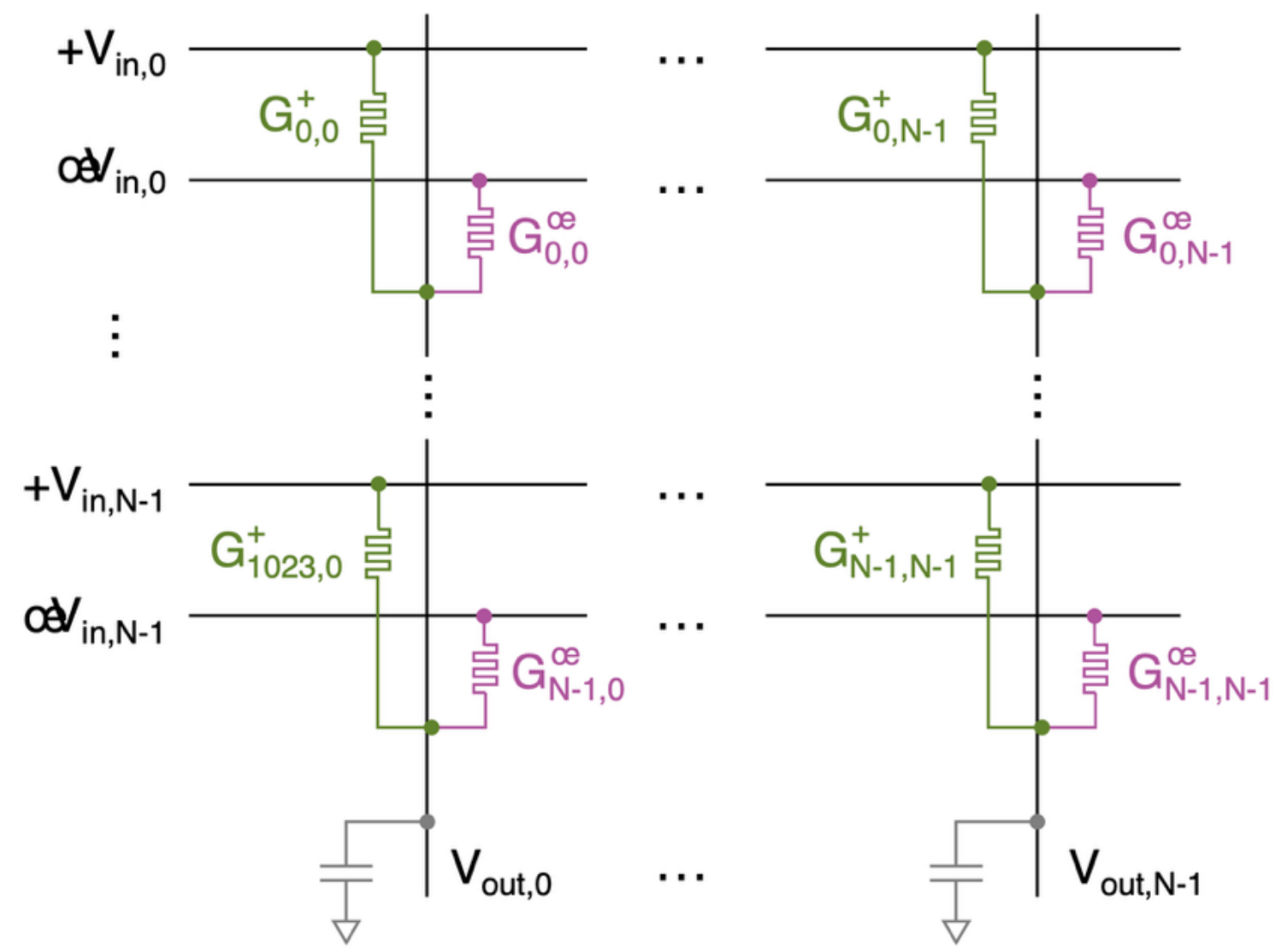

**Supplementary Figure 21**. Diagram of an analog voltage-mode MVM with an N×N matrix. The capacitor is the effective capacitance of the column.

The output voltage $V_{out,j}$ can take on a continuous range of values even if the input voltages $V_{in,i}$ are binary (i.e. digital inputs are applied one bit at a time). Therefore, accurate voltage-mode MVM requires memory devices with highly linear I-V characteristics across all conductance states. ETCRAM has highly linear I-V characteristics across at least five orders of magnitude in conductance, which can enable accurate and efficient voltage-mode MVMs. Voltage-mode MVM has been demonstrated in a memristor-based accelerator[12], but I-V linearity was neither shown nor claimed, which likely contributed significantly to the observed MVM inaccuracy.

I-V linearity also allows the device to perform multiplications with analog voltage inputs, whether via current-mode or voltage-mode MVMs, to enable one-shot processing of multi-bit inputs. Memories like SONOS or PCM cannot accommodate analog voltage inputs because their nonlinear I-V characteristics would cause the conductance (i.e. weight value) to be a function of the applied voltage. Therefore, prior SONOS-based[13] and PCM-based accelerators[17,19,20] apply strictly binary input voltages, and support multi-bit inputs either through bit-serial application of input bits or pulse width modulation. These options would cause the MVM latency and total energy consumption of the peripheral integrator circuits and/or ADC to increase linearly or super-linearly with the input bit resolution. The I-V linearity of ETCRAM can support analog voltage inputs with one-shot processing of all input bits, which ensures low output peripheral circuit energy consumption even with high-resolution digital inputs.

(3) **Large conductance dynamic range**. Converting high-resolution digital inputs to analog voltages can consume significant energy if using a high-resolution DAC, because the energy, area, and/or latency cost of the DAC typically scales super-linearly with input resolution. One solution to this problem is to use lower-resolution DACs and use the dynamic range of the device conductance to support part of the dynamic range of the inputs. As an example, we start with a voltage $V_i$ that is generated by an 8-bit DAC, a device conductance $G_{i,j}$ that encodes an 8-bit weight, resulting in a current that encodes the product of two 8-bit values. Instead, the same current can be achieved by summing the current from two ETCRAM devices:

$$I_{i,j} = G_{i,j} V_{i,\,b4\text{-}7} + \left(\frac{1}{2^4} G_{i,j}\right) V_{i,\,b0\text{-}3}$$

where $V_{i,b4\text{-}7}$ and $V_{i,b0\text{-}3}$ are voltages generated by 4-bit DACs which encode the higher-significance 4 bits and lower-significance 4 bits of the original 8-bit input, respectively. The two voltages have the same range, i.e. use the same 16 voltage levels, and the lower significance of $V_{i,b0\text{-}3}$ is accounted for by scaling the conductance of the corresponding device by a factor of 16. The conductance of the device must now have 12 bits of dynamic range (4096×). This amount of range is not available in most synaptic memory devices such as memristors and PCM, but is well within the full dynamic range of ETCRAM (>$10^6$), without loss of accuracy since the conductance errors $\sigma_G$ in ETCRAM scale proportionally with the conductance – see Fig. 1d. The reduction in DAC resolution enabled by ETCRAM allows high-resolution digital MVM inputs to be supported with zero latency cost, a linear area cost with input bits, and a linear DAC energy cost with input bits. The >$10^6$ conductance dynamic range of ETCRAM can support input and weight precision scaling beyond 8 bits, all processed in a one-shot analog MVM.

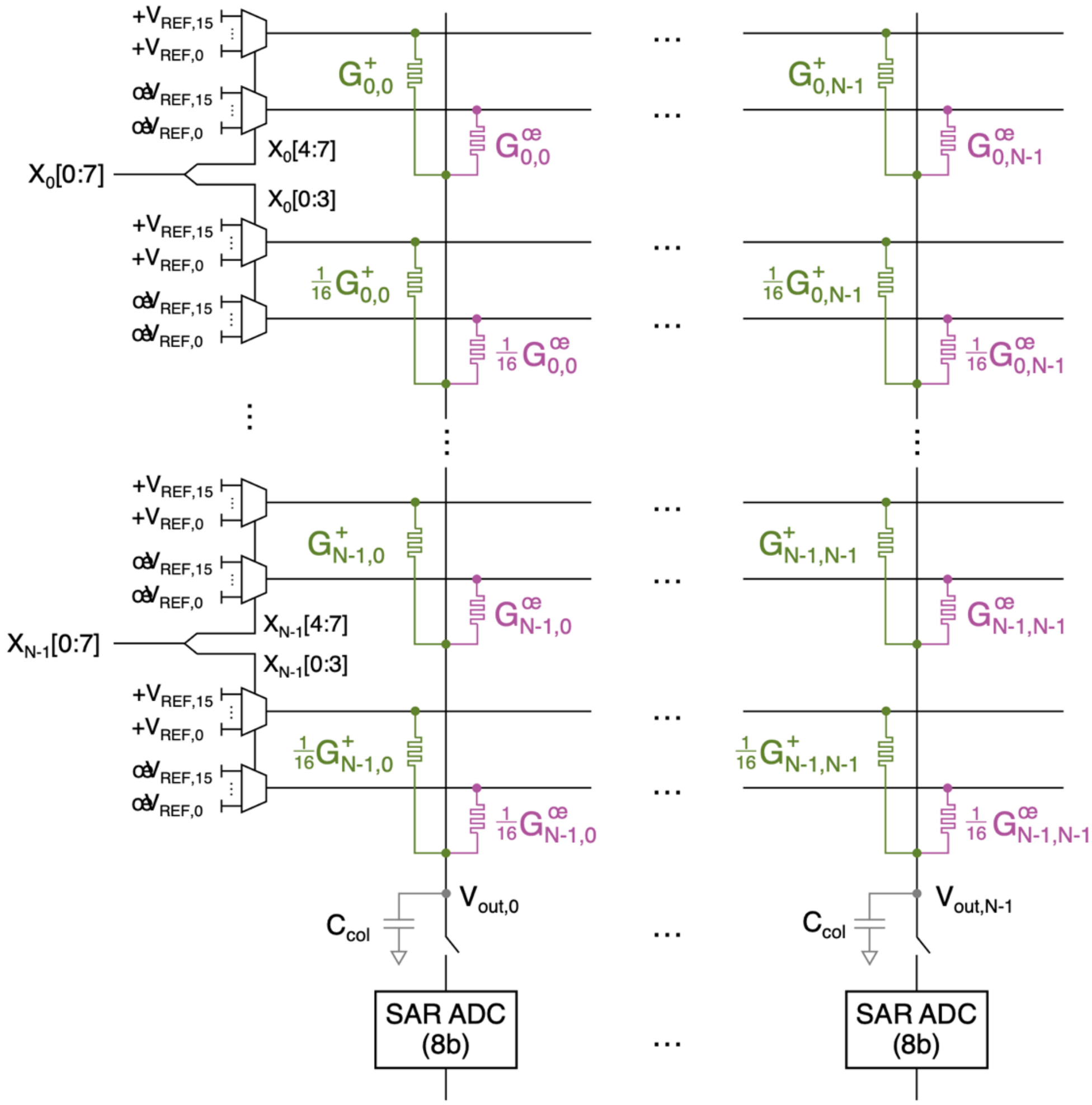

**Supplementary Figure 22**. Voltage-mode MVM implementation with 8-bit inputs and 8-bit outputs. The 8-bit inputs are decomposed into two 4-bit slices.

Supplementary Fig. S22 shows an implementation of a voltage-mode MVM using ETCRAM with 8-bit digital inputs and 8-bit digital outputs. We use the above technique to split the 8-bit inputs into 4-bit slices, and since the 4-bit voltages have the same range, only 16 voltage levels are required to support 8-bit inputs. A complementary set of 16 voltages can be used to support negative weights. For positive weights, $G^- = G_{off}$, and for negative weights, $G^+ = G_{off}$. Using ETCRAM, it is possible to set $G_{off} \approx 1$ pS – see Fig. 1c – so that one branch always draws nearly zero current. The equally spaced reference voltages can be generated by a circuit that is shared by the full array, e.g. a resistor ladder. 4-bit multiplexers select one of the 16 voltages to apply to each row. This method of supplying multi-level voltages for analog MVM was implemented in a recent work[21].

**Energy estimate**: We now estimate the energy efficiency of the voltage-mode MVM using the scheme in Supplementary Fig. S22 on a neural network workload, assuming that the ETCRAM device can be scaled to a moderately advanced process node, e.g. 22nm. We also assume N=2304, e.g. a large 2304×2304 MVM, where ETCRAM can still retain a dot product error of less than 0.5% – see Fig. 2f. We consider three components of the total energy: (1) $E_{array}$, the energy dissipation in the ETCRAM array, (2) $E_{ADC}$, the energy consumption of ADCs, (3) $E_{DAC}$, the energy consumption to provide the multi-level input voltages, and (4) $E_{scale}$, the energy for the per-column digital scaling necessary for voltage-mode MVM, described above.

For voltage-mode MVM, the ETCRAM array is driven by the input voltages and the column voltages settle to stable values. After settling, these voltages are connected to the inputs of 8-bit ADCs to be converted to digital outputs. The array must draw current during this entire duration, therefore:

$$E_{\text{array}} \;=\; 2\times\left(1+\frac{1}{16}\right)\times N^2 G_{\text{avg}} V_{\text{max}}^2 \times (\tau_{\text{col}}+\tau_{\text{ADC}})$$

where $G_{avg}$ is the average ETCRAM conductance, $V_{max}$ = 1V is the maximum input voltage, $\tau_{col}$ is the settling time of the column voltages, and $\tau_{ADC}$ is the latency of the analog-to-digital conversion. The factor of 2 accounts for positive and negative weight components and the 1/16 factor accounts for the devices with lower conductance that support the lower 4-bit slice of the digital inputs. For the conductances, we use $G_{max}$ = 1.6 μS (consistent with Supplementary

Table 4). The weight matrices of neural networks tend to concentrate near low values; e.g. for the ResNet-50 benchmark CNN, the average conductance of a device is only ~2% of $G_{max}$[22], which leads to $G_{avg} \approx 32$ nS.

The 90% settling time of each column's voltage divider is given approximately by:

$$\tau_{col} = 2.2 \times \frac{C_{col}}{\left(1+\frac{1}{16}\right) \times N \times G_{avg}}$$

where $C_{col}$ is the capacitance of the column. Assuming a typical scaled CMOS metal interconnect capacitance per unit length and that an ETCRAM cell with one select transistor can be implemented in a $12F^2$ cell area with F=22nm, we estimate the total capacitance of a column of 4608 cells to be 70.2 fF. We round this up to $C_{col}$ = 100 fF to account for other parasitic capacitances on the column. This leads to a settling time of $\tau_{col} \approx 2.8$ ns.

We assume an 8-bit successive approximation register (SAR) ADC on each column that can digitize an analog voltage with a latency of $\tau_{ADC}$ = 8 ns, i.e. 1 ns per bit. The total energy consumed by the ETCRAM array for this MVM size is: $E_{array}$ = 3.90 nJ.

For SAR ADC energy consumption, we start with an expression for the lower limit of ADC energy consumption, based on a widely known survey of published ADCs.[23] For an ADC with ENOB (effective number of bits) = 8, this limit is 0.28 pJ/conversion. We allow a realistic 8-bit SAR ADC to operate at 4× lower efficiency, or 1.12 pJ/conversion. The total energy of 2304 ADC conversions is therefore: $E_{ADC}$ = 2.58 nJ.

We estimate the energy needed to drive the 4-bit voltage inputs based on prior work by Lee et al, which implemented a similar method in 28nm silicon[21]. Chip measurements by Lee et al quantified the total energy to supply 1152 4-bit reference voltages to be 110 pJ, or 95 fJ per input. The array in Supplementary Fig. S22 requires 4N or 9216 such inputs, for a total energy of $E_{DAC}$ = 0.88 nJ.

The digital scaling correction for voltage-mode MVM can be applied locally, immediately after digitizing the MVM outputs. We conservatively estimate this cost to be 0.2 pJ per column, following the cost of an 8-bit multiplication in a 45nm process.[24] The total energy for scaling 2304 dot products is $E_{scale}$ = 0.46 nJ.

Finally, the energy per operation of the ETCRAM-based voltage-mode MVM is estimated to be $(E_{array} + E_{ADC} + E_{DAC} + E_{scale}) / 2N^2$ where the factor of 2 is based on the convention that each multiply-accumulate is two operations; one multiplication and one addition. Based on the energy values calculated above, the efficiency is **0.74 fJ/op**, or equivalently, **1360 TOPS/W** at 8-bit data resolution.

We note that this energy efficiency is a projection of what is uniquely enabled by the ETCRAM device, using circuit techniques that have been demonstrated in the literature. However, it is not based on a detailed or optimized circuit implementation, where other unforeseen losses of efficiency or accuracy may arise, and other design trade-offs may need to be considered.

## Supplementary Note 18

Determination of temperature rise during ETCRAM programming

To determine the temperature rise of ETCRAM devices, the temperature dependance of the Pt heater resistance can be used to estimate the temperature reached during programming. The temperature coefficient of resistance (TCR) of Pt is commonly used to measure temperature thanks to the excellent linearity of the Pt TCR over wide temperature ranges. Due to the thin nature of the Pt heater (d < 100 nm), the bulk Pt TCR value of $\alpha$ = 0.0039 /°C at 20 °C cannot be assumed to be true here must be independently determined.

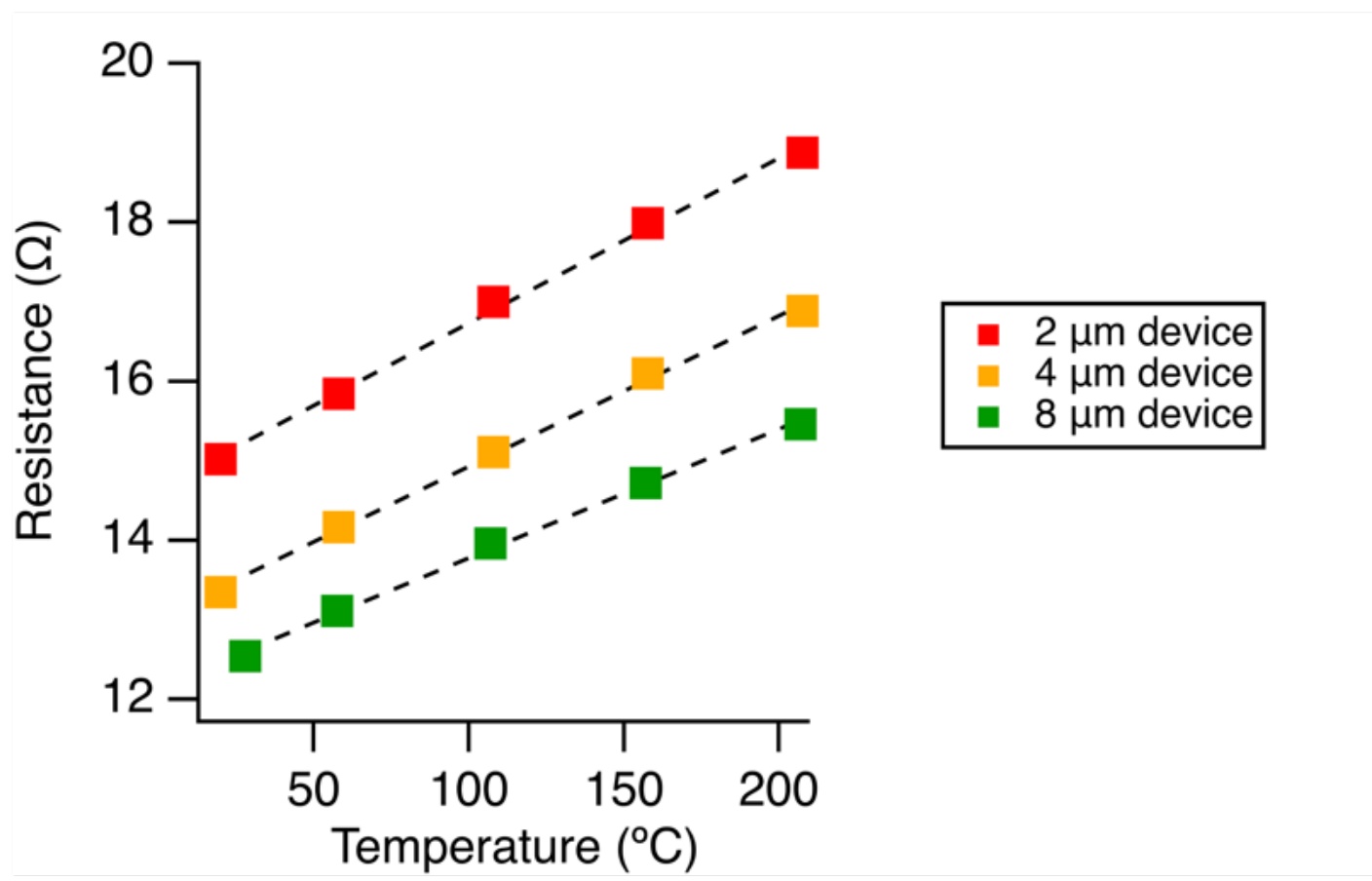

**Supplementary Figure 23**. Pt heater resistance as a function of temperature.

Complete ETCRAM devices with Pt heaters of three different sizes (8 µm, 4 µm, 2 µm) were placed on a Lakeshore temperature stage ($T$ = 293 K - 480 K) and the Pt resistance was measured using the slopes of the characteristic I-V curves. Linear fits to the resistances yield TCR values of $\alpha_{8\mu m}$ = 0.0162, $\alpha_{4\mu m}$ = 0.0190, and $\alpha_{2\mu m}$ = 0.0208 ($\alpha_{Ave}$ = 0.0186 /°C), which show a slight dependence on size presumably due to deviations from ideal dimensions in photolithography. A small increase in the room temperature resistance was also observed when devices are made smaller, despite identical Pt deposition conditions. To estimate the temperature of ETCRAM device during programming, the heater resistance ($R_{Pt} = V_g/I_{Pt}$) was measured with the DAQ *in operando* and the change in resistance was converted to a temperature rise using $\Delta T = \Delta R_{Pt}/\alpha_{\mathrm{Ave}} = (R_{\mathrm{Pt}} - R_{\mathrm{Pt,20°C}})/\alpha_{\mathrm{Ave.}}$ From this analysis, ETCRAM devices are estimated to typically program in the range of $\Delta T$ = 250°C - 450°C above room temperature, depending on the value of $V_g$ and desired update strength $\Delta G$.

**Supplementary Note 19**

Modeling of Joule heating

We model Joule heating by solving the heat conduction equation in two dimensions for the geometry of Supplementary Fig. S24. A Joule heated wire of length *L* and thickness $t$ has a uniform power $P = IV$ flowing through it. Above the wire we have vacuum extending up to a height $H_{vac}$ while below the wire is the dielectric stack of thickness $h$. Below the stack is the Si substrate of thickness $H_{sub}$. The temperature is divided into four contributions $T_1,...,T_4$ corresponding to the 4 domains.

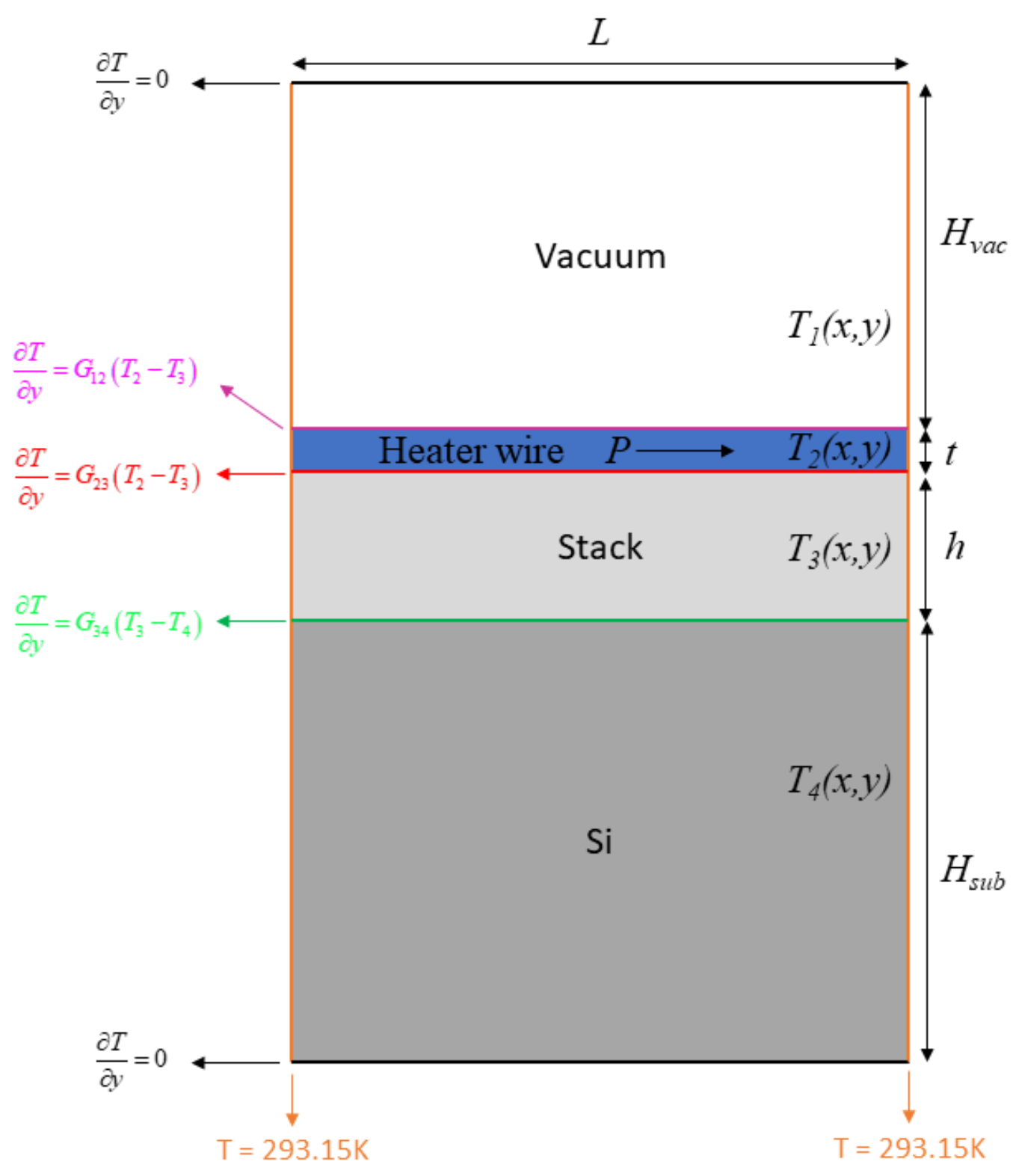

**Supplementary Figure 24**. Geometry for simulations of Joule heating in the device stack. Current of uniform power density *P* flows in the blue wire of thickness *t*. Above the wire, there is heat transfer to vacuum with thermal conductance $G_{12}$. Below the wire, the device stack is modeled as one material of thickness *h* with heat transfer conductance $G_{23}$. The device stack sits on top of Si of thickness $H_{sub}$ with heat transfer conductance $G_{34}$. The temperatures in the four domains are labelled $T_1(x,y)$ to $T_4(x,y)$. At the right and left edges of the domain we use fixed temperature boundary conditions $T = 293.15K$. Above and below the domain we use the conditions of zero temperature gradient.

We solve the heat conduction equation in each of the four domains

$$\kappa_i \nabla^2 T_i = \begin{cases} -Q \text{ in domain 2} \\ 0 \text{ otherwise} \end{cases} \tag{0.1}$$

where $\kappa_i$ is the thermal conductivity in domain *i*, and $Q$ is the power density. For uniform Joule heating with power $P = IV$ in a wire of length $L$, width $W$, and thickness $t$ we have

$$\kappa_i \nabla^2 T_i = \begin{cases} -\dfrac{P}{WLt} \text{ in domain 2} \\ 0 \text{ otherwise} \end{cases}. \tag{0.2}$$

We solve these equations with appropriate boundary conditions at each of the interfaces. The left and right edges of the computational domain are held at a fixed temperature (293.15 K) while at the top and bottom there is a no-flux condition. Between the materials we utilize thermal boundary resistance conditions with the thermal interfacial conductances $G_{ij}$ linking the various domains. The values of the different parameters are shown in Supplementary Table 5. Vacuum was represented with low values of the thermal conductivity and the heat transfer coefficient, with the value of $G_{12}$ similar to that for radiative heat loss. The heat transfer coefficients $G_{23}$ and $G_{34}$ were chosen to be of the same order of magnitude as values previously measured for high thermal conductivity solid/solid interfaces.

| | | | |
|---|---|---|---|
| $\kappa_1$ | 0.1 W/mK | $G_{12}$ | 30 $W/m^2K$ |
| $\kappa_2$ | 10 W/mK | $G_{23}$ | $10^9$ $W/m^2K$ |
| $\kappa_3$ | 1 W/mK | $G_{34}$ | $10^9$ $W/m^2K$ |
| $\kappa_4$ | 148 W/mK | $H_{sub}$ | 600 μm |
| $H_{vac}$ | 500 μm | $t$ | 5 nm |
| $h$ | 250 nm | | |

**Supplementary Table 5.** Values of parameters used in the thermal simulations

The above equations were solved using the COMSOL finite element simulations software, employing the Heat Transfers in Solids module. A non-uniform mesh was used and refined until results were converged. We simulated various channel lengths *L* from 50nm to 500nm, with the width of the wire the same as *L* when setting the power density. For each channel length we found the power needed to give a 300 K temperature increase in the middle of the channel. Supplementary Figure S25 shows the needed power as a function of *L*. Starting from the longer channels of 500 nm we see a linear decrease of the needed power down to about 100 nm in channel length, followed by an increase as the channel length further decreases. At the minimum point the power is 320 μW.

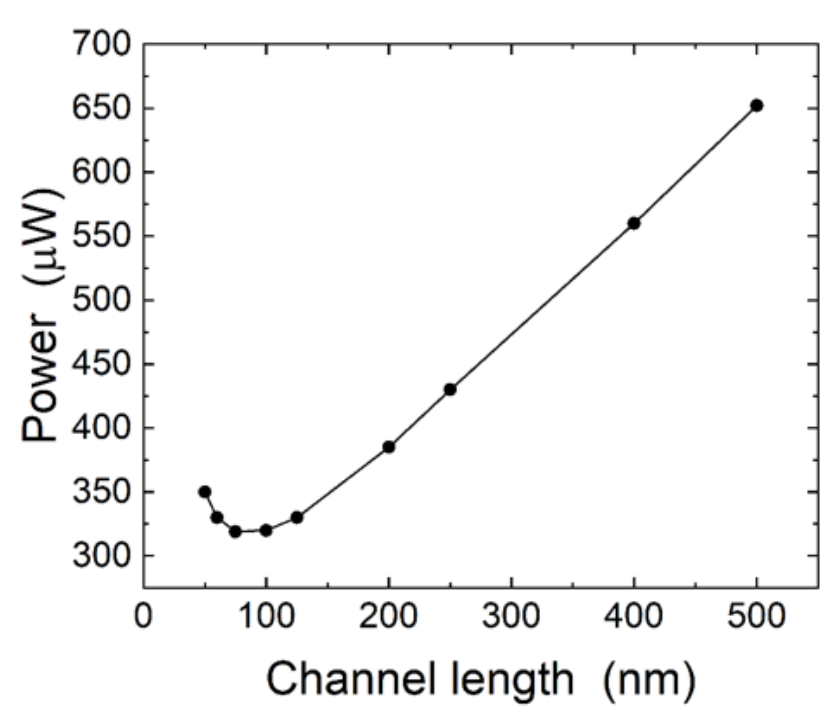

**Supplementary Figure S25.** Needed power to reach a 300 K temperature increase in the middle of the channel as a function of the channel length.

Fig. S26 shows corresponding two-dimensional temperature profiles for three different channel lengths, showing the variation of the temperature profile versus *L*. In all situations, a 300 K temperature increase is maintained at the middle of the channel for all *P* and *L*, and a diminishing of temperature near the device edges. Additionally, a decrease of heat into the Si substrate (e.g., temperature at *d* = 100 nm below the device) is observed as *P* and *L* are reduced. Modeling the full three-dimensional heat transport in COMSOL shows agreement with the above 2D results. The thermal profile in the y-z plane of the *L* = 250 nm device is shown in Fig. S27.

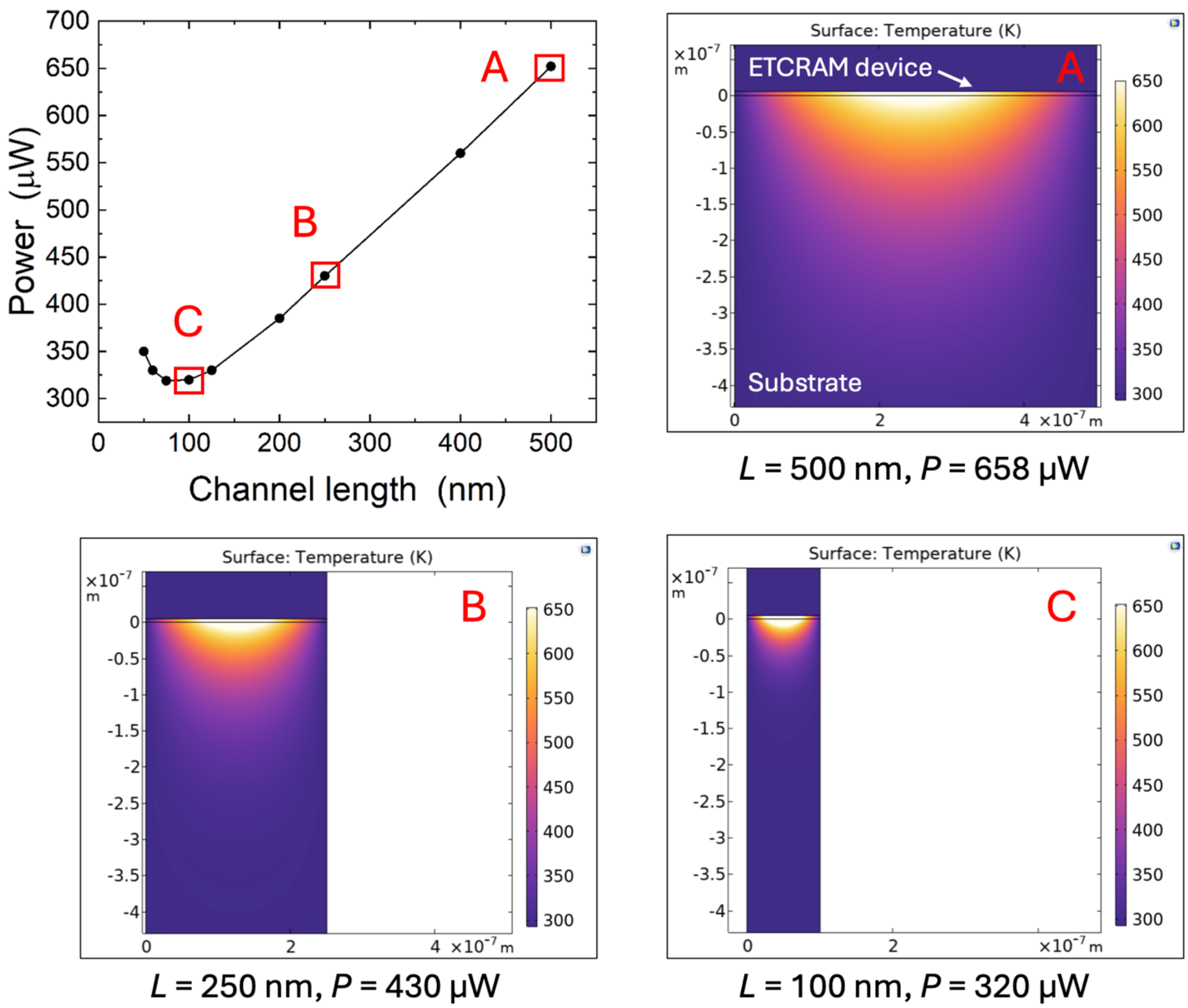

**Supplementary Figure S26.** 2D temperature profiles for three different device dimensions and corresponding powers. The ETCRAM device stack is indicated by the thin rectangle at the top of the profile, and the Si substrate is the large region below.

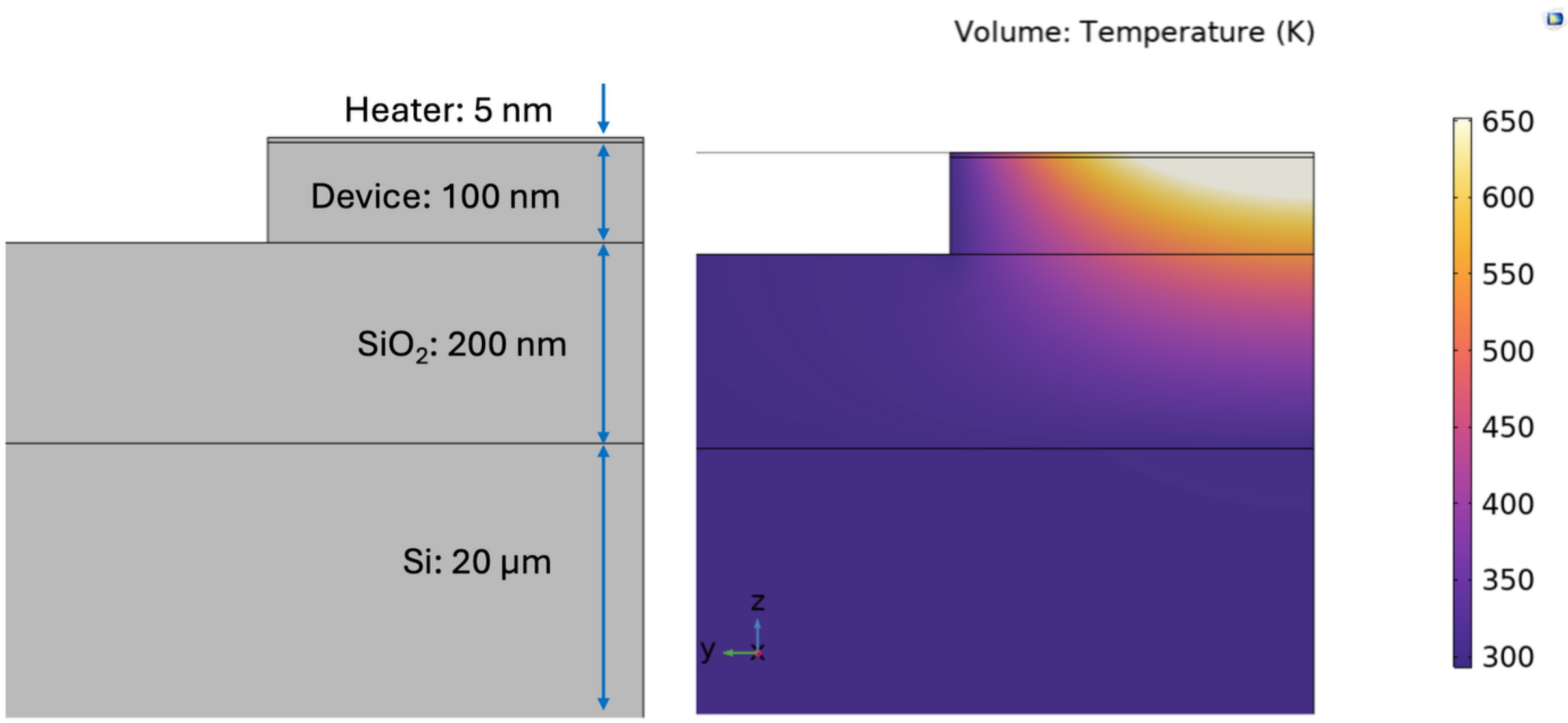

**Supplementary Figure S27.** 3D temperature profiles for an ETCRAM device with *L* = 250 nm. The different layers and their thicknesses are shown on the left, and the temperature through the stack on the right.

**Supplementary Note 20**

Neural network inference accuracy comparison of memory devices

Here, we assess the accuracy of different memory technologies for AI inference applications. We performed end-to-end accuracy simulations of neural networks, which is built up from multiple arrays of memory devices that each implement a convolution or fully-connected layer of the network. We focus on computer vision workloads and select benchmark tasks and CNNs at multiple levels of size, complexity, and error tolerance. The workloads are summarized in the top part of Supplementary Table 6. MNIST[25], CIFAR-10[26] , and ImageNet[27] are three well-known benchmark datasets for image classification, in increasing order of difficulty. We also perform object detection on Microsoft's COCO test set[28] and use the standard accuracy metric of mean average precision (mAP), averaged over bounding box intersection-over-union (IoU) thresholds from 50% to 95% in 5% increments. The number of test images is 10000 for MNIST and CIFAR-10, 50000 for ImageNet, and 5000 for COCO.

Notably, none of the CNNs used for evaluation were prepared using hardware-aware training methods. We used either pre-trained weights from the web (for ResNet-50[10] and YOLOv9-C[29,30]) or trained the weights using standard methods in machine learning (for ResNet-32[31] and LeNet-5[25]) without any specialization to analog hardware. Then, the trained weights are transferred directly to the target values of device conductance.

We simulated analog MVMs for each device technology following the same methodology and assumptions that were described in Supplementary Note 14, with two differences. First, for all of the devices, we included the effects of 8-bit input quantization and 8-bit ADC resolution, with limits that were calibrated to each layer of each CNN. This includes the "ideal device", which represents neural network weights with perfect precision. Second, to keep these large-scale CNN simulations computationally tractable, we did not simulate the effect of parasitic IR drops.

The accuracy results for the different devices are shown in Supplementary Table 6, which report the mean and standard deviation of the accuracy across 10 inference simulations of the full test set with re-sampled random conductance errors.

| | **Tasks** | | | |
|---|---|---|---|---|
| Dataset | **MNIST** | **CIFAR-10** | **ImageNet** | **COCO** |
| Neural network | LeNet-5 | ResNet-32 | ResNet-50 | YOLOv9-C |
| # weights | $6.18 \times 10^4$ | $4.68 \times 10^5$ | $2.56 \times 10^7$ | $2.53 \times 10^7$ |
| Accuracy metric | Classification accuracy (top-1) | Classification accuracy (top-1) | Classification accuracy (top-1) | Object detection $mAP_{50\text{-}95}$ |
| **Device** | | | | |
| Ideal device | 99.53% | 91.88% | 76.31% | 51.73% |
| ETCRAM | 99.53% ± 0.01% | 91.93% ± 0.06% | 76.28% ± 0.04% | 51.65% ± 0.04% |
| SONOS[13] | 99.53% ± 0.02% | 91.70% ± 0.12% | 74.46% ± 0.48% | 46.34% ± 1.40% |
| PCM[3] | 99.44% ± 0.06% | 88.71% ± 1.17% | 23.89% ± 7.53% | 0.81% ± 0.39% |
| Memristor[4] | 98.88% ± 0.36% | 70.70% ± 8.15% | 0.13% ± 0.03% | 0.003% ± 0.002% |

**Supplementary Table 6.** Comparison of simulated inference accuracy using several device technologies on computer vision workloads of increasing complexity.

All four devices can achieve high accuracy on simple tasks such as MNIST. However, in the absence of hardware-aware training, the more complex and realistic AI workloads (e.g. ImageNet and COCO) can only be executed accurately using devices that have higher conductance precision. Only ETCRAM can achieve an accuracy that is close to ideal on COCO object detection, which is both the most difficult and the most commercially relevant of the four AI applications. Based on the results in Fig. 3g, when IR drop effects are included, we expect the accuracy advantage of ETCRAM to be even greater, particularly for large networks that use larger arrays (e.g. ResNet-50 and YOLOv9-C). While hardware-aware training can improve the attainable accuracy using the other devices, it imposes potentially large overheads on machine learning training workflows that are typically already complex.

## Supplementary Note 21

CMOS integration and 2.5D analog processing platform

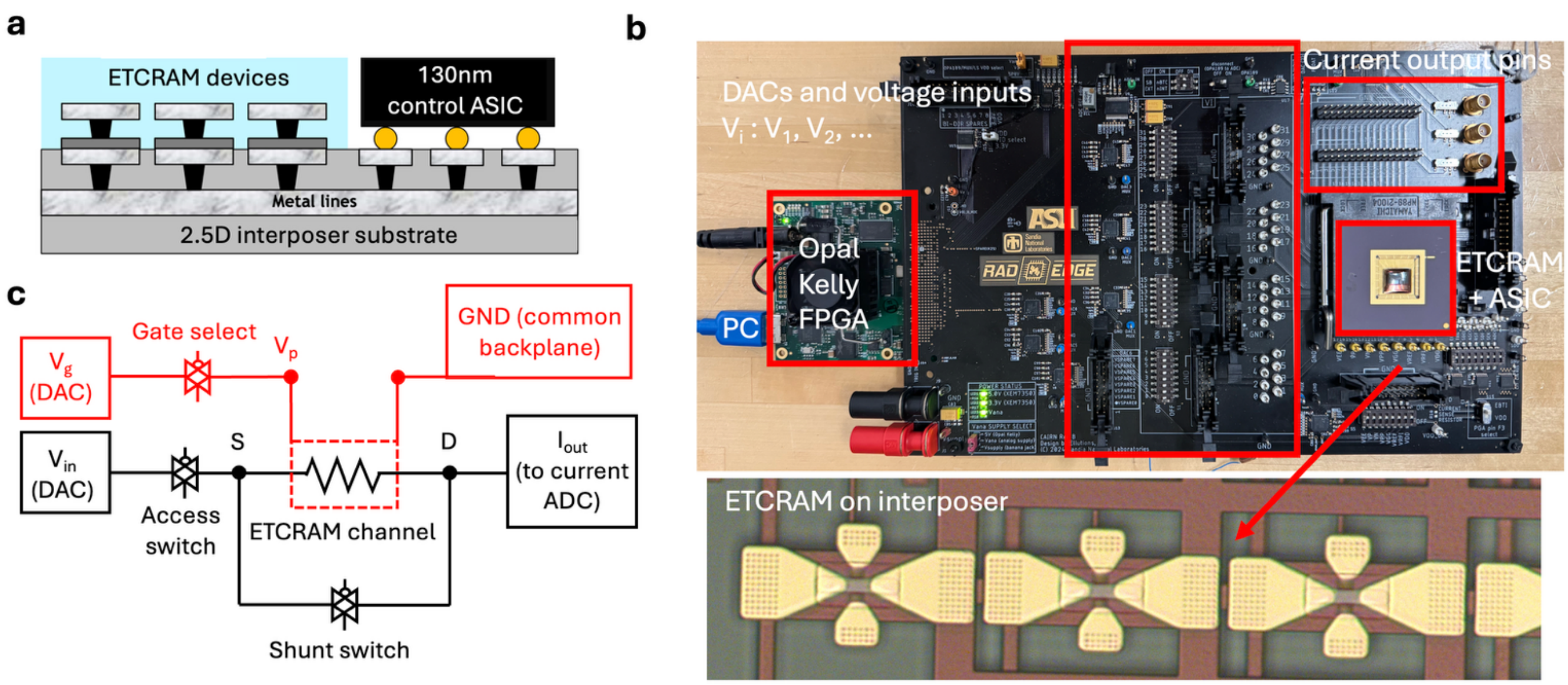

**Supplementary Figure S28.** Integrated ETCRAM test platform. (a) Schematic of the 2.5D integration scheme employed in this work. (b) Corresponding images of the device test platform, with the main components indicated by the red boxes. Several adjacent devices are shown on the interposer substrate. (c) Circuit diagram of an ETCRAM device connected to the test platform, indicating the positions of the ETCRAM terminals relative to the other components on the test platform.

To demonstrate basic functionality of ETCRAM in a CMOS-integrated chip, we have developed a 2.5D analog memory characterization platform (Fig. S28a) that enables electrical characterization and information processing applications of ETCRAM devices. A chip containing multiple individual ETCRAM devices was integrated on a 2.5D silicon interposer substrate with a 130nm ASIC chip, which contains CMOS access transistors and control circuitry. The ASIC was connected to a printed-circuit board (PCB) containing DACs and input/output pins, and an Opal Kelly FPGA was used to interface between the components on the PCB with a PC using a Verilog/Python interface. Fig. S28b shows a photo of the testing board and other main components.

ETCRAM devices in the CMOS chip are similar in design to the four-contact standalone devices in the present work and are based on a 2 μm $MoSi_2$ gate to minimize current draw. Channel source (S) and drain (D) terminals are connected to the ASIC (Fig. S28c), with series access switches and parallel shunt devices connected to each device. Separate circuitry on the interposer (indicated by red elements) is used in the application of DAC-sourced write voltages ($V_g$) to device gates. One terminal of the gate (GND) is shared among all device gates as a common backplane and its value is set by a separate DAC. Each device has an individual gate select transistor in series for isolated writes and to avoid electrical disturb of other devices.

The input voltages applied to S are generated by digital-to-analog converters (DACs) on the PCB (red box) which are coupled into the ETCRAM array through the interposer substrate. The ASIC is bump-bonded to the interposer and the interposer is wire bonded to a chip carrier, allowing for simple plugging in and handling of the ETCRAM chip. The current outputs at D are measured using 9-bit on-chip analog-to-digital converters (ADCs) with current-based ramps that compares the current against a reference current to determine the current through the ETCRAM device. The ratio between the measured current $I_{out}$ and the input voltage $V_{in}$ gives the measured conductance G. Further design details of the 'Cairn' ASIC can be found in the following reference[32]. An Opal Kelly XEM7350 FPGA is used to control the DACs and handle digital communication with a PC.

**<u>Supplementary Note 22</u>**

<u>Multiple device retention measurements</u>

Long-term retention of multiple programmed devices was measured across 100 devices programmed on a single chip using the test platform described above. The average percent decrease in conductance from the initial state after a two-month period was 0.87% ± 1.6% across these 100 devices, based on the distribution plotted in Fig. S29. No change in the spread of conductance distributions was observed after two months. The test chip was coated with epoxy, and the chip was held at ambient conditions between measurements with no programming.

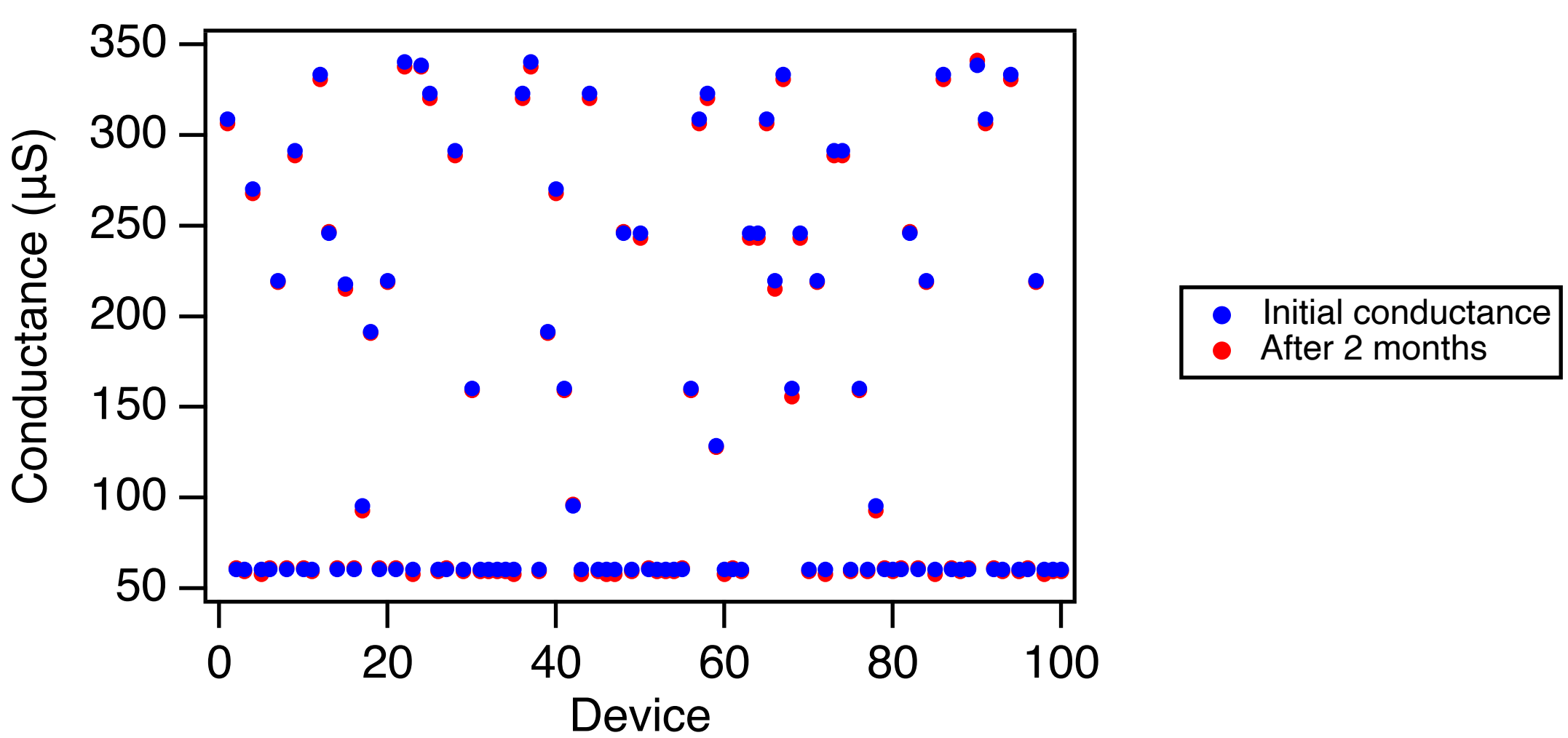

**Supplementary Figure S29.** Distribution of conductances of 100 devices after initial programming and two months later, showing an average percent decrease of 0.87% with standard deviation of 1.6%.